# Physical Principles of Quantum Biology

Nathan S. Babcock

with

Brandy N. Babcock

*For*

*Faithful Enthusiasts*

*of*

*Quantum Biology*

# About the Authors

Dr. Nathan S. Babcock is a distinguished expert in the quantum sciences, with over two decades of experience primarily concentrated in the domain of quantum biology. His academic journey began at two prominent Canadian quantum research institutions: the University of Waterloo in Ontario and the University of Calgary in Alberta. After completing his Ph.D. in Physics, Dr. Babcock further advanced his understanding of the intricate relationship between quantum mechanics and biological processes through postdoctoral research in structural biology at Simon Fraser University in British Columbia. His research then led him to explore spin chemistry, conducting groundbreaking studies on radical electron pair models of avian magnetoreception at the Living Systems Institute at the University of Exeter in the UK. Dr. Babcock continued to refine his expertise in open quantum systems by investigating the quantum mechanical phenomenon of superradiance in microtubules at the Quantum Biology Laboratory at Howard University in Washington, D.C., where his innovative research on quantum effects in microtubules garnering significant attention worldwide. His unwavering determination, comprehensive knowledge of the subject, and infectious enthusiasm for the pursuit of physics has given this book its fundamental core. Through his contributions, he aims to ignite curiosity and foster a deeper understanding of the interplay between quantum physics and the living world.

Brandy N. Babcock is an enthusiastic advocate for the sciences. Her journey began with an initial pursuit of Astrophysics and Theoretical Science at the University of Pittsburgh. Although she shifted her focus to health and wellness, where she thrived as a small business owner and coach for 15 years, her love for science never waned. During that time, Brandy honed her skills in small-scale project management, developed expertise in educational course creation and print media including publishing and contributing to multiple books, and gained extensive experience coaching others to achieve their personal and professional aspirations. Currently pursuing a certification in project management, Brandy is impassioned to re-engage with the scientific community in meaningful ways. Her experience in the holistic health field has made her acutely aware of the prevalence of pseudoscience, particularly the gross misuse of terms like "quantum healing." Alongside her husband Nathan, she is deeply passionate about educating the public on the true nature and significance of quantum biology, ensuring that accurate information prevails in the discourse surrounding this emerging field. Her unique blend of visionary thinking, organizational prowess, talent for coaching, and innate ability to communicate complex scientific concepts has made her an invaluable asset in the co-authorship of this book, where her insights and dedication have been instrumental in shaping its vision from inception to completion.

## Disclaimer

This book is the result of extensive research and collaboration between the authors, and so it is important to clarify the role of artificial intelligence (AI) in its creation. Generative AI was not used in any capacity to write the main content of this book. Nathan utilized AI-enhanced search engines critically in his research, which ironically often asked him to confirm that *he* was not a robot. Brandy employed AI assistance for the development of supplementary materials including the Sample Syllabus and the initial Glossary of Terms (prior to Nathan's revisions).

# Foreword by the Authors

This book was written out of necessity after the first *Gordon Research Conference on Emerging Methodologies to Investigate Quantum Effects in Biology* held in Galveston, Texas in March 2023. It was motivated especially by the comments of the graduate students performing research in quantum biology who attended the conference. As we spoke with them during meals at the event, they remarked that the field lacked cohesion as a whole, which presented challenges to those wanting to enter into research in quantum biology.

At present there is no textbook on quantum biology. This issue was raised at the closing "townhall meeting" at the conference, where it was widely agreed that a technical manual would be needed in order for quantum biology to develop into a mature scientific field. This technical monograph on quantum biology is not a textbook, because at the time of this writing there is not yet a consensus concerning the scope of quantum biology, even among professors and senior researchers who are actively working in the field. Before a textbook suitable to educate students can be written, a general agreement must be established among experts in the scientific field and its critics as to the extent of its subject matter and its physical basis. The purpose of this monograph is to furnish the basis for that consensus, as this book offers an overview of the foundational physics.

A diverse body of literature has emerged debating the importance of quantum mechanics to biology and the question of which quantum effects enable biological function. This lively debate has stimulated a great deal of scientific activity, both experimental and theoretical, by intensifying interest in individual topics as well as in the field as a whole. As stimulating as it has been, this productivity has come at the cost of clarity—presenting quantum biology as a patchwork of disjoint topics and piecemeal definitions, rather than a definite field of inquiry motivated by a unified theoretical framework and a consistent body of rigorously-established experimental results. It is our hope to address this by providing a comprehensive compendium of quantum effects in biology, presented from a fundamental quantum physical perspective.

This arXiv pre-print aims to establish a clear quantum physics basis for quantum biology. The pre-print will be open to suggestions for edits throughout the 2025 year to help foster a communal consensus in that regard. Feedback during that time may be sent to the email below. The book will then stay available for free on the arXiv in perpetuity to encourage the free and unfettered dissemination of ideas and the advancement of the field.

It was important that this monograph and the citations therein be accessible to those entering into the field, therefore it does not include any complex mathematics, numerical examples, or physical derivations to obscure the essential presentation of the concepts. Rather, it is written in a more-or-less colloquial style, with the intention of setting the form and structure for introductory courses on quantum biology. In this sense, it is an essential primer with a corresponding syllabus included at the back of the book, and a forthcoming series of online lectures presented by the first author on his YouTube channel https://www.youtube.com/@drbabcock to provide an expository overview of the essential material. A separate, forthcoming book is being created to impart quantum biology students with the necessary mathematical foundations.

Nathan is grateful to the many colleagues and professors who have inspired, encouraged, supported, and challenged his work on quantum biology over the last two decades including Betony Adams, Margaret Ahmad, Clarice Aiello, Janet Anders, Salil Bedkihal, David Beratan, Hans Briegel, Luca Celardo, Jeremy Choquette, Majed Chergui, Nina Copping, Victor Davidson, Paul Davies, Aurélien de la Lande, Art Du Rea, David Feder, James Freericks, Gilad Gour, Geoffrey Guy, Stuart Hameroff, Lucien Hardy, Stefan Idziak, Daniel Kattnig, Stu Kauffman, Robert Keens, Youngchan Kim, Philip Kurian, Peter Kusalik, Benjamin Lavoie, Tony Leggett, Elliot Martin, Rudy Marcus, Johnjoe McFadden, Alisdair McKenzie, James Murray, Sergei Noskov, Alistair Nunn, Hartwig Peemoeller, Drew Ringsmuth, Chris Rourk, Dennis Salahub, Chris Samson, Barry Sanders, Marlan Scully, Christoph Simon, David Sivak, Michael Skotiniotis, Sharon Strawbridge, Robert Thompson, Luca Turin, Vlatko Vedral, Nathan Wiebe, and to the many others who have kept this work in their thoughts and prayers.

Nathan S. Babcock

Brandy N. Babcock

nbabcock@gmail.com

# Table of Contents

# Part III: Nanomedicine & Biotechnology

# List of Abbreviations

**3D** three-dimensional

**ADP** adenosine diphosphate

**AlkB** alkylation B

**AMP** adenosine monophosphate

**ARDS** acute respiratory distress syndrome

**ATP** adenosine triphosphate

**CALI** chromophore-assisted light-inactivation

**CASSCF** complete active space self-consistent field

**CDT** chemodynamic therapy

**CI** configuration interaction

**CICR** calcium-induced calcium release

**CISS** chirality-induced spin selectivity

**cQED** cavity QED

**CT-QMC** continuous time quantum Monte Carlo

**DFT** density functional theory

**DMFT** dynamical mean-field theory

**DMRG** density matrix renormalization group

**DNA** deoxyribonucleic acid

**ECM** extracellular matrix

**EDT** electrodynamic therapy

**ELF** extremely low frequency

**EPR** electron paramagnetic resonance

**ET** electron transfer

**ETC** electron transport chain

**ETF** electron transfer flavoprotein

**FAD** flavin adenine dinucleotide

**FDA** Food and Drug Administration

**FeS** iron-sulfur

**FMO** Fenna–Matthews–Olson

**GDP** guanosine 5'-diphosphate

**GGA** generalized gradient approximation

**GMF** geomagnetic field

**GTP** guanosine 5'-triphosphate

**HER2** human epidermal growth factor receptor 2

**HF** Hartree-Fock

**HIF** hypoxia inducible factor

**HMF** hypomagnetic field

**HPD** hematoporphyrin derivative

**HSP** heat shock protein

**ICH** intracerebral hemorrhage

**IR** infrared

**ISC** intersystem crossing

**LDA** local density approximation

**LED** light-emitting diode

**LHCII** light-harvesting complex II

**LHN** light-harvesting nanotube

**LLLT** low level light therapy

**MFEs** magnetic field effects

**ML** machine learning

**MM** molecular mechanics

**MO** molecular orbital

**MR** multi-reference

**MSC** mesenchymal stromal cell

**NADPH** nicotinamide adenine dinucleotide phosphate

**NIR** near-infrared light

**NISQ** noisy intermediate-scale quantum

**NMR** nuclear magnetic resonance

**NOX** NADPH oxidase

**OXPHOS** oxidative phosphorylation

**PBM** photobiomodulation

**PCET** proton-coupled electron transfer

**PDT** photodynamic therapy

**PEG** polyethylene glycol

**PEMF** pulsed electromagnetic field

**PSII** photosystem II

**PTT** photothermal therapy

**QA** quantum annealing

**QED** quantum electrodynamics

**QM** quantum mechanical

**QM/MM** quantum mechanics / molecular mechanics

**RDT** radiodynamic therapy

**RET** resonance energy transfer

**RIRR** ROS-induced ROS release

**RNA** ribonucleic acid

**ROS** reactive oxygen species

**RPM** radical pair mechanism

**SCC-DFTB** self-consistent charge density-functional tight-binding

**SCF** self-consistent field

**SDT** sonodynamic therapy

**SR** stochastic resonance

**STM** scanning tunneling microscopy

**THz** terahertz

**TRIM** tissue resonance interaction method

**Trp** tryptophan

**UPE** ultraweak photon emissions

**UV** ultraviolet

**UVR8** UV resistance locus 8

**WMF** weak magnetic field

# Preface

As a quantum physicist, my interest in quantum biology was motivated by the hope that quantum mechanics could bring insight to biology. I was convinced that quantum theory would reveal the nature of life itself. In time, I began to see that the reverse was also true, that life is in turn revealing the nature of quantum physics.

Futurists have touted the promises of quantum mechanics, heralding it as the fate of science and technology. Quantum computing is hyped as the destiny of data processing, quantum cryptography is branded as the ultimate unbreakable zenith of communications security, and quantum information is identified as the fabric of reality itself—a modern *æther*.

Lifeless, cold, isolated systems are widely considered to define the essence of quantum theory. Meanwhile we overlook the open-ended aspects of our universe and our native biology, most notably the essential features of life: growth and regeneration, healing and development, acumen and instinct, optimization and synthesis. These are the biological characteristics that hold the spark to illuminating many of humanity's greatest questions.

Quantum biology challenges us to renew our understanding of the foundations of quantum mechanics. This return to established postulates presents the opportunity to formulate a framework for the foundations of quantum biology—a framework that identifies life as the exceptional feature of quantum physics. Rather than inquiring how isolated quantum systems operate independently from the environment, quantum biology asks the quintessential question, *How does life ultimately work?*

# Part I: Quantum Physics of the Living World

> One can best feel in dealing with living things how primitive physics still is.
>
> — Albert Einstein, quoted in Refs. [1, 2]

## Chapter 1 – Quantum Theory and the New Observables

To establish quantum biology in its proper context within the history of physics, we are recalled to an evening lecture at the Royal Institution in London on the 27th of April 1900, when Lord Kelvin lamented that two nineteenth century clouds hung over the horizon of science [3]. The scientific observations at the time were incompatible with predictions of the leading theories, presenting a crisis for the dynamical theory of heat and light:

> The beauty and clearness of the dynamical theory, which asserts heat and light to be modes of motion, is at present obscured by two clouds. I. The first came into existence with the undulatory theory of light, and . . . it involved the question, How could the earth move through an elastic solid, such as essentially is the luminiferous ether? II. The second is the Maxwell-Boltzmann doctrine regarding the partition of energy.

Kelvin's first metaphorical cloud ('Cloud I') was the motion of material bodies with respect to the ether [4]. At that time, a luminiferous ether was believed to enable propagation of electromagnetic waves such as light or radio waves (analogous to water waves traveling on the surface of the ocean). This ether was believed to exist everywhere throughout all space, leaving the nagging question of how material bodies moved through it without displacing it. Experiments by Michelson and Morley to measure the Earth's movement through the ether had failed, resigning Kelvin glibly to describe Cloud I as "very dense."

'Cloud II' covered the kinetic theory of the time, which could not explain the relationship between temperature and heat. Maxwell and Boltzmann had proposed that, all other things being equal, the energy of a thoroughly-mixed substance would be stored equally amongst its microscopic degrees of freedom. This "equipartition" principle worked well at predicting the amount of heat required to increase the temperature of many materials at high temperatures, but failed scandalously at low temperatures. Moreover, Kelvin felt that Maxwell and Boltzmann's proof of the principle was inconclusive, prompting him to deny the Maxwell-Boltzmann equipartition principle entirely.

The clouds darkening the 19$^{\text{th}}$ century theory of heat and light were dispelled with the advent of modern physics. The density of Cloud I was penetrated by the vision of Einstein, who proposed the principle of relativity that requires the speed of light to appear constant to all observers, regardless of their reference frames. As such, the concept of a three-dimensional ether was overtaken by that of four-dimensional spacetime [5]. Spacetime remains the term used today [6], while the idea of a "relativistic ether" [7, 8] has become taboo among physicists [9].

If that were not enough, Einstein dispelled Cloud II by imposing so-called quantization rules on the molecular motions of atoms in solids [10]. Those motions (i.e., degrees of freedom) become frozen at low temperatures [11, 12]. On the other hand, the spacing between the quantum energy levels becomes negligible at higher temperatures where the classical equipartition theorem was found to work well. Thus, energy quantization rationalized the failure of energy equipartition at low temperatures and in systems where conventional classical assumptions tend to fail.

No adherent to or believer in the classical theory of heat and light could have foreseen, let alone accepted, Einstein's solutions to the problems posed by Kelvin's Clouds. Even the pioneers who developed Einstein's work—Fitzgerald and Lorentz on special relativity and Planck on quantization—initially saw their own works as mathematical heuristics without fully appreciating their deep physical implications. The profundity of those implications was, however, appreciated by the chemist Nerst who called an emergency meeting of Europe's most eminent physicists to confront the emerging crisis in the first of the famous Solvay conferences. Nerst's invitation read [13],

> According to all appearances, we are now in the midst of a new development of the principles on which the classical kinetic-molecular theory was based. The systematic development of this theory leads on the one hand, to a radiation formula that disagrees with all experimental results; from this same theory are deduced, on the other hand, assertions on the subject of specific heats . . . that are likewise refuted by many measurements. It has been shown, especially by Planck and Einstein, that these contradictions disappear if one sets certain limits on the motions of electrons and atoms oscillating about an equilibrium position (the principle of energy quanta); but this interpretation in turn departs so much from the equations of motion used up to now that its acceptance would necessarily and indisputably entail a vast reform of our current fundamental theories.

That first conference in 1911 focused on the disconnect between classical physics and the new quantum theory—a problem which remains a thorn in the side of fundamental physics today. Over the course of the ensuing Solvay Conference series, the emerging quantum revolution stunned its discovers foremost because it did away with traditional notions of what observable quantities represent in science, or indeed if they represent anything at all [14].

Fully embracing the departure from classical concepts, Heisenberg formulated an abstract theory that admitted only observable quantities by discarding all empirical connections to unobserved quantities [15, 16]. Dirac recognized Heisenberg's essential departure from the old concepts and descriptions of physical reality and formalized it into a new system of algebra [17]. Unlike conventional algebra, Dirac's operator algebra was found to be equivalent to an advanced form of matrix algebra developed by Born and Jordan [18], which was later discovered to reproduce the wave theory developed by Schrödinger [19, 20]. So it was that quantum mechanics was born.

Kelvin himself had previously proclaimed that one's knowledge is meagre, unsatisfactory, and unscientific unless one can express it in numbers [21], heralding the need for a theory to numerate observable and unobservable quantities alike. Wildly successful as a measurement calculus [22, 23], quantum mechanics' failure as an objective theory of reality is well known [24–26]. Despite its enormous success, quantum mechanics is increasingly recognized as a source of consternation, rather than an authoritative guide that definitively clarifies the nature of reality [27]. This failure was felt early in the effort to unify quantum mechanics and relativity, as Dirac observed [16]:

> To have a description of Nature is philosophically satisfying, though not logically necessary, and it is somewhat strange that the attempt to get such a description should meet with a partial success, namely, in the non-relativistic domain, but yet should fail completely in the later development. It seems to suggest that the present mathematical methods are not final. Any improvement in them would have to be of a very drastic character . . .

The fundamental incompleteness of quantum theory in the relativistic limit was eventually addressed by the advent of quantum field theory in an overhaul of quantum mechanics that brought problems of its own [28, 29]. Those problems are exacerbated by the fact that quantum mechanics are based on a set of abstract mathematical axioms that are postulated *ad hoc* [30], rather than inferred directly from fundamental principles of Nature. Unlike Newton's laws which govern classical physics, and Einstein's principles of relativity (such as the finite speed of light), there are no universally accepted laws of quantum theory to preclude discoveries of whole new phenomena [31, 32].[1]

Perhaps the closest thing that quantum mechanics has to a physical law is an arcane mathematical property known as 'Hermitian symmetry' [20, 34] which preserves all relevant physical characteristics of a quantum system by requiring it to be bounded and closed in the sense of being completely isolated from its surroundings. Although this symmetry is widely considered fundamental to quantum theory [35, 36], it is rarely satisfied for long in systems that are open to the environment. This demanding condition leaves an enormous gap between the limits of conventional microscopic quantum mechanics and the everyday macroscopic phenomena of the classical world [37].

The gap separating the quantum realm from the scale of the classical world is usually so considerable as to leave no ambiguity between the scale of quantum mechanical particles and that of everyday classical objects. There are exceptions to the rule, some as simple as the ordinary bar magnet which holds its magnetic field by virtue of the quantum "spin" interactions hidden inside [38]. Yet the quantum-classical divide is bridged nowhere more vividly than in living systems—from microtubules and mitochondria to night-migrating songbirds and entire ecosystems—which rely on inherently quantum-dynamical effects to survive. To that effect, Schrödinger argued that the physics of his time was unable to account for living processes, posing the question of how living organisms made of atoms could overcome the tendency for all matter to degrade from order into disorder in his 1943 public lecture series, *What is Life?* [39]. He proposed that classical physics would be unable to explain molecular biology even in principle, contending that a quantum theory of biology would lead to the discovery of new physical laws simply because the construction of biological systems is so different from anything tested previously in the laboratory. In his own emphatic words [39], "The classical physicist's expectation, far from being trivial, is wrong."

By the 1960s, the most vital questions in quantum biology were already well known, although it still was not clear how to relate the abstract mathematics of quantum mechanics directly to its role in biology [40]. The issue at hand reflected long-standing concerns about the deficiencies of conventional quantum theory, dating back to the first Solvay Conference, and in turn recognized by Einstein [41], Schrodinger [42], Dirac [16], and later Deutsch [43], Bell [44], Leggett [45], Gell-Mann [46], and Weinberg [47]. The essential issue is that quantum mechanics only describes closed quantum systems which are theoretically observed by classical measuring devices [48, 49].

Ironically, "orthodox" quantum theory implies that the principles of quantum systems do not apply to measuring devices which are intrinsically governed by classical probabilities [44]. The fundamental distinction between quantum systems and classical measuring devices results from essential differences between open and closed systems, insofar as classical probabilities do not apply to closed systems [50]. This raises questions about whether quantum theory is either consistent [51] or exact [52] in a debate which dates back to original disagreements between Dirac and Heisenberg as to whether quantum mechanics as a theory is open or closed [53].

That apparent paradox presents no problem to conventional quantum mechanics, where there is a definite—if informal—divide separating the observer and the observed (known as the "Heisenberg cut" [54]). For a quantum system that is open to its environment, the problem is conveniently addressed by re-envisioning it as a closed quantum system that is observed by its classical environment. The convention begins to break down, however, for quantum systems that are open to interact with themselves through the back-action of the surrounding quantum continuum [55]. Heralded by Dirac, the failure of quantum theory to account for that "self-interaction" has long posed an obstacle to forming a complete and coherent interpretation of quantum theory [56, 57]. In the same vein, feedback through self-measurement and self-organization are hallmarks of biological systems [58–61], posing a fundamental challenge to quantum theory to provide a consistent description of self-referential systems [62, 63].

In this respect, biology presents fertile ground to investigate a new class of physical systems with properties quite literally at the boundaries of conventional quantum mechanics. In addition to established examples of quantum physics in biology such as the electronic properties of liquid water [64, 65], biological electron transfer [66, 67], nuclear quantum phenomena [68], enzyme-mediated reactions [69], collective spontaneous emission [70], biological (protein-protein) recognition [71], and the ultrafast (fs) dynamics of countless photochemical and phonon-mediated biological processes [67, 72–74], there is growing interest in the role of an expanding number of "exotic" non-Hermitian quantum effects that are involved in photosynthesis [75–77], coherent exciton transfer [70, 78], and photobiomodulation [79].

[1] To the contrary, even though quantum theory was derived by Heisenberg from the physical principle that the theory should not contain "relationships between quantities that are . . . unobservable in principle" [33], this point is rarely, if ever, taught to students as the founding assumption of quantum mechanics or included in standard textbook treatments on the subject.

In an echo of the first quantum revolution, the conventional quantum mechanics of last century are now being extended beyond the domain of isolated atoms and molecules to explore a range of open quantum systems [80, 81]. In general, the theory of open quantum systems describes the interaction of a quantum mechanical entity with its external quantum environment, sometimes characterized as a reservoir or bath [82], often with loss or gain [76, 83]. Invariably, living systems rely on the environment to procure energy and expel waste, ensuring that biological order develops robustly [84]. Recently, advances in open quantum systems have inspired renewed interest in foundational questions in relativity and quantum theory alike [85]. As such, quantum biology affords an ideal test bed for studies of open quantum systems and presents fresh opportunities to break through the cloud cover that still obscures a complete view of the fundamental principles of quantum theory. This view, in turn, promises to illuminate a growing list of rich physical effects with no counterparts in conventional quantum theory [86, 87].

Hence, we find no greater opportunity in physics today than to explore the emerging field of quantum biology, where we encounter the rich open system dynamics of gain and loss, life and death, growth and decay. Likewise, quantum biology poses unprecedented opportunities to observe, characterize, and ultimately understand many vital noise-driven and symmetry-breaking effects that are distinctive of living systems such as reproduction, healing, sickness, ageing and death. A radical new quantum theory of the 21st century—and the living world—promises to uncover new and extraordinary phenomena as life reveals more to us about quantum physics than ever before.

# Chapter 2 – Quantum Electrodynamics: Lighting Up Life

Sunlight provides the energy for life on earth through photosynthesis, yet the relationship between light and life is even more fundamental: the quantum unit of light (or "photon") defines *the* quantity of electron excitation. A bound electron is excited into motion by absorbing a photon of light. In this sense, light is conceptually indispensable from electronic energy [88]. It is not enough to claim that light powers life. Light is inextricable from the electronic energy driving the life force, fixed by the ratio of electronic energy $E$ to the emitted light frequency $f$. The need to fix $E/f$ to the constant $h$ (known as Planck's constant) is *the* "quantum" of quantum physics [89]. This concept was so stunning to the founders of quantum theory that Planck first thought it to be an artifact or a mathematical "trick" [90]. It took the genius of Einstein to realize that it was no trick, and Planck's formula $E = hf$ was there to stay. This had far reaching consequences, expressed famously by Heisenberg in the "uncertainty principle," which marked the departure from deterministic classical physics to the quantum realm of inherent uncertainty [91, 92].

The quantum nature of the electron's motion was observed by German physicist Heinrich Hertz in 1887 when he first noticed that the ultraviolet (UV) light from a spark produced by an electrified surface (i.e., an electrode) could induce a second spark from another metal surface, unlike light produced by other lower frequency sources [93]. Thus, the "quantum leap" was witnessed (via the observation of a first spark igniting a second one at a distance) but was not explained until Einstein subsequently published his theory for it in 1905 [94]. Today, the "photoelectric effect" provides an elementary example of quantum theory that continues to be demonstrated in classrooms and laboratories everywhere. The quantum theory of absorption and emission of radiation was later formalized by Dirac [95], before giving rise to the principles of quantum electrodynamics [96]. According to quantum electrodynamics, a quantum particle such as an electron can pass through a potential energy barrier that would be impassable according to classical physics. According to the quantum theory of tunneling, the wavelike particle can slip through a barrier even when it does not have enough energy to overcome it [97]. This is unlike the case in classical mechanics, where a particle must have enough kinetic energy to overcome an intervening barrier in order to pass across it.

Quantum tunneling is currently the accepted theory of electron transport during biological energy transduction [98]. During decoherent tunneling (sometimes referred to as incoherent tunneling [99]), the long-range coherence of the tunneling particle is destroyed after each quantum jump. Nevertheless, even decoherent tunneling is, strictly speaking, a coherent quantum effect with no counterpart in classical physics [100]. In contrast, resonant tunneling is widely described as "coherent" because it preserves the quantum mechanical phase of the electron after tunneling [99]. Although the question of whether living systems can exploit resonant tunneling has intrigued scientists for decades, the prevailing view is that most biological electron transfer steps take place over long distances (one to two nanometers) in a series of quantum jumps that occur by decoherent tunneling [98, 101]

Quantum electrodynamics enter biology during electron transfer between proteins [102] where electronic transitions deliver the electric force *via* the quantum "jumps" [67, 103, 104]. In other words, biological electron transfer provides the current of life, driving cell respiration as electrons tunnel quantum mechanically between the proteins that power the metabolism of all known organisms [105], ranging from cells to whole ecosystems [106, 107]. Interest in electron transfer reactions blossomed in the 1960s when those reactions were found to present much greater difficulty and complexity than previously imagined [108]. DeVault and Chance soon realized that quantum tunneling was the simplest way to explain the rate of activationless electron transfer in cytochrome proteins [109]. Although classical physics requires that an object must have sufficient kinetic energy to penetrate (or overcome) a potential barrier, quantum uncertainty allows an electron to spontaneously "tunnel" through an insulating barrier if the tunneling time is short enough [110]. This quantum mechanical restriction mitigates risk of electron escape.

Electron tunneling reactions were first characterized by Marcus who received the Nobel Prize in Chemistry in 1992 for that work [111, 112]. Biological electron tunneling has been established "indisputably" [102, 113], with a growing number of biologically-inspired technological applications [114]. The electronic excitation that accompanies light absorption during photosynthesis is just one of numerous examples of light harvesting and sensing in biology where light absorption is governed by the principles of quantum electrodynamics [115]. Photosynthesis has also attracted attention for the role that quantum mechanics plays in the efficiency of its energy conversion. Nearly every photon absorbed during photosynthesis generates an excited electron to perform work [116], garnering interest from physicists interested in rationalizing photosynthesis' extraordinary product yields ($> 95\%$ [117, 118]).

Electron transfer in biomolecules can be surprisingly efficient [119], and experiments have revealed that long-range coherent change transport is feasible in the chiral helix of deoxyribonucleic acid (DNA) [120]. Vibronic (nuclear-electronic) coupling has arisen as an essential factor determining the efficiency of charge transfer in biological and artificial systems alike [121], and model simulations have shown how control of inelastic electron tunneling can be used to gate long-range electron transfer using coupling-pathway interferences [122]. Likewise, protein junction experiments have shown that the shape of the respiratory enzyme azurin can switch conformations along with the kind of electron transfer through it [123]. Dynamical methods have also been used to switch between tunneling-mediated and solvent-controlled electron transfer in cytochrome *c* [124]. Resonant tunneling observed in electron transfer through azurin has revealed an active role for azurin's Cu(II) metal cofactor in coherent electron transport, illustrating the crucial influence of quantum mechanical effects on biological electron transport [125].

Quantum tunneling during respiration is just one example of the crucial role that quantum coherence and entanglement play in enabling the existence of life itself [67]. Coherence (*i.e.*, collective synchronization) is found ubiquitously in living cells [126] where quantum coherence is essential to allow long-range tunneling during electron transfer [127]. It is exactly the coherence of the electronic wave function—delocalized in space—that allows it to tunnel from one site of the electron transfer chain to the next. In other words, if the quantum coherence of the tunneling electron were completely destroyed then no tunneling would occur. This long-range coherence creates the state of an electron superposed across two distant locations, constituting an established form of quantum entanglement [128]—another hallmark of quantum mechanics [129]. Quantum noise also plays a role in the emergence of collective behavior [130, 131] with applications for complex systems biology [132]. The anti-reductionist idea that "more is different" [133] spawned work on quantum resonances [134] and related non-Hermitian effects (i.e., that violate Hermitian symmetry) [135]. This overall line of thought has likewise spurred a growing body of research on quantum dynamics in biology [136–142] with potent applications for medicine and the health sciences [143–147].

Effects of other quantum mechanical phenomena such as electronic spin are now recognized to have a profound effect on vital processes such as biological electron transfer [148]. Cellular respiration drives homeostasis [149], relying on the proton-coupled electron transfer of the electron transport chain (ETC) [103] (*i.e.*, for chemiosmosis [150]). In eukaryotes, a chain of respiratory complexes carry out cell respiration on the inner mitochondrial membrane [151]. High-resolution images of the respiratory complexes have now been obtained by cryo-electron microscopy [152–154], revealing highly-organized arrangements of the electron transfer chain complexes in mitochondria [155]. Life-giving functions such as photosynthesis in plants and nitrogen fixing in soil bacteria are also being increasingly recognized as processes that depend on quantum mechanical correlations due to quantum spin-exchange interactions [156].

Quantum theory is indispensable to the description of many biological effects, including the light-harvesting processes of biological chromophores excited during photosynthesis [157], light-sensitive chemical reactions that are responsible for vision [158–160], clinical light-based therapies [161], biological pigmentation [162, 163], ergosterol (vitamin D) production [164, 165], and a variety of bioluminescent effects such as the chemiluminescent mechanisms involved in firefly flashing [166–169]. Despite this extensive evidence for the existence of coherent quantum effects in biological systems, the so-called "warm, wet, and noisy" conditions of the biological environment are sometimes presumed to suppress non-trivial quantum phenomena in the cell [170]. This is ironic because those warm, wet, and disordered cellular environments reproduce the conditions for Brownian motion [171, 172], the topic of Einstein's second *annus mirabilis* paper of 1905 [173]. In another masterstroke, he used those very same features of the warm, wet, and noisy microenvironment to predict the existence of atoms.

This became a major triumph for quantum theory because the atomic model was still under dispute at that time. Prominent chemists and physicists including Ostwald and Mach opposed the theory of the atom so vehemently that the teaching of the atomic hypothesis was even banned in France [174]. It was Perrin's confirmation of Einstein's theory of Brownian motion—the random movements of microscopic particles in a liquid—that convinced many scientists of the reality of atoms while winning Perrin the Nobel Prize in Physics in 1926. Far from being conceptually trivial, the structural properties of liquid water remain among the most challenging of material properties to predict. This is primarily due to the ubiquitous presence of non-local quantum mechanical effects [64], where the complexity of the problem is such that several elementary chemical properties of water have only recently been solved [65].

Marcus' theory is crucial to understand the very high quantum efficiency of photosynthetic organisms [175]. The predictions of Marcus electron transfer theory differ dramatically from conventional transition state theories of reaction rates, qualitatively and quantitatively [111]. Notably, Marcus' theory predicts the "inverted" effect of an electron transfer reaction rate that decreases with increasing Gibbs free energy of product formation. This is in contrast with conventional reaction rate equations which predict only an increasing reaction rate with an increasing driving force. Marcus' inverted effect arises in the rate equation for a weakly-coupled chemical reaction [176]—derived as a consequence of Dirac's formulation of quantum electrodynamics [177]—where time-dependent perturbation theory is applied to obtain the reaction rate to second order (*i.e.*, using Fermi's "Golden Rule #2").

Nonadiabatic electron transfer dynamics are crucial to the efficacy of energy transduction processes in life, which are believed to control the rate limiting step of the metabolic quinol cycle (Q cycle) [178]. The Q cycle is indispensable to the process of aerobic respiration, wherein molecular oxygen ($O_2$) is reduced to form water ($H_2O$) in the production of adenosine triphosphate (ATP), the universal energy currency of life [179]. In fact, it has been established that proton uptake controls electron transfer in the enzyme cytochrome *c* oxidase [180], which catalyzes the energy-generating reaction step converting oxygen (O) and hydrogen (H) into water. In addition to the four protons that are used to convert molecular oxygen into water, cytochrome *c* oxidase transports another four protons across the respiratory membrane, increasing the acidity (pH) imbalance across the membrane. The resulting transmembrane protein gradient is exploited in turn by the the enzyme ATP synthase, which uses it drive the synthesis of ATP.

Respiratory enzymes optimize intervening media to enhance interprotein electron tunneling dynamics [181], and quantum interferences between optimized tunneling pathways have been identified in biological and biomimetic electron transfer systems [182]. When respiratory enzyme cytochrome *c* [183] binds with its redox partner, the binding mechanism modulates its electron transfer rate by modifying the transfer pathway [184]. Its partner enzyme, cytochrome *c* oxidase (*i.e.*, complex IV) uses redox state-dependent solvent organization to control a key proton transfer step. Advances in time-resolved laser spectroscopy have opened the possibility of observing electron tunneling in real time [185]. Likewise, ultrafast measurements of electron transfer dynamics in the electron carrier protein flavodoxin have revealed fundamental mechanisms of non-equilibrium dynamics that may be critical to enabling biological functions (including DNA repair) [186]. Thus, biological electron transfer mechanisms remain a topic of dedicated research [187], with renewed interest in mechanisms involving electron spin and molecular chirality [188, 189].

In physical terms, the electronic energy used to drive ATP synthesis is not any different than the energy of an electron excited by the photoelectric effect at frequency $f$. In this light, one might consider a photon being absorbed by a photosynthetic organism (such as a plants, algae, or photosynthetic bacterium) that uses sunlight to generate bioelectric power making sugar for food. While it might seem that electron and proton transfers must occur on different timescales due to the vast difference in inertia between them (the proton being over a thousand times heavier than the electron), simultaneous photo-excited electron–proton transfers have been observed [190] and are, in fact, essential to a plethora of diverse energy conversion processes in both chemistry and biology [191].

Thus, proton and electron transfers are intimately connected in biochemistry, where proton-coupled electron transfer (PCET) is an essential charge transfer mechanism [192], and electron and proton transfer steps may occur consecutively or in concert [193]. In mitochondria, for example, long-range PCET drives cellular respiration in the ETC [103, 194], where electron transfer cofactors are arranged into three proton-translating complexes (I, III, IV), along with succinate-dehydrogenating complex II [195] and ATP-synthesizing complex V [196]. The organization of these distinct complexes reveals the subtle quantum engineering principles [197] that govern the precise arrangement of electron and proton transfer sites [198], in turn controlling key physiological effects such as reactive oxygen species (ROS) production as a byproduct of electron transfer [155] when oxygen is not completely reduced into water.

Mitochondrial complex I is the largest and most complicated of the respiratory chain complexes [199]. As an excited electron traverses the ETC, it is first injected into mitochondrial complex I, where transits a column of iron-sulfur clusters in a proton-translocating mechanism that pumps protons across the mitochondrial membrane into a rich solvent matrix. Complex I couples quinone reduction with conformational changes across its four proton pumps [200]. Simulations have shown how a chain of $Fe_4S_4$ clusters in complex I enable its tunneling efficiency [201], but a consensus on the mechanism of proton pumping in the membranous part of complex I is lacking [202–204]. Real-time observations have revealed proton-coupled electron transfer via ubiquinone [205] in complex I [206–208], and a blueprint is emerging to explain coupling between electron transfer and proton pumping [202].

Electron bifurcation enables aerobic respiration in the Q cycle where it is crucial to oxidative phosphorylation, the process of synthesizing adenosine triphosphate (ATP) from its precursor adenosine diphosphate (ADP). During respiration, Coenzyme $Q_{10}$ ($CoQ_{10}$, otherwise known as ubiquinol) shuttles electrons from mitochondrial and bacterial respiratory Complexes I and II to Complex III, where an electron bifurcation reaction enables the simultaneous reduction of cytochromes *b* and *c* by ubiquinol [209, 210]. Cytochrome enzymes have been a topic of research interest for many years due to their central importance to energy transduction in all forms of life [211].

Proton-translocating, membrane-bound complexes I, III and IV interact with each other to form respiratory supercomplexes known as respirasomes [212]. Although the mechanism and purpose of respirasome formation is not entirely clear [213, 214], respirasomes have functional implications for the properties of the respiratory chain. Beyond the immediate kinetic advantage of proximity afforded by supercomplex formation, respiratory complex assemblies help to mitigate the formation of ROS during proton-coupled electron transfer [215]. The respiratory chain is a major source of ROS-generating electrons in the cell [216], and oxygen-based free radicals play a critical role in cell signaling, immunity and growth. However, excessive ROS production can cause a myriad of diseases [217].

Ubiquinone (a.k.a. coenzyme $Q_{10}$ [218]) is a major source of ROS with concurrent antioxidant properties [219]. It serves as an electron shuttle between complexes I, II and III of the mitochondrial respiratory chain [220, 221], where excess ROS production can result from impaired charge transfer from the electron transfer flavoprotein (ETF) enzyme to ubiquinone during fatty acid metabolism [222]. Ubiquinone deficiency constitutes a rare yet debilitating disorder [223] that is associated with various forms of disease, both mitochondrial and otherwise [224]. ETC dysfunction decreases the membrane potential, which compromises oxidative phosphorylation needed for ATP synthesis and impairs redox homeostasis, dysregulating an assortment of metabolic processes implicated in human disease [152].

Today, the main structure and core subunits of cytochrome *c* oxidase, its active site, and the transfer paths of electrons, protons, oxygen, and water are all reasonably well understood [225]. However, there are still uncertainties in certain aspects of the proton translocation mechanisms, including the exact nature of the proton pump's proton-loading site and the path the proton follows to exit the membrane [225]. Physical models of proton-pumping in respiratory and photosynthetic ETC are being developed to include aspects of many-particle and self-synchronization effects to fully account for the high quantum yield and thermodynamic efficiency of biological proton pumps [226]. Recently, a microscopic description of the Q cycle mechanism based on an open quantum system formalism has been proposed to account for its thermodynamic efficiency under physiological conditions [227].

Beyond its crucial role in respiration, biological electron transfer is essential to metabolic processes including enzymatic reactions controlled by biological photocatalysis [228, 229]. For example, the enzyme cytochrome P450 plays a essential role in many biosynthetic and metabolic reactions [230, 231]. Cytochrome P450 reductase serves as a key electron donor during the oxygen activation by monoxygenase enzymes, but the electron transfer mechanism has been difficult to decipher, requiring quantum mechanical calculations to determine the role played by the local electric field in enhancing biological electron transfer [232]. PCET reactions may appear to occur synchronously (*i.e.*, in one uniform reaction step without forming thermodynamically-stable intermediates), but on ultrafast timescales they entail asynchronous non-equilibrium dynamics [233]. This presents the inherent challenge of modeling PCET systems dynamically using electronically and vibronically non-adiabatic simulations with quantized protons [234]. Biocatalytic reactions often depend on the conditions of the biochemical surroundings including ionic interactions, van der Waals forces, and allosteric effects, making it necessary to incorporate comprehensive information from structural biology, enzyme kinetics, and spectroscopy to devise working models of biological catalysis [235].

Chiral molecules are ubiquitous in biology [236], and chirality-induced spin selectivity (CISS) effects also show promise to utilize the coherent properties of electron spin [237], where the spin polarization induced by this effect occurs dynamically and simultaneously with charge polarization [188]. These effects are biologically relevant, *e.g.*, in proteins containing heme where heme-enhanced spin polarization is predicted [238, 239]. CISS has been observed in various contexts that include photoinduced electron transfer systems, quantum dots, and biological molecules [240]. CISS effects have broad implications for catalysis, enantiomer distinction, long-range electron transfer, and biological recognition [241]. CISS occurs whenever molecular chirality controls the spin-polarization of electrons, presenting implications for spin transport at physiological temperatures ad functional spin control in dynamic environments such as the cellular ETC [240]. Coherent spin control may be essential for ROS regulation in the ETC at the point of ROS formation where magnetic effects govern the relative yields of ROS such as superoxide ($O_2^{\bullet -}$) and hydrogen peroxide ($H_2O_2$) [242]. CISS effects are typically rationalized by the quantum chemical interaction between electron spin and orbital angular momentum [243] during wavelike electron transfer [244]. As a consequence, these results also carry significant implications for chemical reactions involving radical species more generally [245]

Insights taken from quantum biology have begun to enable applications of quantum coherent processes in vitro such as artificial photosynthesis [246], yet fundamental advances in the theory of biologically active electron transport will continue to be needed for the development of present-day and future protein bioelectronics applications [247]. These findings highlight the essential character of the many quantum electrodynamic processes involved cell respiration and metabolism, and raise questions about the fundamental physiologically mechanisms underlying common therapies intended to target the mitochondrial ETC (including simple, over-the-counter health supplements such as coenzyme $Q_{10}$ [248–250]. The influence of these coherent, coupled electronic and vibronic processes on vital biological electron transfer reactions should not be underestimated.

# Chapter 3 – Definitions of Non-Triviality for Quantum Biology

The question of whether non-trivial quantum effects play a role in biology has emerged as a topic of extensive debate, although the meaning of the word "trivial" can vary widely in this context. For example, even though all the electrons in the universe are intrinsically and perpetually "entangled" due to the effect of being identical to and indistinguishable from one another, this entanglement is trivial in the sense that it has no observable consequences for most electrons that are far apart [251]. When two electrons come close together, their intrinsic entanglement becomes significant (*i.e.*, non-trivial) with observable consequences found in the structure of chemical bonds. Non-trivial entanglement between nearby electrons provides the basis for fundamental research in the field of quantum chemistry [252], where it is considered in the context of density functional theory using the mathematical formalism of exchange-correlation functionals [253, 254]. Nevertheless, the word "trivial" is sometimes used differently in quantum biology, where the role of quantum effects in determining biological structure has been overlooked [136].

Whereas discussions of quantum biological triviality have inspired a wide body of topical literature [136], a general lack of agreement concerning the meaning of "non-trivial" has been a cause of ambiguity, precluding any rigorous or tangible outcome to this debate. Incongruous or vague and inconsistent terminologies have obscured the preeminence of quantization, quantum coherence, and entanglement in determining practically every aspect of biological structure and function. The resulting equivocation has obscured the meaning and importance of quantum biology as a contemporary research area, with deleterious consequences for progress in the field [255]. It cannot be overemphasized that quantum mechanical effects such as coherence and entanglement determine the structure of biological macromolecules and enable the function of electron tunneling processes during cell respiration [66].

Writings on quantum biology are sometimes premised on the intuition that quantum effects in biology are primarily trivial and that implications for crucial non-trivial effects are rare [256]. However, this perspective is at odds with the extensive body of physical chemistry literature documenting the limitations of semiclassical principles used to model condensed biological matter and the myriad of challenges associated with implementing quantum mechanical corrections to semiclassical models [72]. The structures of biological macromolecules have long posed a challenge to computational chemistry [257]. Likewise, the challenges associated with modeling quantum chemical dynamics in biological matter are well known [72]. Whereas those phenomena designate non-trivial problems in quantum chemistry, quantum biology is far more than a subfield of physical chemistry. A growing body of work has increasingly implicated an ever-broadening range of quantum phenomena in living matter, drawn widely from research on quantum optics, photonics, and open quantum systems theory [75–77, 258–260].

In this respect, it is imperative to define "triviality" for quantum biology in a meaningful and concrete way to make contact with the sense of the word as it appears in other fields and to designate a set of best practices for its usage in quantum biology. To this end, let us draw inspiration from the concept of triviality as it is found in quantum field theory and connect it with existing concepts in quantum chemistry. Although solving for properties of biomolecules may be "trivial" in the sense that the characteristics of a quantum mechanical system of interacting electrons can be reproduced exactly by an artificial model of non-interacting ones under the influence of a fixed "exchange-correlation" potential, the quest to identify such a universal potential remains a central challenge of quantum chemistry [261]. By this metric, the hardest problems in quantum computing would also be "trivial" [262], including the complicated task of finding exact solutions to the time-independent Schrödinger equation (i.e., full configuration interaction calculations [263]) using a non-local exchange-correlation potential [264]. The non-local correlations of quantum theory are trivialized in return for solving a highly non-local potential function instead.

The burden of this dilemma is well known in the annals of quantum chemistry [265] where quantum chemical approximations are ordered in a ladder of increasing complexity, from effectively-trivial classical potentials based on electron densities to complex functionals incorporating non-local exchange effects [266]. As a model's complexity increases, its range of parameters expands from the local electron density alone to account for such other factors as spin, density gradients, kinetic energies, and quantum correlations imposed as contributions from physically-unrealistic empirical parameters [264]. In practice, the applicability of using a semiclassical local density approximation (LDA) is extremely limited [267]. Therefore, when assigning degrees of triviality in quantum biology one must consider model complexity as laid out in quantum chemistry and physics, first in terms of time-independent approximations and subsequently in terms of time-dependent dynamics of increasing complexity.

So-called "trivial" quantum effects that determine ground-state electronic structures of biomolecules often reflect globally-entangled and highly-correlated (coherent) quantum states of matter [268]. The enormous challenge of simulating biomolecules has prompted increasing interest in the use of quantum computers to solve many key problems in structural and functional biology [269]. Even though structural questions of quantum biology represent highly non-trivial open problems for other fields, they've often gone unrecognized in defining an essential subfield of quantum biology [136], where instead they have been designated as "trivial" in the following technical sense.

One definition of triviality in quantum biology designates the case when the quantum *dynamics* of the system are trivial (i.e., stationary or negligible) [270], even if underlying quantum correlations are not. This is convenient, from a classical point of view, because it allows one to devise a static (or slow-moving) picture of a biomolecule using quantum mechanics, which can subsequently be modeled using only classical trajectories. In this approach, one solves for the physical properties of a biomolecular system using quantum mechanics in order to estimate values of the physical quantities that subsequently define parameters of a classical model. Quantum theory is "central but trivial" insofar as the model system's dynamics are classical even if its parameters are not [271].

That dynamical concept of triviality is inadequate to characterize quantum biology because many static examples that it designates as "trivial" define notoriously hard problems or even grand challenges of other areas [272]. As a prime example, consider a common biological molecule such as caffeine [273]. Although the electronic structure of a caffeine molecule may be considered trivial in the sense that its quantum electrodynamics are stationary, the full quantum simulation of the caffeine molecule represents a major outstanding problem at the limit of present-day quantum information processing applications [274]. Quantum simulations of the structure of atmospheric nitrogen-fixing enzyme nitrogenase, while trivial in the same electrostatic sense [271], are far beyond the capabilities of present-day classical and quantum computers.

To develop a concept of triviality more suitable for quantum biology, let us draw from quantum field theory where the idea of triviality was proposed by Landau and coworkers in the 1950s [275]. In that context, a system of many interacting particles is considered "trivial" if its properties can be reproduced by an equivalent system of non-interacting particles [276]. The many-particle theory becomes trivial in the sense that its predictions are no different than an equivalent one-particle theory. This "mean field" approach already provides the most common method for simulating the electronic structure of matter, known in quantum chemistry as density functional theory [277].

Density functional theory is based on the assumption that the many-electron density function describing the electronic structure of a system may be replaced by an equivalent density of non-interacting electrons. This mean field approach to solving a many-body problem—by considering the average of an ensemble of isolated one-body systems—was first adopted as a formal method in quantum chemistry by Hohenberg, Kohn, and Sham [278], where the electronic structure of a system of interacting electrons is represented as a function of its density only [279]. Given that the ground state of the system and all of its properties may be uniquely determined from its electron density [280], a system of interacting electrons is re-imagined as a fictitious system of non-interacting ones suspended in a classical force field that reproduces the same electron density as the true many-electron system. This results in a set of one-electron solutions to the many-electron problem, also known as the Kohn–Sham equations [281].

In principle, the one-electron Kohn–Sham equations provide an exact solution to the many-electron problem. This would appear to suggest (as Hopfield and others have [136, 271]) that biochemical structure models are "trivial" according to the formal definition of the term provided by quantum field theory. However, this naïve view overlooks the fact that even though the Kohn-Sham equations can in principle provide a semiclassical solution to the exact quantum mechanical many-electron problem, that in practice this method relies on the construction of a fictitious potential function that will precisely reproduce the features of the exact quantum mechanical correlations [281]. However, finding a semiclassical approximation to reproduce the exact many-electron density function is now known to be as difficult as solving the hardest verifiable problems in quantum computing [262].

To lowest order, quantum mechanical methods used in electronic structure calculations typically begin with mean field theories that do not account for long-range electron correlations, despite the fact that these correlation effects are necessary for realistic simulations of many condensed matter systems [282]. Numerous efforts have been made to reproduce quantum mechanical exchange and correlation effects using effective classical potentials, with very limited success [283]. Rather, accurate quantum chemical studies of biological systems have proven intractable because the size and complexity of biological molecules prohibit the elucidation of large-scale biological structures and processes using full-scale quantum-mechanical calculations [257]. As such, the study of exchange-correlation functionals came to represent a central problem in molecular structure theory because there is no general solution to the problem of finding a classical function that reproduces all the consequences of quantum mechanics [265].

The electrostatic case where quantum dynamics are neglected from the molecular model is known as the Born-Oppenheimer approximation, which is derived using the quantum adiabatic theorem in the limit of the zeroth-order approximation in time [284]. In this framework, stochastic transitions (*i.e.*, quantum jumps) between adiabatic "Born-Oppenheimer" quantum states are the first-order corrections to the zeroth-order limit. Jumps are modeled using a time-independent likelihood that predicts only a transition rate (i.e., the probability of the quantum jump occurring in time [115]). In biological matter, these transitions are not trivial according to any of the definitions of triviality given above. Nevertheless, quantum jumps are not always recognized as coherent quantum effects because the jump rate is described using classical probability theory—once the Born-Oppenheimer potential surface has been solved using time-independent quantum mechanics *and* the simple hopping probabilities have been obtained to first order by Fermi's Golden rule. This is the basis for Marcus' Nobel Prize-winning work [112].

Quantum biological effects include electron transfer steps between proteins in the cell respiratory chain [285], where Marcus theory plays an essential role in accounting for the free energy optimizations in electron transfer proteins [286]. For example, electron transfer drives proton translocation during respiration, transferring hydrogen nuclei across the membrane to generate a proton gradient [200]. However, biological energy transfer is not limited to spontaneous surface hopping [287], and electron transfer can involve complex correlations between electronic and nuclear motions [288]. Thus, higher-order quantum electrodynamic effects are often accepted as non-trivial because they involve explicit time-dependent quantum dynamics [270]. This reveals how triviality is in the eye of the beholder, where it is not so important how one draws the (ultimately arbitrary) line between "trivial" and "non-trivial." Instead, it is crucial to define a set order of approximations to describe quantum effects in biology. Ordering phenomena from most to least trivial is essential to characterize the complexities of quantum effects when extrapolating beyond the low-order semiclassical approximations that underpin most biochemical models.

Modeling biological systems presents a challenge foremost because of the ubiquitous presence of weak binding and charge transfer effects which are not properly described by conventional methods in density functional theory (i.e., local electron densities or generalized gradients) [289], although some approximations developed for biological systems have successfully determined the structure and dynamics of assorted peptides, ligands, and DNA [289, 290]. Nevertheless, these approximations fail to describe most long-range dispersion effects in biomolecules [290], where non-covalent molecular interactions such as dispersion are key to determine many important biochemical properties [291]. More recently, molecular orbital methods have allowed efficient quantum mechanical calculations of electronic states of macromolecules [292], enabling predictions of many life science phenomena *a priori* with applications to drug design, molecular recognition, and structural biology (such as characterizations of metalloprotein interactions) [293].

Combined approximation methods have achieved predictions of complete protein structures with reasonable accuracy [294], though few studies have been able to deliver quantitative accuracy when predicting biomolecular energies [295, 296]. As such, structural models often fail to deliver accurate information about the biochemical transition states, protein domain movements, and conformational changes that enable enzymatic reactions [297]. Consequently, applications of quantum chemistry methods to biological systems are supplemented by advanced methods in quantum mechanics, typically beginning with Møller-Plesset perturbation theory [298]. For example, fragment-molecular-orbital density-functional tight-binding calculations have enabled structure-based analyses of molecular interactions associated with the deadly SARS-CoV-2 virus. Far from trivial, these calculations are carried out on vast super-computers which incorporate high-order quantum mechanical corrections (e.g., fourth-order Møller-Plesset theory) to account for the contribution of quantum coherence and electronic correlations to vital biological processes [299].

In closing, the concept of "trivial" as inconsequential, negligible, or stationary does not reflect on the scientific significance or life-sustaining importance of quantum effects in biology [138]. Even though many seemingly trivial effects (such as the existence of stable nuclei, atoms, and molecules) may be defined according to classical electrostatic models, they still cannot be predicted by classical theory *a priori*. This seemingly-classical electrostatic picture of molecular biology is still fundamentally dependent on quantization, coherence, and nonlocality (i.e., single-particle entanglement [128]). With this classical picture in mind, we stress the importance of trivial quantum effects. These effects constitute an indispensable sub-field of quantum biology because they lay the foundation for describing all other non-trivial effects. In quantum biology, as elsewhere in physics, trivial effects are *most fundamental*.

# Chapter 4 – Photosynthesis & Open Quantum System Dynamics

Sunlight is the ultimate source of energy for life as we know it [300]. From elementary phototrophic archaea, cyanobacteria and phytoplankton to the most complex plants, the process of photosynthesis enables life on Earth by converting solar energy into chemical form [301, 302]. Like cell respiration, photosynthetic processes are described by the principles of quantum electrodynamics [259]. Although the structural components and regulatory mechanisms of photosynthesis are well characterized, physiological consequences of the fundamental processes that control photosynthetic light-matter interactions and electron transport phenomena remain challenging to predict [303].

Experiments on green algae dating back to the 1930s showed that many chlorophyll pigment molecules would be required to harvest enough light to produce a single oxygen molecule from water during photosynthesis [300]. This prompted the idea of inter-molecular exciton transfer of photosynthetic quanta [304], later formulated into the contemporary picture of an aggregate of light-absorbing pigments that act collectively as a light-harvesting antenna [305]. The high efficiency of the process inspired a theory that quantum coherence could enable energy transfer between the antenna pigments before funneling it to the photosynthetic reaction center [306–308].

Interest in the role of coherent quantum effects in photosynthetic processes was spurred by findings of wavelike energy transfer in a photosynthetic pigment system known as the Fenna–Matthews–Olson (FMO) complex [309]. The FMO complex is a pigment-rich protein found in light-harvesting antenna systems of green sulfur bacteria [310]. During bacterial photosynthesis, the FMO complex transfers exciton energy from the light-harvesting antenna to the reaction center, making the FMO complex an exemplary system in which to investigate the quantum mechanics of the pigment-protein structures that occupy life's photosynthetic complexes [137, 310]

Coherent oscillations were initially discovered in the FMO complex under cryogenic conditions [311, 312],

and were later confirmed to exist at physiological temperatures as well [313]. A series of high-profile studies ensued, bolstering an idea that coherent wavelike energy transfer could enable highly-efficient bacterial photosynthesis [312–318]. Subsequent studies provided estimates of the electronic structure and population dynamics of the complex, confirming the delocalization of excitons across multiple pigments [306] and establishing that the coupling of nuclei to electronic excitations initiated long-lived coherent nuclear vibrations in the FMO complex [319, 320].

This revealed that observations of long-lived coherences were not due to superpositions of excitonic states after all, but were induced by vibrational–excitonic (vibronic) couplings [321, 322]. It is now recognized that observations of coherent quantum dynamics longer than 600 fs are far in excess of known estimates of exciton lifetimes which are no longer than 100 fs [323, 324]. This established conclusively that long-lived coherences in the FMO complex result from hybrid vibronic states involving highly-coordinated interactions between photo-induced excitons and nuclei [325], thus implicating an unexpected role for hybrid quantum dynamics during photosynthesis [320].

Interest in the influence of coupled electronic and nuclear motions on photosynthetic energy transfer [157, 326] grew with acceptance that the coherent oscillations observed in the FMO complex persisted much longer than the lifetimes of superpositions of individual exciton states (i.e., excitonic coherences) [325, 327, 328]. In the FMO complex, interactions between initially-excited electronic states and the surrounding nuclear vibrations have a screening effect, preserving the coherent dynamics from the destructive effect of the surrounding environment and suppressing the rate of decoherence with respect to that of the unscreened excitons [329]. The coherent mixing of excited electronic (exciton) and nuclear vibration (phonon) modes of motion creates hybrid (polaron) quantum states in a far more complex and nuanced picture of the nuclear/electronic phenomenon than in the original theory of long-lived exciton oscillations [329, 330].

Instead, the long-lived oscillations are now interpreted as vibronic coherences, induced by the initial excitonic impulse through a resonance process that transforms the nuclear and electronic subsystems into an inseparable (*i.e.*, entangled) quantum superposition [137, 331]. Substantial coupling between excitonic states and the vibrational environment enable dissipation that drives energy transport in photosynthetic systems [332, 333]. Rather than suppressing environmental dissipation, photosynthetic systems were shown to exploit environmental interactions to reduce the impact of structural variations on energy transport efficiency [334]. Further studies showed that specifically-tuned vibronic interactions (mediated by cysteine amino acid residues) could control the degree of resonance between photo-activated excitons and pigment vibrations in order to funnel potentially-harmful excess energy to quenching sites on the periphery of the protein [117, 310]. Similar results were found in studies of photosystem II (PSII)—another pigment-protein complex found in cyanobacteria, red and green algae, and plants [302]—where an energy-dependent quenching mechanism governs the photosynthetic response to varying sunlight levels by regulating light-harvesting [307]. This dedicated quenching mechanism mitigates the risk of light-induced damage by judiciously controlling the amount of solar energy that is dissipated as heat during photosynthesis [335, 336].

Just as ultrafast spectroscopy experiments have ruled out the presence of long-lived electronic dynamics in the FMO complex, vibronic coupling was also found to facilitate ultrafast energy transport in light-harvesting complex II (LHCII), a key component of photosynthetic light-harvesting systems in green plants [337]. Although models of exciton dynamics may produce reasonable estimates of experimental spectra by including thermal fluctuations, exact simulations of photosynthetic light-harvesting typically require much more detailed models of system-environment coupling to account for the full spectral density of an antenna system embedded in its quantum environment [330].

In practice, derivations of quantitative macroscopic properties from underlying microscopic processes often resort to a wide variety of methods that range from multi-scale theoretical modeling to machine learning and brute-force trial and error [338]. The failure to apply quantization principles correctly may result in an inappropriate classical description (e.g., by violating Newton's Laws [339]) [340]. However, when performed correctly, studies of quantum dynamics underpinning vibrationally-assisted energy transfer can inform classical models of the photosynthetic energy funnel [327, 341]. For cases where underlying quantum dynamics can be approximated classically (e.g., using molecular mechanics [342]), results obtained from classical approximations can be sensitive to the choice of a classical force field because effective classical models may not uniquely recover the results of quantum mechanics. In-depth knowledge of quantum theory may then be required to derive the appropriate classical limit from the underlying quantum dynamics.

Beyond quantum mechanics' broadly established role in enabling vibrationally-assisted exciton transfer [339] and the growing appreciation for its importance to vibronically-driven excitation processes [343], organisms may also harness quantum effects to gain adaptive advantages in the ecosystem [138]. For example, the quantum efficiency of long-range photosynthetic electron transfer in cyanobacterial photosystem I (PSI) depends critically on long-range tunneling-mediated electron transfer by way of the Marcus 'inverted effect' wherein the charge transfer rate counter-intuitively decreases as the force driving transfer grows [175]. As an often sought-after example of a quantum biological process without any classical counterpart [344], inverted electron transfer also typically involves nuclear quantum effects which cannot be ignored [67, 345]. Cyanobacterial PSI relies on the inverted effect to enhance photosynthetic efficiency by preventing the unproductive back-transfer (recombination) of the excited electron to its initial state, unlike many other photosynthetic systems [175]. Results of this kind reveal why coherent vibronic interactions are increasingly recognized as essential to adaptations that enable biological energy transduction [343].

These examples also show how the high efficiency of photosynthetic light-harvesting is not due to the efficiency of the electron transfer itself [334, 346], but its directionality: Nearly 100% of the excitons reach the reaction center (rather than escaping or recombining). Different photosynthetic apparatus achieve this in different ways, either by promoting forward transfer vibronically (as in FMO complex and LHCII) or by discouraging back transfer by the inverted effect (as in cyanobacterial PSI [175]). Quantum dynamics are also implicated during light-harvesting in photosynthetic purple bacteria [347] where they are relevant to exciton migration [348]. Electron transfer reactions in purple bacteria exhibit a variety of deviations from Marcus theory, which have been attributed to conceptual and technical problems with predicting the electron transfer factor [349]. Problems with predicting electron transfer factors are compounded by influences of nuclear vibrations and coherent relaxation processes, motivating increasing attention to the advantages that collective coherent quantum dynamics can impart to these processes [350].

Coherent relaxation processes like those found in the FMO complex [306] are just one example of a broader class of phenomena found in open quantum systems, as exemplified in the physics of "superradiance" (i.e., collective spontaneous emission) [351]. Superradiance has been observed in the photosynthetic light-harvesting systems of purple bacteria [352] and the chlorosomes of green bacteria [353], where collective coupling can improve the likelihood of energy capture by enhancing the rate of electron/energy transfer to the reaction center [354]. This improves prospects for the initial light capture through non-local pigment-excitation effects in the related process of superabsorption (i.e., collective light absorption [355]) [300, 306]. Superabsorptive light-harvesting was shown to benefit from the presence of environmental noise which can be tuned to enhance energy trapping and storing [355], just as molecular vibrations can enable the distribution and storage of energy during photosynthesis [344].

Photosynthetic processes have taken on an exemplary role in the research of open quantum systems [75–77], where intrinsic molecular vibrations found in biological light-harvesting systems do not only influence the optical responses of these systems, but also drive exciton transport on which they rely [332]. Far from being detrimental, noise and losses can be coordinated to amplify spectral intensities, suppress fluctuations, and enhance coherence in resonator systems reminiscent of photosynthetic pathways [356]. The same dissipative processes are increasingly becoming recognized as quantum resources that can be engineered to enable key quantum information processing tasks such as quantum state preparation, stabilization, and measurement in quantum computing applications [357]. In other words, the environmental interactions that were once believed to destroy quantum coherence in biological systems are now being recognized as non-trivial quantum effects in themselves.

# Chapter 5 – Light Receptors, Spin Chemistry, & Cryptochrome

The reception of electromagnetic signals is common in the biological world, where photoreception extends far beyond phototropism and light-harvesting (such as photosynthesis) to include a wide range of light-receptive processes including vision, circadian photoentrainment, photobiomodulation, plant photomorphogenesis, and plasmodial phototactic responses [358, 359]. Optical photoreceptors found in the eye carry out the first step in a visual phototransduction process that is fundamentally described by principles of quantum electrodynamics [360, 361].

Rod and cone cells are the two types of photoreceptor cells that allow vision in vertebrates, enabling low light and color vision, respectively [362]. Despite key functional distinctions, rod and cone cells have comparable photo-efficiencies and active lifetimes [363]. Until the year 2000, rod opsins (rhodopsins) and cone opsins (photopsins) were the only two types of opsins known to exist in the mammalian retina. Beginning in 1998 with the discovery of a new type of opsin in the light-sensitive skin cells of African clawed frogs [364], a series of findings led to recognition of a third opsin receptor in the eye. Dubbed melanopsin for its role in modulating skin pigmentation, this blue-light photoreceptor has also now been identified in the retina, brain, blood, and adipose tissue [365].

Opsins are G-protein-coupled light receptors in visual and non-visual light-sensing systems of animals [366], where they typically activate guanine nucleotide-binding proteins (a.k.a. "G proteins") via the photo-isomerization of retinal. More than one thousand opsins have been categorized into seven subfamilies. Rhodopsin, in particular, is considered a prototype opsin for its role in low-light sensing and peripheral vision, enabled by high photo-absorption efficiencies that make it capable of single-photon detection [158, 363]. Rhodopsin gains its light-receptive properties from the prosthetically-bound pigment molecule retinal, which initiates phototransduction upon photoabsorption, electronic photoexcitation, and photoisomerization that triggers a neuronal signalling cascade by separating retinal from the surrounding opsin [360]. As such, retinal photoisomerization in rhodopsin is the topic of fundamental studies that combine computational techniques such as atomistic modeling and hybrid quantum mechanics/molecular mechanics (QM/MM) with ultrafast spectroscopy experiments to infer biological design principles [159].

Rhodopsin has also been suggested as a structural model of olfactory receptors [367]. Although about half of all known olfactory receptors are G-protein-coupled receptors (GPCRs) like rhodopsin, high-resolution structural models of olfactory receptors have been experimentally difficult to obtain. GPCRs are cell surface receptors that detect particles outside the cell, in turn activating cell responses. Each GPCR consists of seven $\alpha$-helical protein segments that fold back and forth through the cell membrane in a series of six loops, with three intracellular loops interacting with G proteins and three extracellular loops interacting with external ligand molecules. Olfaction is believed to rely on the formation of ligand-receptor hydrogen-bonding networks where the odorants provide ligands that serve as electron donors and/or acceptors in the bonding network [367], thus enacting a mechanism by which to relay external signals into the cell. Like other light-dependent GPCRs, melanopsins are pigments that activate their associated G proteins when exposed to light [368]. Similar to other opsins, melanopsin contains a light-sensitive vitamin A aldehyde, 11-cis-retinal, which is photoisomerized to form all-trans-retinal [365].

Unlike the ocular pigments rhodopsin and photopsin, melanopsin is not involved in visual perception. Instead, melanopsin modulates a range of other non-image-forming processes that include circadian regulation, sleep cycles, and pupil dilation [369]. Originating with the Latin expression *circa diem* [365], circadian rhythms are present in most living organisms [370] where they coordinate day-to-day metabolism and physiology [371]. Complex and intrinsically connected to a wide range of biological functions including psychomotor coordination, sleep, digestion, and mood, the circadian clock system has been linked to practically all aspects of health and disease [365]. However, melanopsin does not work in isolation when entraining the circadian clock to daily cycles of light and darkness. Discoveries made in vitamin A deficient mice [372] prompted the hypothesis that an unrelated class of pigments known as "cryptochromes" may also play a central role in circadian light entrainment in mammals [373, 374]

Cryptochrome, a sensory photoreceptor protein, is the primary candidate magnetoreceptor in animals because it is one of the only vertebrate proteins known to generate the reactive radical pairs needed for the chemical operation of an inclination compass [375]. Cryptochrome's central role in circadian regulation inspired the hypothesis that it could act as magnetoreceptor because circadian rhythms were known to respond to magnetic field variations [376]. Following work of Schulten *et al.* [377, 378], Ritz *et al.* hypothesized that the recombination of photo-generated radical pairs could enable magnetic field reception in cryptochrome [376]. That model of magnetically-sensitive radical-pair dynamics in cryptochrome became known as the *Radical Pair Model of Magnetoreception* [379–381].

That model rose to prominence after it predicted the disruption of the avian compass sense by $1 - 10$ MHz radio frequencies [382–387]. This was borne out in findings that birds' magnetic orientation with respect to the Earth's magnetic field could be disrupted the by radio frequency field noise in the 0.1 to 10 MHz range [383–385, 388–391]. A vast body of theoretical and experimental work was subsequently carried out to develop and test models of radical pair-mediated magnetoreception. A cryptochrome-dependent magnetic sense was demonstrated in fruit flies [392–397], cockroaches [398], and plants [399–402]. Extensive studies of magnetoreception were also performed on birds, employing comprehensive experimental controls to test numerous aspects of avian magnetic faculties [382–385, 387, 391, 403–422]. Further studies suggested that weak broadband electromagnetic fields in the MHz range are more disruptive to avian magnetic compass than strong narrow-band fields [423], whereas 0.1 to 100 kHz noise was found not to disrupt the avian compass sense.

Ritz *et al.*'s theory prompted a feasibility analysis indicating that in principle any detection limit could be satisfied by a sufficiently-sensitive radical-recombination reaction rate [424]. This theory therefore became a popular way to interpret animal magnetoreception because it could rationalize the influence of a magnetic field as weak as that of the Earth on a physiological reaction [379, 425]. As a result, magnetic senses in birds and other animals are now widely believed to be mediated by the interconversion between excited "singlet" and "triplet" electronic states of charge-separated radical pairs produced in cryptochrome by photo-excitation. The magnetic field dependency results from differences in the singlet (radical recombination) and triplet (free radical escape) yields of the reaction product. Singlet-triplet interconversion rates are modulated by variations in the external field because the effective strength of each radical's interaction with the field is controlled by its local magnetic environment (*e.g.*, via local anisotropic hyperfine [426, 427], spin-orbit [428], dipolar [429], and/or exchange interactions [430]).

Today, the most widely accepted theory of the avian magnetic sense is based on a specific radical pair reaction, known in spin chemistry as the radical pair mechanism (RPM) [379]. The RPM is a magnetically-sensitive chemical mechanism that is light-driven, insensitive to magnetic field polarity, responsive to a range of field intensities, and vulnerable to radio frequency (rf) field noise [429], consistent with the findings of experiments on the avian magnetic sense. Three decades after the RPM was proposed in 1969 [431, 432], it became associated with the photoreceptor protein cryptochrome because cryptochrome in birds eyes can generate the radical pairs that theoretically enable the avian visual magnetic sense [376]. Ritz *et al.* invoked the RPM to form the hypothesis that the avian compass sense depends on a delicate balance between singlet and triplet quantum states in the cryptochrome radical pairs. Electromagnetic field noise disrupts this balance by driving transitions between the radical pair quantum states.

The RPM describes the hyperfine-mediated chemical effect of a nuclear magnetic field on the rate of radical pair recombination for a *geminate* pair of radicals "born together" by photoexcitation in either a singlet or a triplet state. If at least one radical contains a nucleus with spin, the nuclear magnetic field modulates the rate of interconversion between the singlet electronic state (which can recombine) and triplet states (which cannot) [433]. The external influence of an weak magnetic field has a symmetry-breaking effect which splits the energies of the otherwise-degenerate singlet and triple states [434]. If the radicals are born as a singlet, then the influence of a weak magnetic field is to decrease the rate of recombination by transforming some singlets into triplets which escape as free radicals. For triplet-born radicals, a weak field enables recombination instead. The field-induced energy splitting between these states make them vulnerable to interference by radio frequencies in the MHz range, near the Larmor frequency of electronic spin precession about the Earth's field.

The conversion between excited singlet and triplets is known as an intersystem crossing (ISC). When ISC occurs, the presence of an ambient (*i.e.*, external) weak magnetic field will lift the degeneracy between the three triplet states while in turn modulating the strength of the coupling between $S_0$ and $T_0$ [434]. Modulation of the rate of exchange between $S_0$ and $T_0$ spin states by a weak external magnetic field is the defining feature of the RPM and other similar mechanisms in chemistry [435], where radical pairs are typically formed with the breaking of a chemical bond. If the singlet $S_0$ state of a bound pair of electrons is severed to release a pair of radicals, each containing an open shell spin-half electron, then the two electrons recombine in a charge transfer step that recovers the original bound singlet configuration. However, the two free electrons cannot directly recombine after transitioning to any of the three triplet states (due to the Pauli exclusion principle).

Despite its success [436], the RPM has not been adequate to interpret many magnetic field effects (MFEs) in biology [437]. The effects of dipole-dipole radical and electronic exchange interactions, ignored in most models of the RPM, tend to render the effect of the RPM negligible under realistic conditions by suppressing its anisotropy [429]. However, anisotropic MFEs involving cryptochrome can also be markedly enhanced in the presence of scavenger radicals [438], prompting proposals that the viability of the RPM as a magnetic sensing scheme could be recovered by exploiting these radical scavenger-mediated enhancements [429, 439]. A number of other magnetic effects that operate completely independently from the RPM have also been suggested as alternative mechanisms, such as magnetite compasses, electromagnetic induction, radical scavenging reactions, or level crossing effects [440–442].

The role of interference by nuclear spin in the conventional RPM has also prompted the idea of a role for the RPM in the anaesthetic action of noble gases, where the effectiveness of the anaesthetic has been correlated with the nuclear spin [443]. This suggests that nuclear spin coupling may interfere with the anaesthetic effect. Noble gases have large spin orbit couplings [444], indicating that a robust model of may need to incorporate multiple radical pair-mediated ISC mechanisms (e.g., hyperfine, spin-orbit coupling, etc.) before it can provide a satisfactory interpretation of noble gas anaesthetic action. Similar mechanisms have been proposed to explain the isotopic distinction between $^6$Li and $^7$Li isotopes found in rat responses to lithium treatments [445], although the cellular pathway underlying the observed behavioral responses remains unclear [446, 447]. In a similar vein, radical pair dynamics have been proposed to explain observations of MFEs and lithium effects on the circadian clock [448].

Although the RPM provies an established basis for the magnetic sensitivity of many chemical reactions [435, 449] and magnetoreception in fruit flies and human cells has been shown to depend on the presence of cryptochrome, it remains to be demonstrated that the RPM can constitute the mechanism for a working chemical compass *in vivo* [425]. Difficulties associated with demonstrating a working proof-of-principle are linked to the problem of correlating biophysical models directly to behavioral data [450]. This issue has precluded the determination of the exact mechanism(s) underlying magnetic faculties in birds and other species, where concurrent receptor signaling and amplification provides a topic of ongoing theoretical and experimental interest [451–453]. In fruit flies, the magnetic sense was found to depend only on the presence of the cryptochrome C-terminal tail, and modest effects were observed even without the part of the cryptochrome that produces radical pairs [395]. Likewise, the cryptochrome C-terminal tail is critical to enable neuronal magnetic-field sensitivity [454]. Cryptochrome's intrinsically disordered C-terminal tail is known to play an essential role in the mammalian circadian regulation [455], where it controls circadian timing by regulating cryptochrome's association with transcription factors and master genes [456].

Thus, the RPM's significance to cryptochrome's magnetosensitive dynamics remains unclear [440, 457, 458]. Many results are still contentious, and even the existence of a magnetic sense in fruit flies is still hotly debated [459–461]. Moreover, bird magnetoreception was shown to rely on a light-activated process followed by a light-independent magnetosensitive step, ruling out most magnetoreception models based on photoreduction-generated radical pairs [422]. The lack of an accepted theory of biological magnetoreception amidst a plethora of inconclusive results has spawned a proliferation of competing models [441], and a growing number of alternatives to the conventional RPM have been proposed that include radical pair processes based on spin-orbit coupling [428, 437], spin-vibronic coupling [462], electron-electron dipole-dipole interactions [429, 463], radical scavenging [441], or combinations thereof.

Those competing models share a common chemical reaction scheme in which the magnetic field influences coherent electron spin dynamics to modulate an observable electron-transfer reaction rate [441]. Thus, each mechanism is initiated by the formation of two or more radicals (i.e., "open-shell" chemical species [464]), where the coherent relaxation dynamics of the radicals are traditionally modeled using the Haberkorn master equation [465], although more advanced methods derived from open quantum systems theory have also been explored [466, 467]. In spite of many outstanding theoretical and methodological discrepancies, there is a weighty body of evidence supporting the existence of magnetically-sensitive photochemical reactions in cryptochrome *in vitro* in the presence of weak magnetic fields in the milliTesla (mT) [468–470] and sub-mT range [471, 472]. Magnetic compass effects were also observed in the cryptochrome-related molecule photolyase in the mT [473] and sub-mT range [474]. Magnetic modulation of light-induced decay signals in photolyase [473] were followed by demonstrations of magnetic responses to an Earth strength ($\sim 50 \mu$T) in a synthetic molecular system as a prototype chemical compass [475]. These findings were reinforced by observations of spin-correlated flavin-superoxide radical pairs in cryptochrome [476].

It is remarkable that weak magnetic fields on the order of 25 mT [477], 100 - 500 $\mu$T [478], or tens of nT [479] can have a marked effect on cell physiology or animal behavior, because the interaction of the Earth's magnetic field (30 – 65 $\mu$T) with an individual molecule is at least a million times less than typical thermal energies $\sim k_{\mathrm{B}}T$ at cell temperatures [379]. This makes the energy of the magnetic signal much smaller than typical thermal-fluctuation energies, so that the expected signal-to-noise ratio of the magnetic stimulus becomes negligible [480, 481]. As a consequence, the physics of physiological processes that transcend such seemingly-insurmountable ambient noise have attracted sustained research interest [379], motivating research on nonlinear amplification effects [439, 482].

Cryptochrome photoreceptors are found in all known kingdoms of life [483, 484], where they regulate growth [485], circadian rhythms [456], morphogenesis, phototaxis [486], DNA transcription [487], and other various physiological responses to blue light [488]. Likewise, the related photolyase molecules employ light-activated radical pair dynamics to repair carcinogenic damage to DNA by ultraviolet light across all domains of life (except placental mammals [489]) [490, 491]. In contrast with claims that quantum effects are washed out by decoherence in biological systems, research has shown that sizeable MFEs are compatible with fast singlet-triplet dephasing in cryptochrome [492]. Although many questions remain unanswered about the role of radical pair / electron spin dynamics *in vivo*, fundamental spin chemical mechanisms like the RPM provide critical intuition needed to understand a wide range of essential biological processes with broad biomedical and biotechnological implications.

# Chapter 6 – Dynamic Control of DNA Repair in Photolyase

Gene repair is crucial to healing, genetic regulation, and cellular replication. Photolyase is an enzymatic photoreceptive protein that absorbs blue light to inject an excited electron into UV-damaged DNA [491], reversing the DNA damage while releasing thermal energy [493]. Direct evidence for magnetosensitivity in these enzymes first emerged when magnetic modulation of light-induced signals were demonstrated in *E. coli* photolyase where photo-activation initiates catalysis [473]. The dynamical evolution of photolyase activity has now been mapped out, including the timescales for its multiple catalytic steps [494]; from photo-induced electron transfer, to the creation of a radical pair of electrons, to the recombination of electrons after breaking up DNA lesions. Spectroscopic analyses have shown that the high efficiency of DNA repair by photolyase is due to a synergistic optimization of key steps in the photo-repair process, rather than the isolated optimization of a single photo-induced event [495].

Photo-induced oxidation is a primary source of DNA damage, which may be catalyzed by reactions with singlet oxygen ($^1O_2$), removal of a hydrogen atom (forming a free radical), or the loss of an electron from an aromatic base (forming radical cation) [496]. Photolyase binds to DNA in a light-independent step before catalyzing the repair of DNA lesions upon illumination with 300 – 600 nm light [497]. Beginning with the photolyase enzyme docking to its DNA substrate, the key steps in DNA repair involve the absorption of a photon by a light-activated pigment known as flavin adenine dinucleotide (FAD). During photo-activation, the FAD pigment couples to a network of tryptophan (Trp) amino acid residues to generate a radical pair of entangled electrons. A sophisticated mechanism controls the dynamics of the radical electrons as they separate, target the damage, and ultimately recombine in a complex multi-stage DNA repair reaction that transcends the limits of medicine today.

During enzymatic photoactivation, the FAD cofactor residing inside the photolyase photoreceptor becomes excited by an incident photon of blue light, triggering an electron transfer cascade from a chain of three or four tryptophan residues [498]. That electron cascade in turn creates a charge-separated state that is believed to result in the formation of a radical pair if the FAD cofactor is initially prepared in its oxidized ($FAD^{ox}$) state before photoactivation [473]. Observations of a similar photo-activated electron cascade have also been carried out in cryptochrome [499, 500]. The cascade produces a spatially-separated but still-entangled radical electron pair which can then undergo coherent singlet-triplet oscillations and may become sensitized to the ambient magnetic field by the presence of a third (e.g., nuclear or electron) spin [463]. The outcome of this process is the birth of the $FAD^{\bullet-}$ / $Trp^{\bullet+}$ radical pair which forms the basis for the canonical RPM [436, 501]. In this picture, photo-absorption (first by DNA and then by photolyase) is the initiator of both genetic damage and DNA repair [502, 503].

Like cryptochromes [504], photolyases are blue-light photoreceptors that are widely synthesized in plants and animals, as well as in prokaryotes and simple eukaryotes [505]. The cryptochrome/photolyase protein family [486] is broadly categorized into two distinct types of cryptochrome photoreceptors (plant and animal), and three distinct types of photolyase enzymes that are categorized by what kind of DNA damage they repair; either (6-4)-pyrimidine-pyrimidine photoproducts (6-4PPs) or cyclobutane pyrimidine dimers (CPDs) in double-stranded DNA, or CPDs in single-stranded DNA [506]. Cryptochromes generally have disordered C-terminal tail extensions which prevent them from repairing DNA [455], and as such they were widely considered incapable of repairing DNA [483, 507] until DNA-repair activity was demonstrated in a fungal cryptochrome *in vitro* [508, 509].

As primary sequence homologues, photolyases and cryptochromes share similar structures with molecular weights in the range of 50–75 kDa [484, 510, 511], as well as a common molecular cofactor FAD [512]. FAD is photo-active coenzyme involved in many important metabolic processes that are vital to cell respiration and homeostasis [513]. In addition to a common FAD prosthetic group, these light-activated proteins often contain a second cofactor that varies between species [514, 515], such as an additional light-harvesting folate or pterin chromophore [516–518]. The biological function of FAD is closely related to changes in its molecular shape [519], and it is believed that the hairpin-like "U-shape" conformation [520] adopted by the FAD cofactor enables photo-induced electron transfer in cryptochromes and photolyases [521]. FAD is ideally suited as a magnetosensitive cofactor because it undergoes single electron transfer steps through a stable semiquinone radical intermediate [522], unlike its related coenzyme nicotinamide adenine dinucleotide (NAD) which acts primarily as a non-magnetic two-electron donor or acceptor in metabolic enzymes [523].

Tryptophan has the highest photoabsorbance of all the amino acids, making it a common subject of studies of photoinduced processes in biomolecules [524]. It is the primary electron donor in the photoactivation of photo-lyase [525]. Conservation of the Trp chain across the whole cryptochrome/photolyase family is associated with its role in the photo-reduction of the excited FAD cofactor [526] which is enabled by three conserved Trp residues [526, 527]. These residues help to quench the excited FAD singlet state after photo-activation, through a rapid multi-step chemical reduction. In some species of cryptochrome [498, 499, 528], a fourth Trp extends the Trp chain from the inner FAD cavity to the protein surface. This generates a surface-exposed $Trp^{\bullet+}$ radical upon FAD reduction which may be reduced by the solvent or involved in chemical signaling [529].

Trp chromophores have also been identified for their unique role in enabling UV-initiated dimer monomerization for light sensing in the photoreceptor known as UV resistance locus 8 (UVR8) [530]. This plant photoreceptor is believed to be the first light perception-and-harvesting system discovered to use a network of Trp amino acids as a funnel to enhance its light-perception quantum efficiency [531]. This utilization of a network of intrinsic amino acids for light sensing and harvesting marks a departure from other photoreceptor motifs which rely on a separate cofactor (such as flavin adenine dinucleotide in cryptochrome and photolyase) or pigment (such a retinal in rhodopsin) to enable light detection. Growing recognition of Trp-mediated photodetection in UVR8 opens up a new horizon to expand our understanding of collective light-protein interactions in vast chromophore arrays such as those found in the microtubule Trp networks and other extended biomolecular structures. Natural light-harvesting systems are renowned for the transport properties that rely on the presence of organized structural scaffolds [532, 533].

Tryptophan metabolism is implicated across a range of disease processes including cancer and neurodegeneration, where it has been identified as a promising diagnostic and therapeutic target [534, 535]. Trp is practically ubiquitous in photosensitive proteins [536], and findings of chains of Trp (as well as photoactive tyrosine) residues in many diverse proteins indicate that these chains form a common link between the internal protein structure and the surrounding biochemical environment [310, 325]. Coherences measurements can provide insight into the excited-state dynamics of the structures which are central to biological light-harvesting systems, revealing aspects of the electronic structure that are far beyond the level of detail captured by simplified classical models [325]. Deciphering the principles that enable highly efficient energy transfer in biological systems may be expected to facilitate the design of nanotechnological energy transfer mechanisms for biosynthetic systems [325].

To restore the DNA integrity, photolyase relies on a structural scaffold of Trp residues to control light-activated electron tunneling from the FAD cofactor of the repair enzyme into the DNA lesion [229, 537], followed by back-transfer of the electron to photolyase in a complete catalytic cycle [538]. The amplitude and inclination of a weak magnetic field has been shown to influence the rate of DNA lesion repair by using both photolyase and modified cryptochrome enzymes [474]. Advanced time-resolved absorption spectroscopy experiments have shown that photolyase's flavin chromophore can switch between its semi-reduced and fully-reduced forms under physiological conditions [539, 540]. Proton transfer during charge recombination can then follow one of two possible mechanisms which switch near pH 6.5 [541], reminiscent of pH-dependent quantum yields and color changes observed during firefly luminescence [542]). Molecular dynamics simulations have also informed insights into distinct roles for electrons in two (promoting *vs* inhibiting) bond-cleavage steps relating to lesion repair [543].

Recently, ultrafast X-ray crystallography was used to observe the coordinated structural changes that stabilize radical pairs and optimize electron dynamics during the electron-transfer cascade that occurs across an arrangement of tunneling electron transfer pathways during photolyase photo-excitation [229]. Distinct electron tunneling pathways are critical to the control of the electronic parameters which determine the catalytic efficiency of DNA repair [544] and optimize its operation with respect to its noisy surroundings [545]. In these cases, the control of electron dynamics is critical to ensure a high DNA-repair quantum yield close to 100% [494]. In others, subtle control mechanisms regulate DNA-repair operation and efficiency by exploiting arrangements of multiple electron transfer pathways [546]. Advances in time-resolved imaging methods have enabled studies of DNA repair processes at the atomic level, in order to draw new inspirations for DNA-protective drugs and synthetic DNA repair systems from existing methods that organisms already employ to mitigate DNA damage [547].

Today there is growing interest in harnessing aspects of the photolyase DNA-repair mechanism for therapeutic and cosmetic applications [548]. Pathologies associated with DNA damage can lead to tumors and metabolic disease [549]. The development of novel medical treatments based on DNA-repair enzymes will require a comprehensive understanding of the complex sequence of structural and functional dynamics that enable DNA photo-catalysis [550, 551]. Fundamental insights into the quantum dynamics of photolyase and related enzymes will continue to open up new avenues of research for biochemistry and new modes of catalysis for novel enzyme systems [552]. These hold promise for revolutionary advances in the health and biomedical sciences.

# Chapter 7 – Enzyme Catalysis: Quantum Fundamentals

Flavins play essential roles in countless fundamental biological processes such as biological electron transfer, bioluminescence, blue light reception, circadian regulation, vitamin biosynthesis, antioxidant defence, redox sensing, gene expression, and light-driven DNA repair [553, 554]. Flavin's central role in photo-absorption, electron bifurcation, signaling, and catalysis have made it a topic of intensive computational biochemistry investigations [555]. The extraordinary adaptability of flavins to act as chromophores, redox cofactors, and free radicals reveals their pivotal importance to a diverse range of fundamental physiological processes. This adaptability is generally attributed to the highly correlated and delocalized electron structure of flavins' definitive isoalloxazine ring system [553].

Flavins are organic molecules that contain the tricyclic heterocyclic compound isoalloxazine, such as riboflavin (vitamin $B_2$), flavin mononucleotide (FMN), and flavin adenine dinucleotide (FAD). Flavins are highly versatile, light-sensitive, electron carriers found in many enzymatic systems. The isoalloxazine ring likewise provides the electronic structure needed for the photo-generation of spin correlated radical pairs in flavoproteins such as cryptochrome and DNA photolyase [556]. As such, flavoproteins are prototypical systems in which to study the role of fundamental quantum mechanical effects in diverse areas of molecular biophysics that range broadly from light-activated gene repair to geomagnetic field sensing in plants and animals. Unlike related nicotinamides which are primarily two-electron carriers, flavins can transfer one or two electrons at a time. The capacity to transfer individual electrons impart flavins with unique spin-chemical properties which further open them to a range of radical reactions. Most notably, photoactived flavins are known to generate spin-correlated radical pairs which make them subject to MFEs which have set them apart as a leading candidate magnetoreceptor in biology [556].

Its role as a redox sensor, antioxidant, and free radical generator place flavin crucially at the center of cellular immune function, energy transduction, and morphology which all depend fundamentally on the redox status of the cell. Cell redox control is also linked to cell cycle control via the initiation of cell replication and apoptotic triggers [557, 558]. This suggests a central role for flavin and other photo-active redox cofactors centrally in cell signaling and homeostasis, where quantum effects can have a subtle dependence on the surrounding electrostatic environment. For example, there are no observed changes to the UV–visible spectrum of flavin when its active site tyrosine becomes deprotonated in vitro, whereas simplified models that do not account for flavin's electrostatic environment predict a significantly-altered flavin spectrum upon tyrosine deprotonation. This may seem trivial, but the lack of change in the flavin spectrum is only predicted in quantum mechanical simulations where the solvent-and-ion reorganization due to tyrosine protonation is fully taken into account [559]. The essential role played by these flavin molecules, combined with their subtle-yet-fundamental and often poorly-understood redox chemistry, make them a critical scientific target for developing a biomolecular basis for the regulation and control of major cell processes.

Just as the protein environment tunes the flavin's photoactivity [559], so too can the flavin cofactor influence the protein-folding mechanism as a newly-transcribed amino acid chain envelopes it [560]. These biophysical processes are essential to innumerable aspects of molecular-scale physiology, and critical to all major forms of biological energy generation (namely, aerobic and anaerobic respiration, photosynthesis, and denitrification) [561]. Yet the nature of the environmental interactions that control flavin biophysics remain ambiguous even after many decades of flavoenzyme research, and its environmentally-dependent functional properties continue to elude prediction [562]. For example, research elucidating the mechanism of flavin oxidation by $O_2$ remains very limited, and the interaction mechanism of flavin with oxygen remains poorly understood [563]. This is due in part to the fact that there are no structural rules to predict when or how a flavoprotein will react with oxygen in a given setting. Rather, a number of subtle factors such as electrostatic pre-organization, charge distribution, protein dynamics and active-site solvation contribute to the balance of interactions that control reactivity with oxygen in flavin-containing proteins [564]. Flavin reactivity can also be adjusted by covalent structural modifications, either synthetically, or naturally as covalent modifications found in about 10% of native flavins [553]. Simulations of flavin photophysics require an array of complicated quantum mechanical approximations represented in an assortment of computational methods, and there is only limited consensus regarding which methods are appropriate [561].

Accurate quantum mechanical modeling of flavin photophysics therefore hinges on the need to rigorously establish an appropriate set of modeling techniques with robust physical benchmarks. Flavins represent challenging candidates for quantum simulations for the same reason that they are important to study: their versatile and tunable photophysical, spin-chemical, and redox properties emerge from the highly correlated and delocalized elec-tronic structure of their distinctive isoalloxazine ring systems. Quantum mechanical properties of biomolecules like flavin are modulated by complex interactions with the surrounding cellular matrix including the protein, solvent, and electromagnetic field in which the flavin functional group resides. This amounts to a fundamental quantum control task, as the protein environment evidently manipulates the delocalized electronic state of the biomolecule to achieve specific physiological goals. Physiologically, this molecular scale quantum control becomes critical to ensuring metabolic regulation and establishing homeostasis. Thus, modelling the biophysical control of key molecular properties like the redox activity of organic cofactors can require in-depth quantum mechanical analyses which account for the electron correlation and delocalization effects that determine many molecular properties [553, 555, 561, 563]. These may include individual electron or proton transfers, fundamental energy transduction processes, or structural modifications to enzymes and other molecules.

Of the myriad functions of flavin in biology, photo-activated electron transfer is arguably its most fundamental role. During photosynthesis or respiration, an aromatic quinol (either ubiquinol or plastoquinol, respectively) is oxidized by a cytochrome complex that separates and distributes the two electrons from the oxidation site [565]. Studies have revealed an intricate mechanism governing this essential biological process, enabling it to occur reversibly and spontaneously [566]. As a consequence of this complexity, the essential mechanism underlying bifurcating enzymes eluded characterization for decades, precluding the design and synthesis of artificial electron-bifurcating enzymes [567]. The thermodynamics of respiratory electron bifurcation puzzled chemists for decades because it enables the thermodynamically unfavorable reduction of cytochrome *b* by coupling it to the more favourable reduction of cytochrome *c* via an iron-sulfur (FeS) cluster [568]. The key to this reaction is its reversibility, allowing the process to operate either forward or backward interchangeably (without significant energy loss) by coupling a thermodynamically "downhill" reaction to an "uphill" one [569].

Electron bifuration allows crucial yet thermodynamically-costly reactions to occur spontaneously [570] by way of enzymatic reactions in which pairs of electrons (from a two-electron donor) are distributed separately over distinct electron transfer pathways which correspond to different chemical reactions [571]. Thus, electron-bifurcating enzymes optimize the use of free energy by coupling thermodynamically-unfavorable ("endergonic") reactions to thermodynamically-favourable ("exergonic") ones, enabling a variety of chemical reactions that have key implications for cell physiology [572, 573]. Electron bifurcation is therefore now considered one the primary energy conversion mechanisms in biology [570], along with ATP hydrolysis and ion gradient-driven processes which provide the driving forces in living systems by enabling thermodynamically unfavorable reactions.

Quantum effects of electron bifurcation during photosynthesis have drawn attention from researchers interested in cyanobacteria, algae, and plants where cytochrome serves as the primary electronic coupling site during photosynthesis [574], connecting light-harvesting chlorophyll molecules to photosystems I and II [575]. Simulations derived from crystallographic data and electron transfer models have now been employed to predict the electron bifurcating function of the cytochrome $bc_1$ complex from first principles [576]. Recently, detailed studies have shed light on characteristic features of the free energy landscapes that enable high-efficiency electron bifurcation [567], leading to the development of a general theory of bifurcation processes [577]. A number of chemical gating schemes were proposed to rationalize the absence of any electronic short-circuiting during the charge-separation step of electron bifucation, but those schemes could not adequately explain the existence of charge bifurcation because they did not address its reversibility (a crucial aspect of short-circuit suppression in the bifurcating systems) [567].

Electron bifurcation was considered unique to Mitchell's Q cycle for forty years before Buckel and Thauer discovered that flavin-based electron bifurcation is carried out during anaerobic metabolism in microbes [568, 578]. A viable explanation for how the Q cycle carries out highly-efficient electron bifurcation was not proposed until recently, once electron bifurcation became broadly recognized as a key feature of enzyme complexes that perform thermodynamically-costly reduction/oxidation (redox) reactions [579, 580]. Inspired by Keilin's ground-breaking work on cytochrome systems in the cellular respiratory chain [150] and building on the Wikström-Berden model of electron transport through complex III [581], Mitchell was the first to identify electron bifurcation as the mechanism underpinning the Q cycle during oxidative phosphorylation, the process by which 95% of all energy is obtained in aerobic organisms [582, 583].

Electron bifurcation steps are crucial to the function of respiration and photosynthesis in both prokaryotes and eukaryotes [574] where the respiratory enzyme cytochrome *c* catalyzes the transfer of a pair of electrons from the electron-carrying quinol to distinct electron acceptors (an iron-suflur cluster and a *b*-type heme) [566]). The crucial action of an electron bifurcating enzyme is to catalyze a symmetry-breaking reaction that distributes the energy shared between a pair of electrons unequally between them, exciting one electron at the expense of the other and sending them separately over two electron pathways (*i.e.*, one along a higher-energy pathway and the other along a lower-energy pathway).

Correlated electronic motion is an essential feature of electron bifurcation processes [584] because mean field theories fail to capture essential correlations that enable biological energy transduction processes [584]. Inherently quantum mechanical effects such as spin-spin coupling and electron transfer are implicated in the fundamental mechanisms of electron bifurcation that underpin energy transduction in all domains of life [580]. Protein structural rearrangements serve as a control mechanism for electron dynamics along electron transfer pathways through many diverse electron bifurcation enzymes [585]. Hence, the underlying quantum dynamics of the correlated electron motion are further complicated by a multitude of protein conformational states which synchronize transitions between electron transfer and bifurcation states of electron transfer enzymes, coordinating structure-function relationships to optimize catalysis [297].

Spectroscopic observations of the partially-reduced form of bifurcating ETF revealed a sharply-peaked band around 726 nm which gradually appeared and disappeared during FAD reduction [586]. That unprecedented finding indicated the presence of a delocalized charge-transfer species involving both FAD cofactors, distributing electrons coherently across the flavins in a degeneracy-breaking effect reminiscent of that of the RPM [434]. The two-flavin arrangement found in bifurcating ETF is therefore likely to contribute to the signature efficiency of the electron bifurcation process. Experimental observations have also associated fast, efficient charge separation with delocalized electron transfer in organic electronics [587]. In those systems, electron delocalization breaks the symmetry of the system, creating a level-repulsion effect that divides the electrons into characteristically-distinct energy states [588].

Electron-bifurcating flavoproteins constitute an exemplary class of enzymes to use in studies of quantum biology because of their high tunability, nuanced reactivity, air tolerance, and relative simplicity [573]. Exploratory studies have revealed a wealth of quantum mechanical effects in bifurcating flavoproteins that range from kinetic isotope effects to electron tunneling, delocalization, and exchange effects [580, 589]. The wide range of quantum processes exhibited by bifurcating ETFs present an exceptional staging ground for investigations into open quantum system dynamics and the interplay between the quantum mechanical principles which govern their stability.

The trend that we find in all quantum biological systems studied in Part I is the characteristic of quantum mechanical delocalization in determining biochemical kinetics, whether it be in the elementary delocalization effects that produce "incoherent" tunneling, vibronic delocalization in photosynthesis, nonadiabatic charge transfer, singlet-triplet interconversion in the radical pair mechanism, other forms of intersystem crossing in photolyase, and finally electron pair delocalization in the reversible kinetics of electron bifurcation. The importance of charge delocalization is not limited to problems in conventional quantum theory. Just as charge delocalization is critical to efficient charge transfer and radical pair dynamics in organic molecules, exciton delocalization is crucial to the separation of timescales found in the lifetimes of resonant (i.e., superradiant and subradiant) states in open quantum system models of interacting pigments in biological systems. All these exemplary processes represent different cases of the same overarching phenomenon: quantum control of biological kinetics by charge delocalization and relocalization. These phenomenon not only exist, but ultimately thrive in the open quantum systems of biology.

These foundational investigations help bring focus to quantum biology by framing formal concepts for it. This conceptual framework is universally applicable across biological systems, similar to the development of the theoretical frameworks used to define existing fields such as electromagnetism (according to Maxwell's equations), relativity (based on Einstein's principles), and classical physics (using Newton's Laws). The advent of a consistent theoretical framework for quantum biology expands the scope of quantum physics from isolated quantum systems to encompass integrated quantum environments.

# Part II: Coherent Quantum Effects in Biology

> It can scarcely be denied that the supreme goal of all theory is to make the irreducible basic elements as simple and as few as possible without having to surrender the adequate representation of a single datum of experience.
>
> — Albert Einstein, 1933 [590]

## Chapter 8 – Ultraweak Photon Emission & Cell Processes

Although the treatment of disease with sunlight is as ancient as medicine itself, even a century ago there was no real understanding of how sunlight influences health and disease [591]. As such, the formal study of the role of radiation in biology is purely a modern development [592], sparked by the discovery of the photoelectric effect and the characterization of the blackbody spectrum that together gave birth to quantum mechanics. Today it is well known that light is an essential environmental factor for regulating circadian rhythms, metabolic rates, and cell growth [593].

Light is broadly divided into three categories centered around the 380–720 nm spectrum of visible light, where wavelengths of light longer than 720 nm constitute infrared (IR) spectra and wavelengths shorter than 380 nm define the UV regime. IR light is widely considered to have beneficial effects, offering protection against a number of chronic diseases which may be linked to improvements in mitochondrial function and ATP production. Visible light influences many aspects of vitality with an impact on retinal function, sleep, cancer, and persistence of mental health disorders [593]. Visible light from artificial lighting has been shown to have photon energy-dependent toxic effects which can reduce longevity while increasing oxidative stress, neurodegeneration, and chronic conditions such as obesity and diabetes [593]. Though exposure to UV light is well known to induce cancer, promote aging, and decrease longevity, moderate doses of UV light can be beneficial [593].

The study of the impact of UV light on living cells was pioneered by Gurwitsch in his definitive onion root experiments of 1923, which first revealed that living cells emit a faint spectrum of radiation [594, 595]. These faint radiation emissions, sometimes described as biophotons [596], are now recognized widely as ultraweak photon emissions (UPE) [597, 598]. The wavelengths of UPE span the UV, visible, and IR light spectrum, where the UVA-visible-IR range of UPE are documented especially well [599, 600]. However, what most distinguished Gurwitsch's discovery of UPE was his claim that certainly wavelengths of the faint light could promote cell replication. According to Gurwitsch [601, 602], UV UPE primarily in the 190–250 nm wavelength range [603, 604] could also promote growth in nearby cells in a phenomenon that he described as the "mitogenetic effect" [605, 606]. Although the mitogenetic effect described by Gurwitsch still remains to be decisively established [605], there is an ever-growing body of evidence that a diverse range of photon energies are involved in biological processes [595, 607–611].

Quantum effects of light are distinguished by their wavelength dependence, where different colors of light can produce different effects in living organisms with substantial implications for health and medicine [612]. This is due to the discrete nature of the light quanta themselves (*i.e.*, the photons) [96]. Consequences of the influence of different wavelengths of light on living processes were studied and applied quite famously by Finsen who came to be regarded as the founder of modern phototherapy after he pioneered light-based treatments for smallpox and *lupis vulgaris* (skin tuberculosis) in the 19$^{\text{th}}$ century, in turn winning the Nobel Prize in Physiology or Medicine for his efforts in 1903. That same year, Rollier established the first clinic for solar therapy, known as heliotherapy, to treat tuberculosis using high-altitude sun baths which were implemented in clinics around the world [613].

The implications of quantum theory for biology and medicine were considered so great that, by 1932, Bohr himself was invited to give the opening address to the *International Congress on Light Therapy* in Copenhagen that year [614]. His lecture marked his first concerted attempt to extend the concepts of quantum physics to the life sciences, challenging the notion that the principles of life itself could be reduced to pure physics and chemistry [615]. Bohr's sentiment was not new, but had been anticipated a quarter century earlier in the philosophy of Bergson [616] who too had won a Nobel Prize in Literature in 1928 [617]. Bohr's lecture was so impressive that it provided the impetus for Delbrück, whom Bohr had invited to attend, to switch fields from physics to biology [618].

Delbrück went on to study the impact of X-ray irradiation on genetic mutations, concluding that the fundamental nature of genetic material must be molecular [619]. He would go on to receive the Nobel Prize in Physiology or Medicine with Luria in 1969 for discovering the genetic structure and replication mechanism of viruses. Drawing on insights from Delbrück and Bohr [615], Schrödinger prepared the 1943 lecture series "What is Life?" [39] in which he put forward the premise that the preservation of genetic inheritance could not be accounted for by classical statistical laws because the ever-present statistical noise found on the molecular scale. He reasoned that any microscopic genetic information would be washed out by ever-present disorderly heat motion, precluding the possibility that life could rely on the function of classical genetic material. In his own words [39],

> All the physical and chemical laws that are known to play an important part in the life of organisms are of this statistical kind; any other kind of lawfulness and orderliness that one might think of is being perpetually disturbed and made inoperative by the unceasing heat motion of the atoms.

Schrödinger's work was so influential that it inspired Watson and Crick to pursue research on the underlying structure of life, with Watson himself noting that Schrodinger's lectures "very elegantly propounded the belief that genes were the key components of living cells and that, to understand what life is, we must know how genes act" [139]. It is a matter of history that Watson and Crick—in collaboration with Franklin and Wilkins [620]—discovered the structure of the DNA double helix in 1953, for which they received the Nobel Prize in Physiology or Medicine 1962. That same year, Delbrück invited Bohr to deliver the opening lecture at the dedication of the Genetics Institute in Cologne. He took the opportunity to revisit the theme of his address to the Light Therapy Congress in Copenhagen thirty years before, delivering a lecture titled, "Light and Life Revisited," which was to be his final public address [618]. In that speech, he re-framed his original thesis, reflecting that,

> ... it appeared for a long time that the regulatory functions in living organisms, disclosed especially by studies of cell physiology and embryology, exhibited a fineness so unfamiliar to ordinary physical and chemical experience as to point to the existence of fundamental biological laws without counterpart in the properties of inanimate matter studied under simple reproducible experimental conditions. Stressing the difficulties of keeping the organisms alive under conditions which aim at a full atomic account I therefore suggested that the very existence of life might be taken as a basic fact in biology in the same sense as the quantum of action has to be regarded in atomic physics as a fundamental element irreducible to classical physical concepts.

In that 1962 address [621], Bohr updated his stance to convey his view that the regulation of living organisms exhibits a kind of sensitivity that is not found in conventional physical or chemical systems. Like Schrödinger, he believed that this pointed to fundamental biological laws without counterparts in the properties of inanimate matter. Thus, he proposed that the task of biology could not be that of accounting for the motion of every individual atom in an organism, as no clear distinction could be made between the life-sustaining mechanisms in an organism and the functions they fulfill. Rather, he proposed a kind of complementarity between the description of the individual atoms of a living system and its activity as a whole. In this specific sense, he proposed that the action of the whole organism in its natural setting could not be effectively reduced to that of its particular molecular components.

Delbrück carried the legacy of Bohr's search for complementarity in biology in an endeavor to express the concept rigorously and in concrete terms [622]. In what would become the third in a trilogy of talks initiated by Bohr, Delbrück delivered an address at the Centennial of the Carlsberg Laboratory in Copenhagen in September of 1976 titled, "Light and Life III." That lecture was dedicated "literally to Light and Life, to wit, to the question of how and when the various basic photochemical reactions involved in life have come about and what they do" [618]. Beginning by introducing a colorful cast of photo-enzymatic "characters" (chlorophyll, protochlorophyll, retinal, phytochrome, cryptochrome, and DNA photolyase), he quickly and adeptly proceeded to survey a list of fundamental topics that remain prototypical research subjects in quantum biology today, including the biomolecular photoexcitations, biological electron and proton transfers, reactive oxygen species, and organic molecular-bridge systems [618].

This sentiment was echoed in much more technical terms by Wiener who developed a similar line of reasoning in his masterwork on *Cybernetics*, wherein he synthesized the complementary notions of Newton's deterministic classical mechanics and Bergson's vitalistic creative evolution, which he summarized as follows [623]:

> ...the many automata of the present age are coupled to the outside world both for the reception of impressions and for the performance of actions. They contain sense organs, effectors, and the equivalent of a nervous system to integrate the transfer of information from the one to the other. They lend themselves very well to description in physiological terms. It is scarcely a miracle that they can be subsumed under one theory with the mechanisms of physiology. ...Vitalism has won to the extent that even mechanisms correspond to the time-structure of vitalism; but as we have said, this victory is a complete defeat, for from every point of view which has the slightest relation to morality or religion, the new mechanics is fully as mechanistic as the old.

Drawing on principles of quantum theory while incorporating insights from Szent-Gyorgyi and Haldane [623], Wiener observed that the stability of living matter, statistically speaking, must be a consequence of its high internal resonance or "degeneracy" [623]. That is to say, he concluded that the stability of a quantum state depends on its resonance structure insofar as the most stable states will be those which can be transformed into themselves by the largest number of isomorphic transformations with "the ability to make small causes produce appreciable and stable effects" [623]. Wiener's identification of resonance phenomenon in living matter [623] echoed Gurwitsch's original conceptualization of the mitogenetic effect in terms of fundamental resonance processes [595]. Reflections of those proposals were recapitulated in subsequent ideas developed by Frölich later in the 20$^{\text{th}}$ century [624, 625].

By the 1990s, after five decades of research on biological light emissions using photomultiplier tubes, UPE from living organisms had been demonstrated conclusively [626, 627]. This body of work included observations of UV UPE in the 190–250 nm wavelength range predicted by Gurwitsch [628], although attempts to reproduce the effect of UV UPE-induced mitosis were set back by methodological failures from research groups that did not follow the necessary experimental protocols laid out by Gurwitsch [595, 605]. Meanwhile, biological photon emissions in the IR range were also brought to light as having potential effects on living systems by way of a rudimentary form of cellular "vision" characterized by a comprehensive series of experiments by Albrecht-Buehler [629–633]. That body of work was exemplified by the finding that the seat of cellular IR detection is contained in the centrosome [630] and observations that pulsed near-IR signals could perturb the stability of radial microtubules around the centrosome in fibroblast cells [632]. The long-range attraction between aggregating mouse embryonic fibroblasts was also found to be mediated by near-infrared light (NIR) light [633], revealing a key role for long-range electromagnetic interactions between living cells. Those findings were corroborated by evidence that NIR laser light can induce the formation of cell protrusions in fibroblasts and neurons by enhancing cellular actin recruitment [634].

The production of UPE has been demonstrated in practically all forms of life [600]. It is widely understood that mitochondria produce the majority of UPE in eukaryotic cells [635], although DNA was also recently proposed as a significant UPE source [636]. The ubiquity of UPE in living tissue and its link to oxidative biochemical reactions have made it a recurring subject of interest for medical diagnostic and therapeutic applications. For example, the optical stimulation of white blood cells has been shown to induce respiratory bursts, not only in target neutrophils, but also in a second population of chemically-separated but optically-coupled cells, supporting claims of the existence of long-range optical coupling between living cells [637]. Endogenous (*i.e.*, non-thermal) UPE from living organisms are well-documented [638] in the context of oxidative stress. Stress-induced UPE carries a spectral signature that is distinct from that of spontaneous UPE [638], implicating biophotonic modulation under physiological conditions in an organism and presenting a potential role for photonic signalling. However, the links between UPE, cell metabolism, ROS generation and microtubule cytoskeletal dynamics remain far from clear [639]. This leaves a critical gap in contemporary knowledge of cytoskeletal structure and function, where cooperative photonic interactions may be essential to coordinate collective behavior [640].

To consider how UPE are likely to affect cellular order and disorder, one may examine certain key characteristics of cancer. Cancer cells exhibit unique UPE signatures, as well as disturbed microtubule structural dynamics. Furthermore, microtubule-stabilizing drugs are commonly used as chemotherapeutics. Taken concurrently, documented evidence for UV and IR microtubule photodynamics suggest that microtubules may provide a mechanism for optical (up and/or down) conversion between IR and UV frequencies. Further research is needed to tease out the link between optical-frequency radiation, microtubule structural dynamics, and diseases like cancer wherein the primary pathology is the manifest failure of the cellular communication, regulation, and control mechanisms.

Dynamical cellular interactions in the terahertz (THz) regime have been known for decades, highlighted in the work of Albrecht-Buehler on cell behavior in response to NIR [631]. Long-range electrodynamic (dipole-dipole) interactions have also been shown to mediate attractive forces between distant proteins when driven by an external light source [641]. These findings provide a proof-of-principle that non-equilibrium collective oscillations can enable long-range dynamical effects to promote distant protein-protein coupling for aggregation, and has been reproduced using THz resonance spectroscopy [642]. The discovery followed from lines developed in previous work on non-equilibrium phonon-condensation [643, 644]. This is consistent with other observations of long-range orientation in solvent ordering involving molecular dipole-dipole interactions [645].

Microtubules were found to reorganize structurally under the application of UV light [646, 647], suggesting the possibility of photochemical control of microtubule dynamics [648–650] which are likewise associated with ROS regulation in living cells [651, 652]. ROS were shown to control cytoskeletal dynamics with glutathione [653], and studies of anticancer drugs have linked ROS production with microtubule dynamics [654, 655]. Microtubules exhibit superradiance [640] and make good light harvesters [78], similar to quantum optical phenomena observed in light-harvesting nanotubes [656]. This is technologically promising, because the presence of superradiance (also known as superfluorescence) can be accompanied by superabsorption and subradiance effects which are advantageous for light harvesting and precision sensing applications. Structural arrangements of chromophores have important functional implications in biological macromolecules. Ring-like arrangements of chromophores, typical of biological light-harvesting complexes, are structurally similar to arrangements of quantum emitters used in quantum optical experiments on cooperative quantum phenomena using advanced light–matter platforms [657].

Strong light–matter coupling has been proposed as a means to overcome long-standing problems precluding efficient optical emissions from carbon nanotubes by enabling a way to suppress light-quenching effects, presenting a viable route to exciton brightening in these systems [658]. Curiously enough, the innate structural organization of microtubule chromophores has been shown to enable the same brightening effect in the lowest-excitonic state [659], which has likewise been proposed to enable observations of enhanced fluorescence quantum yields in tubulin when aggregated into microtubules [640]. Nanotubes have also been proposed as ultrafast light-harvesting systems [660], echoing the discovery of light-harvesting properties in microtubules [78]. Expansive applications of carbon nanotube-based chemical sensors have also been proposed [661], similar to the established sensory properties of biological microtubules *in vivo* [662, 663].

Although light does not conventionally interact with electromagnetic fields, the formation of polaron-polariton quasiparticles can enable the direct manipulation of hybrid light–matter states using electric and magnetic fields [664]. However, in spite of much speculation [665], microtubules have not yet shown promise for the demonstration of a nonequilibrium Frölich condensate [666]. Nevertheless, biophotons have been proposed to play an essential role in brain function [667] with myelinated axons potentially serving as photonic waveguides [668]. Neuronal biophoton emission is facilitated by high potassium ($K^+$) ion concentration and were impeded by the removal of extracellular calcium ($Ca^{2+}$) ions, indicating that measurements of UPE could be useful for monitoring redox physiology, metabolism and stress [669, 670].

Observations of strong light–matter coupling in nanotubes and microtubules highlight the implications of QED effects and suggest a key role for hybrid light–matter polariton states in these systems. Exciton-polariton lasing has been demonstrated using biologically fluorescent proteins trapped between laminate sheets [671]. Polaritonic states hold special promise for the prospect of biologically-generated coherent light because of their capacity to enable electrically-pumped inversionless laser-like beams at a fraction of the power consumption needed for conventional laser mechanisms [672]. Effects of quantum electrodynamics (QED) have been observed in chlorophyll placed in a plasmonic nanocavity [673], and methods have been proposed to observe cavity-mediated excitonic correlations in photosynthetic quantum aggregates using phonon-driven dephasing as a means to probe correlations associated with long-range exciton migration in light–matter systems [674].

Birefringence experiments using oscillating electromagnetic waves have shown that the microtubules respond to electromagnetic fields differently from tubulin [675], and radio frequency fields in the 900 Hz range were shown to influence microtubule structure. UV light inhibits the assembly of tubulin into morphologically normal microtubules [676], and UV radiation was found to induce breaks in cytoskeletal microtubules in the cell cortex [677]. Advances relating the electromagnetic fields generated by cell organelles and supercomplexes to the dynamics of cell mitosis and meiosis have revealed how synchronized oscillations of microtubules, centrosomes, and chromatin fibers may facilitate cell division and reproduction, with profound implications for the diagnosis and treatment of cancer [678]. Prospects for photochemical control of microtubule dynamics [646–650] have been extended to demonstrate optical control of the microtubule cytoskeleton [679] using light-induced microtubule stabilization tech-niques [680]. The application of direct morphological control to the cytoskeleton will have substantial implications for biotechnology, including cytoskeletal manipulation for agriculture [681], embryology [682], immunology [683], and medicine [684].

# Chapter 9 – Electromagnetic Oscillations in Biostructures

Microtubules have attracted considerable attention for their mechanical and electrical properties [685, 686], as well as their capability for ion transport [687], charge storage [688–690], and signal amplification [691]. Given these features, microtubules likely act as intracellular electromagnetic computing elements [692, 693], in line with known computational abilities of single cells [694] and motivating novel information processing schemes [695]. Microtubules have been considered as electric field generators [692, 696], and ion condensates along microtubule surfaces were identified using local pH measurements [688]. Tubulin has an unusually negative charge on its C-terminal tail, imparting it with a range of signaling capabilities [685]. Fano resonances were detected in tubulin and microtubules by Raman spectroscopy [697], in line with a qualitative model of cell metabolic spectra [698].

Microtubules are main elements of the cytoskeleton along with actin and intermediate filaments [699, 700]. Comprising the backbone of the eukaryotic cell, their dynamics enable cell motility, cytoplasmic transport, mitosis and meiosis [701]. Microtubule dynamics are crucial to cytoskeletal regulation for tissue homeostasis, where microtubule destabilization has been correlated with cellular disregulation and carcinogenesis [702]. Microtubules form complex networks of bundles in neuronal tissue, where the reordering of cytoskeletal microtubules is essential for learning and memory [703–705] and disregulation can lead to the occurrence of neurodevelopmental disorders and other pathologies [706, 707]. Microtubule polarity is crucial to the formation of complex cellular structures including neuronal axons and dendrites, though the microscopic mechanisms that govern the formation of these complex hierarchical structures remain largely unknown [708, 709]. A diverse range of factors are known to govern polarization effects in microtubules, although a comprehensive picture of their interactions remains lacking [710].

Microtubules are cylindrical tubulin aggregates which self-assemble by a non-equilibrium process described by the term dynamic instability [711]. Guanosine 5'-triphosphate (GTP) hydrolosis drives dynamic instability during microtubule self-assembly [712], resulting in rich collective behavior and stochastic switching between polymerization by self-assembly and depolymerization by sudden collapse (catastrophe), as well as its reversal (rescue) [713]. This enables cell reorganization and remodeling for mitosis, differentiation, transport, exploration, and apoptosis [714]. Atomic models of dynamic instability have been implemented in multi-scale simulations [715], and there has been progress controlling non-equilibrium microtubule assembly and disassembly using programmable dynamics [716].

Whereas textbook depictions of microtubules may appear static, like samples stabilized by taxol *in vitro*, actual microtubule dynamics are vibrant, dynamic, and transient. This inherent dynamism is often referred to phenomenologically as "dynamic instability," wherein microtubules undergo a continuing and unstable process of growth, catastrophe, and recovery (*i.e.*, rescue), although this moniker does no justice to the actual, much richer cytoskeletal phenomena during cell homeostasis, division, differentiation, growth and regeneration. It is becoming evident that the inherent dynamics of microtubules provide a highly-optimized noise source for the powerful sensory-motor arrangement which may comprise such diverse phenomena as ionic flows, excitonic waves, and solitons, in a complex system of adaptive sensory response and control.

Each microtubule is composed of a repeating array of tubulin protein subunits in an ordered arrangement of heterodimers, each dimer containing a pair of tubulin monomers labeled $\alpha$ and $\beta$ [685]. As well as being mechanically durable, tubulin is an electrically tunable protein suitable for sensing, sorting, and computing applications [717] in flexible optoelectronics [689]. Tests of quantum mechanics in tubulin have likewise been proposed [696, 718]. Various post-translational modifications can modulate microtubule structure and function via polymerization dynamics [699], but these do not fully explain the dynamical properties of microtubules, such as resistance switching [719]. Microtubule disassembly was effected using intense pulses of THz frequency radiation [720], and repeated assembly and disassembly of microtubules by thermal hysteresis revealed signs of protein ageing [721].

Spontaneous electrical oscillations are an innate property of microtubule bundles found in the brain [722], which likely govern such cellular features as cytoskeleton-regulated ion channels and neuronal electrical activity. Microtubules can operate as biological transistors with nonlinear transmission capabilities to support electrochemical signal amplification and propagation [723]. Microtubule oscillations are likely to contribute to cellular electric fields, which in turn facilitate cell function [692]. Electromagnetic fields generated by microtubule oscillations [724] synchronized in concert with chromosomes are likely to effect key cellular processes during mitosis and meiosis [678].

Honeybee brain oscillations were observed to be mediated by microtubules, revealing a general oscillator mechanism to enable brain function and synchronization [725]. Likewise, bundles of mammalian brain microtubules have been used to demonstrate oscillatory entrainment and synchronization, revealing the capacity for microtubules for the generation, propagation, and amplification of coherent electrical signals in vitro when the bundle preparation method avoided the use of microtubule-stabilizer and anti-tumor drug taxol [726, 727]. Simulations of microtubule oscillations have been used to reproduce experimental data based on a kinetic model of microtubule nucleation, elongation (by assembly with GTP-loaded tubulin dimers), disassembly, and dissolution [728]. Microtubules have permanent longitudinal electric dipole moments, which cause them to align parallel to strong electric fields [729], the application of an oscillating electric field has been shown to increase microtubule polymerization rates [730].

In a leap beyond their properties as apparent electrodynamically oscillating stochastic resonance sensors, microtubules have been proposed to operate as quantum optical cavities according to principles of cavity quantum electrodynamics (cQED) [731]. This cQED microtubule model has been developed to propose mechanisms for generating mesoscale coherent quantum states [696, 732]. In particular, long-range laser-like quantum phenomena have been proposed to enable optical signalling in microtubules, free from losses and thermal noise [733]. These proposals are reminiscent to established atom-optics experiments that realize cQED effects using fibre optic networks [734, 735]. The decoherence time scale in microtubules has been estimated at 1 fs–100 fs [736], which may be used as a reference for other decoherence processes in biology [737, 738] and is consistent with decoherence time estimates of up to 100 fs in photosynthetic systems [323]. This implies an ultrashort timescale for coherent processes in cells, where signals traveling at the speed of light nevertheless have adequate time to span the length of a typical eukaryotic cell (with a diameter ranging from 10 to 100 $\mu$m).

This picture of microtubule network as a fibre-optic cQED system is obscured, however, by an outstanding controversy concerning the refractive index of a tubulin [739]. Several prominent works have placed the refractive index $n$ of tubulin in the range of 2.4—2.9 [718, 740, 741], suggesting that tubulin is at least as refractive as diamond (if not more so). This is in stark contrast with established dielectric theory predicting the refractive index of protein to be within a range $n = 1.5$–$1.7$, as well as competing experimental work indicating a value of $n = 1.6$ [739]. An accurate characterization of the refractive index of tubulin is essential to determine its fibre-optic properties.

The wide gap between estimates of the refractive index of tubulin is intriguing because substantial variations in observed values of tubulin's index of refraction could be attributed to plasmon and/or exciton effects in tubulin [718, 739], particularly in light of recent advances demonstrating plasmonic [742], polaronic [743, 744] and excitonic [745] control of refractive indices in photonic matter. Motivated by work on the excitonic properties of microtubules [646, 659, 746, 747], Nganfo *et al.* have examined the dynamical properties of exciton-polaron pairs in microtubules. When an electron in a tubulin dimer is excited, the electronic excitation itself leaves a charge hole where the interaction between the excited electron and hole form an exciton. Consequently, this results in an exciton-phonon quasi-particle known as a polaron [746]. Although a large refractive index $n > 1$ is preferable to confine electromagnetic modes to an optical cavity (by total internal reflection), a smaller refractive index $n \approx 1$ can enhance electronic dipole-dipole coupling in the cavity (by its $1/n^4$ dependence). Balancing this trade-off may be critical for quantum biology where the quality $(Q)$ of cavity confinement must be weighed against the strength of excitonic coupling, in contrast with common experimental cQED set-ups that achieve strong light-matter coupling with lone atoms or dilute gas. Typical cQED set-ups require very large $Q \gg 1$, unlike condensed matter systems where the sheer number of molecules coupled to the field can ensure strong coupling even for modest $Q$. This presents profound implications for the use of light to control chemistry [748] and rapid cell-wide signaling for physiological coordination [749].

Polarons can form easily in polarizable materials by coupling extra electrons and/or holes with polar vibrations, and can lead to profound changes in material properties and functions [750]. Studies have shown that exciton behavior is modulated by polaronic interactions with phonons in an electronically excited medium [751], where modified transport properties and lattice deformations due to exciton-polarons have been studied in depth in the context of structural dynamics [752]. Moreover, recent studies have shown that nonlinear cQED effects can be dramatically enhanced through the tailored use of exciton-polaron interactions [753]. The self-assembly of tubulin into microtubules was found to be enhanced by the application of MHz-band oscillating electromagnetic fields, where a cavity electrodynamic effect was employed to enhance microtubule growth by inducing synchrony and arresting spontaneous microtubule disassembly [754]. Cavity-induced electromagnetic interactions have been used increasingly in chemistry to modify reactivity, conductivity, and relaxation pathways in organic matter [755]. Microtubules present an ideal testbed for investigations of natural and synthetic cQED effects.

# Chapter 10 – Functional Chemical Dynamics in Living Cells

Mitochondria function as critical integrators of cell signaling cascades, due to their central role in metabolism, redox chemistry, and life/death decisions [756]. The mitochondria carrying out these functions can undergo structural and dynamical changes via carcinogenesis [757, 758], and various forms of tumors have revealed dysregulated mitochondrial dynamics [759]. As a result, cancer cells suffer from a range of impaired functions that include metabolic regulation, energy transduction, calcium homeostasis, ROS production, and apoptosis [760]. Mitochondria have thus been designated as a "metabolic switch" of oncogenesis [761] in a portrait of mitochondrial ROS as endogenous oncogens [762, 763]. Aberrant mitochondrial morphology may lead to increased ROS production and consequently decreased mitochondrial health, further exacerbating a self-perpetuating cycle of escalating oxidative stress [764]. This has prompted a range of proposals for new mitochondria-targeted cancer therapies [765].

The bioenergetics of mitochondrial respiration have emerged as important factors in determining whether cancer cells either succumb to or evade apoptosis [766]. A cell's stress response to the harsh tumor environment comes with the acquisition of common malignant traits [767]. Two hallmarks of cancer are dysregulated mitochondrial oxidative phosphorylation (OXPHOS) and the use of glycolysis to support energy production (*i.e.*, the Warburg effect). Both of these contribute to carcinogenesis [757] in a process that has been exacerbated by the emergence of drug-resistant cancer stem cells [768–770] which allow cancers to relapse following initially-successful treatments by conventional radiation or chemotherapy treatments. The theory of ROS-as-oncogen is corroborated by recent work showing how cells placed in a magnetic field weaker than the Earth's—*i.e.*, a hypomagnetic field (HMF)—undergo metabolic shifts towards dysregulated oxygen use (*i.e.*, the Warburg effect) found in the progression of cancer metabolism [771]. Comorbid stiffening of the extra cellular matrix (ECM) during the reverse Warburg effect [772] results in inadequate mechanical transduction and may promote the development of further malignancy [773].

Growing evidence indicates that cancer cells suffer increased ROS stress—over and above that of normal physiological ROS-generating processes—due to oncogenic stimuli, mitochondrial dysfunctions, and metabolic dysregulation that can all contribute to high ROS levels in cancer cells [774]. Further systematic work deciphering the influences of external magnetic fields on ROS in different biological tissues may be needed to formulate a viable mechanistic model of field-induced biological effects, before it will become feasible to devise magnetic field-based clinical applications to treat diseases in which ROS have established pathophysiological roles such as cancer [775]. Although moderate amounts of ROS mediate cell signaling [776], proteolysis, and microbial defense [777], excessive amounts can lead to mitochondrial and DNA damage, lesions, and death [778]. Redox stress is correlated with age-related diseases [779, 780] including cardiovascular disease, chronic obstructive pulmonary disease (COPD), kidney disease, neurodegeneration, sarcopenia, frailty [780], and ischaemic stroke [781]. ROS influence pathogenesis in a wide range of diseases that include autoimmune disorders, congenital defects, diabetes, and cancer [782, 783].

Mitochondrial dysfunction and disturbed cell coherence have thus been described as a "gate to cancer" [784], prompting interest in cellular exposure to coherent electromagnetic fields as a potential cancer therapy [785, 786]. Indeed, tumors in rodents exposed to a constant $60\,\mu$T magnetic field combined with an alternating $100\,$nT magnetic field (with six frequencies) experienced tumor growth suppression with increased survival rates and enhanced average lifespans [787]. Nevertheless, the exact link between mitochondrial dysfunction and cancer remains unclear [788]. Microtubules monitor calcium and ROS signals that in turn control microtubule dynamics in a feedback loop modulating these signal transduction waves in collaboration with calcium ion ($Ca^{2+}$) channels. In this process, crosstalk between microtubules, ROS, and $Ca^{2+}$ ion concentrations enable the propagation of electrochemical signals in response to stress [789]. Although many studies have revealed wave-like flows of electrons and ions along cytoskeletal elements, it was only recently demonstrated that ion flows along the cytoskeleton are optimized for signal transmissions, whereby microtubules carry coarse information to the centrosome and nucleus with minimal data loss [790]. Similarly, ion channels are known for their signal detection and amplification abilities [791].

Non-equilibrium microtubule dynamics are fueled by hydrolosis of GTP into guanosine 5'-diphosphate (GDP)

in a process where practically all of the energy freed by hydrolysis is stored in the microtubule lattice [792]. GTP hydrolosis is involved in regulating most cellular processes, yet like many effects in quantum biology the mechanism of enzyme-catalyzed GTP hydrolosis has been controversial [793]. Free energy input is required for the reaction-diffusion process of microtubule self-organisation that is associated with emergence of complex phenomena in the form of macroscopic traveling waves capable of long-range transport [794, 795] and functional deformations of the cell membrane during the growth of microtubule networks [796]. The rich nonlinear dynamics of microtubule networks have made them a fertile subject for research in solitary waves [797], including kink-like energy transfer mechanisms [798], bell-shaped solitons [799], as well as dark, bright, and singular solitary waves [800]. Solitary waves were proposed as a vital mechanism of information transfer in a nonlinear model of microtubule dynamics that incorporate dipole-dipole interactions there [801]. Microtubules play essential roles in cell motility, signalling, and transport, and a cavity QED (cQED) model of microtubule dynamics has been put forward in order to hypothesize dissipationless energy transfer as well as the theoretical possibility of biological quantum teleportation in microtubules [732].

The timely control of reactive oxygen species and other free radicals is essential to homeostasis, as a sufficient number of ROS are needed for metabolism, signalling, and immunity—but an excess amount can spell death. The problem is complicated by the electronic structure of molecular oxygen ($O_2$). Diatomic oxygen, as it is found in the air we breath, is uniquely configured with a triplet electronic ground state (*i.e.*, $^3O_2$). This means that the lowest energy state of molecular oxygen has non-zero spin, in contrast with the usual electrostatic Born-Oppenheimer approximation. Cellular ROS are known regulators of microtubule dynamics [802], and studies have shown that microtubule dynamics are diminished in the presence of ROS *in vitro* [651]. Taken with findings that microtubule network destabilization, depolymerization, and reassembly could be triggered using an oscillating magnetic field [803], this indicates that microtubule dynamics entail magnetically-sensitive chemical processes that are quenched by ROS. This was anticipated by the hypothesis that perception of the geomagnetic field could be enabled by magnetosensitive chemical reactions that compete with radical scavenging to enable chemical amplification [804].

Cellular ion channels have also been proposed to enable gating effects that are modulated by ambient fields using stochastic resonance [805], and ion channels are recognized as active amplifiers in the cell [806]. Stochastic resonance was considered in the context of ion channels and excitable membrane dynamics [807], and cytosolic calcium oscillations have been linked to 50 Hz electromagnetic field effects in living cells [808]. Simulations indicate that the most sensitive step in a calcium oscillation chain reaction is the release of stored $Ca^{2+}$ ions by calcium-induced calcium release (CICR) [809], similar to mitochondrial ROS-induced ROS release (RIRR) [810]. This finding reveals how experimentally observed magnetic resonance effects may be explained using non-linear dynamics, where the dynamics' extraordinary sensitivity to periodic waves of cellular $Ca^{2+}$ oscillations indicate a role for chaos in cells' detection of weak electromagnetic signals. Various models have been proposed for the purpose of discerning the mechanisms of ionic signal transduction and chemical bifurcation by which calcium oscillations arise in cells [811]. A recent meta-analysis revealed that low frequency magnetic fields are likely to influence calcium homeostasis in cells *in vitro* [812], where moderate-intensity static magnetic fields may influence physiology by varying the action of $Ca^{2+}$ flux through ion channels [813], indicating a key role for ion fluctuations in the cellular magnetic sense.

Biological ion channels can generate coherent oscillations from inherently noisy stochastic components for precision sensing by using stochastic resonance [814], similar to microtubules [790]. In general, stochastic resonance is possible in bistable systems with inherent noise [815], such as tubulin dimers assembled into microtubules. Microtubules are likewise the nearest biological equivalent to carbon nanotube structures [675, 816]. Initially geared toward tuning the optical properties of materials, strong light-matter interactions were revisited over the last decade to control chemistry and material properties through collective strong coupling [748]. Notably, structurally-organized water molecules have been identified as bright light emitters when confined to nanoscale cavities [817], revealing a potential role for light-emitting cavity-mediated effects in solvent-confining nanostructures such as microtubules.

Similar to findings that $K^+$ and $Ca^{2+}$ concentrations modulate biophoton levels in neurons, cellular nitric oxide (NO, a messenger molecule which mediates neural and hormonal communication in various tissues) levels were found to modulate firefly respiration and bioluminescence in a light-dependent switching effect [818]. Further, synchronized action potentials were found to originate from heterogeneous subcellular subthreshold $Ca^{2+}$ signals in the cardiac sinoatrial node, *i.e.*, the chief pacemaker of the heart, during the entrainment of spontaneous excitations in the individual pacemaking cells [819]. The complex interplay between ion and oxide oscillations, physiological synchronization, and biological luminescence [820] present an interconnected series of biochemical characteristics that suggest the need to develop a new paradigm for characterizing synchronization effects in biology [821].

An extensive body of research has emerged to associate physiological ROS modulation with both static and time-dependent MFEs at field strengths on order of mT or smaller. ROS are critical for self-renewal in stem cells, but the ways that ROS signals are integrated into self-renewal mechanisms remain unclear. Although ROS are a well-known bi-product of mitochondrial respiration, recent insights have also revealed specific biochemical regulatory pathways that invoke NAPH oxidase, initiating and amplifying ROS signals that activate transcription factors in a positive feedback loop that drives cellular regeneration [822]. ROS signals also contribute to the proliferation and survival of many cancers [823], making the precision modulation of mitochondria/cell redox communication a promising avenue for therapy and oncology. Moreoever, the multitude of roles played by ROS in homeostasis, carcinogenesis and cellular renewal have revealed a subtle and cryptic system of redox regulation.

The appropriate balance between chemical reduction and oxidation is essential to healthy cellular physiology, where disregulated redox balance often occurs with the production of excess oxidative species, leading to DNA damage, lipid peroxidation and abnormal post-translational protein modifications, causing injury, disease and cell death [802]. Despite the inherent risks associated with reactive oxygen, it is critical to cell function where it regulates cell dynamics including growth, healing, and regeneration. Monitoring and control of cellular ROS levels have become topics of steadily increasing biomedical interest, because ROS are essential to immunity and cell function [824, 825]. ROS include superoxide ($O_2^{\bullet-}$) and hydroxyl ($HO^{\bullet}$) radicals as well as non-radicals including singlet oxygen ($^1O_2$) and ozone ($O_3$) [826]. NADPH oxidase (*i.e.*, nicotinamide adenine dinucleotide phosphate oxidase) is the prevailing source of ROS in many cells [217, 780, 827]. NADPH oxidase, or "NOX", constitutes an ETC connecting NADPH to the phagocytic vacuole where microbial killing occurs in phagocytes (*i.e.*, white blood cells) [828]. Phagocytes are incapable of killing bacteria and fungi without NADPH oxidase. A deficit in this enzyme can result in life-threatening infections [829]. Phagocytic NOX has been called the prototype of the NOX ETC systems.

Dysregulated complex I activity is associated with defects in its assembly which lead to metabolic disorders in humans [830]. About half of human mitochondrial disorders are associated with complex I mutations [103]. Complex II has also gained attention as a ROS generator [831] with mutations correlated with cancer and neurodegeneration [832, 833]. Thermodynamic analyses of mitochondrial complexes have shown that vulnerabilities in complexes III and IV can emerge even under resting conditions. This highlights the importance of metabolic control of these complexes as well [834]. Mitochondria support innate immunity by producing ROS in response to cell damage and stress [835], shaping the orientation and character of the cell's response to stimuli [836]. Control of the electron transfer dynamics which produce ROS and free radicals are thus crucial to maintaining cell homeostasis [105, 837].

Mitochondrial redox stress is another key regulatory factor in immune responses controlling ion homeostasis, ROS signaling, cytoskeletal dynamics [802], mitophagy, and apoptosis [838]. Governing redox homeostasis constitutes a vast undertaking in dynamical quantum sensing and control. When redox homeostasis is included in the cellular model, it extends the problem of regulating electron tunneling for respiration in the ETC. ROS regulation evidently entails an enormous scheme to control quantum spin dynamics in the vast many-electron systems of cells, tissues, and organs. In the NOX ETC, electron transfer cofactors are arranged to tune the transfer of $O_2$-reducing electrons from inside the cell to the ROS-generating site on the outside of the cell membrane in cooperation with structural elements that control NADPH binding, tightly managing ROS generation [839]. In turn, NOX action is coupled to the activity of voltage-gated ion channels [840] and reactive chemical intermediates such as peroxynitrite [841], highlighting its influence on osmotic balance and redox homeostasis. NOX regulation is associated with hypertension, making it a major source of cardiovascular and renal redox stress [842]. For example, neutrophils sequestered in the pulmonary vasculature can release ROS which are key substances in inducing endothelial dysfunction and disruptions responsible for the primary clinical manifestations of acute respiratory distress syndrome (ARDS) [843]. These factors suggest a promising prospect for the pharmacological and dietary manipulation of the cellular NOX activities, which are commonly implicated with oxidative stress in neurodegenerative disorders [844].

In eukaryotic cells, mitochondria produce ROS, regulate cellular redox stress, maintain $Ca^{2+}$ homeostasis, synthesize and break down high-energy metabolites, and regulate cell death—all while producing about 90% of the cellular energy needed to survive [845]. In white blood cells, a number of chemicals are known to amplify ROS availability by modulating NADPH oxidase (NOX) activities [846]. Similarly, hypoxia-induced ROS amplification has been found to trigger mitochondria-mediated apoptosis, accelerating the rate of intracellular ROS production under oxygen-starved conditions [847]. Sodium ($Na^+$) ions have also been found to control hypoxic signalling by the mitochondrial ETC in an unanticipated interaction with phospholipids [848]. Ionic signaling between cells has now been observed in many different cell types to coordinate cellular regulation and controlled cell death [615]. This form of intercellular chemical signaling can become a double-edged sword in light of the fact that cancer cells create a microenvironment of oncometabolites that perpetuate cancer growth in a vicious chemical-signaling cycle [849].

Mitochondria play a central role in determining health [850]. Impaired mitochondrial function can cause debilitating disease [851]. In eukaryotic cells, up to 90% of the ROS are generated in the mitochondria during ATP synthesis [852] wherein mitochondrial complexes I and III are considered primary generators of $O_2^{\bullet -}$ and other ROS [853–855]. Disruptions in the electron transfer efficiency of respiratory carriers such as electron transfer flavoproteins (ETFs) and their partner enzymes are an important source of ROS in cells [856]. Salient features in the structures of the quinone binding sites of the ETC reveal an ideal framework to inspire programs in rational drug discovery [857].

ROS play a central role in autoimmune diseases, inflammatory conditions, cancers and bacterial infections. This has made ROS amplification an attractive effect for use in the design of therapeutic interventions. Practical implementations of ROS amplification are problematic because interventions that produce systemic enhancement in oxidative stress tend to cause severe side effects [846]. Targeted, cell specific control of redox chemistry is therefore a preferable method to enable biomedical applications, but cellular regulation of ROS, metal ions, and metabolites is extraordinarily complex and interconnected. Quantum mechanical effects are implicated in many aspects of the vast network of microtubules, ion channels, and ETCs that enable biochemical regulation and immunity in living cells. More in-depth studies of the fundamental physical mechanisms that govern these systems are needed.

# Chapter 11 – Magnetic Biomodulation: Biodynamic Control

The Earth's magnetic field influences a wide variety of animal behaviors, ranging from migration to herd behavior. Many creatures across various taxa have demonstrable abilities to perceive and respond to magnetic fields, especially birds (known for their long migrations) including European robins, silvereyes, garden warblers, homing pigeons [858] and zebra finches [859]; but also cockroaches [398], fruit flies, honey bees, mice [860], mole rats, lobsters, salamanders [861, 862], salmon, sea turtles, sharks [858], and even humans [863]. Moreover, detailed studies have revealed the importance of the Earth's magnetic field to biological function. Geomagnetic field deprivation was found to reduce neurogenesis and cognition in mammals [864], to disrupt appetite and feeding in insects [865], and to induce abnormal morphological development in amphibians [866, 867]. Insights into the mechanisms underlying these physiological disruptions have come from studies showing that the assembly of microtubules from tubulin is disordered in the absence of the geomagnetic field [868], and that geomagnetic field shielding leads to disordered actin assembly kinetics that results in reduced cell motility of human cells [869]. These findings have inspired a proposal to use hypomagnetic field effects as a means to test the RPM in biology [870]

Long-term exposure to a HMF much weaker than the Earth's geomagnetic field (GMF) was shown to impair nerve growth and development in the mammalian brain [864]. This effect was demonstrated to be ROS-mediated when normal neurogenesis and cognition were recovered by pharmacologically restoring ROS levels in the HMF-exposed mice. The careful controls used in this study further demonstrated the functional relationships between magnetic fields, ROS levels, neuron growth and cognition [864]. Defective neurogenesis and cognition were mitigated in the HMF-exposed mice by returning them to the geomagnetic field (GMF). This restored elevated ROS levels, the benefit of which could again be eliminated by chemically lowering hippocampal ROS levels in the GMF-exposed mice. This study clearly demonstrates a key role for Earth-strength magnetic fields in regulating physiology.

HMFs have unfavourable effects on biological processes including embryonic development, morphogenesis, and behavior [871]. Despite the body of evidence demonstrating the influence of subtle magnetic fields on physiology, substantial progress developing magnetotherapeutics has been stymied by a lack of widely accepted mechanisms to explain how electromagnetic fields induce physiological changes in living cells [872]. Without a unified theory of MFEs on biology, a range of hypotheses have emerged based on diverse mechanisms including radical scavenging and recombination reactions, voltage gated ion channels, kinase signaling pathways, and/or heat shock protein (HSP) [853]. The finding that low-intensity magnetic field-induced ROS generation requires the presence of the flavoprotein photoreceptor cryptochrome in human cells [873] supports a hypothesis that carcinogenesis associated with electromagnetic field noise, electromagnetic field-induced ROS production, medical treatments using pulsed magnetic fields, and magnetic sensing in vivo may share a common mechanistic basis [457].

Questions about the influence of $\sim 50\mu$T geomagnetic fields of physiology have been motivated by models of animal magnetoreception and suggestions that even weaker anthropogenic low frequency electromagnetic fields ($< 1\mu$T) may have important implications for human health. Magnetic fields much smaller than 1 mT can influence rates of radical pair reactions in chemistry, although the resulting product yields are often on the order of just a few percent or less [874]. These marginal product yields have left important questions open about a role for weak field effects in biochemistry that pertain to sensing, metabolism, and cellular growth. Thus, substantial work has been done to characterize the chemical amplification of MFEs due to the presence of scavenging molecules [439, 482] interacting through radical termination reactions [471]. This work has further motivated the question of whether a small magnetically-induced change in the rate of one step in a multi-step reaction could effect a disproportionately large change in the overall rate of the biochemical reactions [874]. Such an autocatalytic nonlinearity would appear to be necessary to rationalize the substantial effects that weak magnetic field (WMF)s can have on cell metabolism.

Weak magnetic fields of less than 1 milliTesla (mT) have been shown to influence various biological processes in hundreds of studies, yet the exact mechanism(s) underlying most of these effects remain unclear [436]. Magnetic fields were shown to influence redox metabolism in human blastoma cells [771], where the application of magnetic shielding induced cancer-like metabolism, upregulating anaerobic glycolysis and repressing ROS production (*i.e.*, the Warburg effect). The application of a HMF via magnetic shielding was also shown to reduce ROS production in blood cells [875, 876], whereas applying a modest magnetic field could produce a moderate increase in ROS levels [877]. Magnetic fields have been proposed as a possible wound therapy based on results from experiments on animals and clinical trials [878]. A subtle balance of homeostatic ROS levels is essential for effective wound healing [879], and it has been documented that changes in static magnetic fields can alter both ROS and calcium ion concentrations, acidity (pH), and growth rates of cancer cells [880].

In another study, a combination of static and oscillatory magnetic fields produced a stimulating effect on the fission and regeneration of planarians, where removing the static component could reverse the effect [881]. Likewise, the application of a low-frequency, pulsed magnetic field was shown to stimulate collagen production in fibroblasts cultured from chicken tendon, indicating that pulsed magnetic fields can also increase vertebrate collagen production [882]. However, the exposure of mouse embryonic fibroblasts to an oscillating magnetic field caused a significant increase in autophagy in vitro, correlated with a simultaneous increase in ROS without the activation of the central autophagy signaling pathway involving mammalian target of rapamycin (mTOR). The applications of static mT-scale magnetic fields were shown to have an effect on the growth of three types of bacteria, decreasing growth rates of *Escherichia coli* and *Staphylococcus aureus*, while increasing the growth of *Bacillus subtilis* [883].

WMFs $< 1$ mT were shown to be able to modulate stem-cell mediated growth and regeneration in planarian flatworms [478]. Endogenous ROS concentrations were shown to be modulated by control of the ambient magnetic field, where ROS levels and associated growth rates were together either augmented or diminished depending on the strength of the applied magnetic field. The correlation between ROS variations and growth was confirmed by demonstrating that growth-inhibition by the magnetic field could be counter-acted by restoring ROS levels biochemically. This magnetostatic field effect provides further evidence indicating that WMFs can influence redox homeostasis and cell metabolism, with a significant impact on tissue growth after an injury.

Magnetic manipulation of ROS levels can alter redox homeostasis, metabolism, and growth of white blood cells. Coordinated sequences of pulsed WMFs have been applied to control metabolism in neutrophils to enable chemical ROS amplification via metabolic resonance using respiratory burst synchronization [777]. In these experiments, electromagnetic resonance with NADPH oscillations in human neutrophils was achieved by synchronizing externally-applied pulsed magnetic fields with the innate frequency of flavoprotein redox oscillations in the neutrophil cells, where magnetic pulses were phased to coincide with the NADPH fluorescence minima. Production of nitric oxide (NO) was found to follow the production of ROS closely as the NAD(P)H oscillation increased cumulatively with the application of each pulse. Production of ROS and NO could likewise be terminated by pulsed WMFs, leading to neutrophil depolarization, reminiscent of NO-mediated control in firefly flashing [169]

In fact, an extensive body of work has been carried out by Novikov and coworkers varying magnetic field conditions to modulate respiratory (*i.e.*, ROS) bursts in neutrophils (*i.e.*, white blood cells). These include studies of the following effects: the effect of hypomagnetic conditions on neutrophil ROS production [871], the effects of weak combined static and oscillating magnetic efffects of neutrophil respiratory bursts [884, 885], the influence of oxygen on neutrophil respiratory bursts upon exposure to a WMFs [886], the role calcium ions and hydroxyl radicals on respiratory bursts in neutrophils [887], the impact of combined static and oscillating magnetic fields on lipid peroxidation in neutrophils [888], the comparison of the responses of human and mouse neutrophils to weak combined static and oscillating magnetic fields [889], the action of weak combined static and low-frequency oscillating magnetic fields on luminol-mediated chemiluminescence in mammalian blood [890], and the quenching of magnetic field-enhanced chemiluminescence by free-radical scavengers in human blood [891].

Even nanoTesla-scale oscillating magnetic fields have been shown to induce biological changes, and the exposure to appropriately tuned fields of this kind can significantly increase ROS levels in biological systems, altering metabolism [892]. The complexity of these effects suggest that several receptors may be responsible for WMFs in biological systems [893] and the effects can seem paradoxical. For example, very low-frequency electromagnetic fields can accelerate wound healing by modulating cytokines and metalloproteinases [894] or induce increases or decreases in cellular DNA methylation levels [895], suggesting that these frequencies correspond to energy separations between molecular states in external magnetic fields [896]. Whole-body exposures of rats to WMFs could be used to selectively induce seizures and reduce learning, or to improve performance at a learning task, depending on the specific magnetic field pattern employed [897]. Studies of physiologically-patterned weak magnetic-field treatments have indicated that the intensity of the field required for the treatment to be effective may decrease substantially with the degree of pre-programmed field patterning [897]. To this effect, new classes of physiologically patterned electromagnetic signals are now being investigated for magnetobiology research and therapy [898].

Quantum mechanical properties of microtubules have been a longstanding topic of scientific interest [731]. This attention has motivated a picture of the microtubule as a quantum electrodynamical cavity with potential quantum information processing capabilities [696]. Microtubules have been proposed to play an essential role in mediating cognition, as anesthetics bind to the microtubule building-block protein tubulin and have been linked to microtubule depolymerization [899], and so may account for selective action of anesthetics on consciousness and memory [900]. Radical pair reactions have also been proposed to play a role in general anesthesia [901], implicating radical pair dynamics in microtubule-based cell perception. Plant microtubules, for example, exhibit sensation relevant to the perception of mechanical stress under various conditions, where sensory microtubules are integrated into the plant architecture—maximizing its robustness and dynamicity [662]. This sensory faculty of microtubules relies on their capacity to exhibit *dynamic instability*, a non-equilibrium phenomenon of ongoing transitions between microtubule growth and shrinkage driven by the hydrolosis of GTP into GDP [902]. The influence of quantum effects in non-equilibrium microtubule dynamics was implicated in the discovery that the assembly of tubulin (into microtubules) is disordered in under the HMF [868]. The failure of microtubule growth in the absence of the GMF led to the hypothesis that radical pairs may be involved in microtubule assembly [903]. The proposed model is problematic because it entails a weak MFE invoking a singlet-born radical pair which is inconsistent with observations of well-organized microtubule growth under high magnetic fields [904, 905]; and because the singlet-born weak-field MFE tends to be reversed in the high-field limit [434].

The effect of magnetic fields on reaction rates and kinetics have been known to chemistry for decades [435], motivated by electronic and nuclear spin polarization effects on chemical reactions discovered in the 1960s [431, 432]. The influence of a WMF on a chemical reaction has also been called the low field effect [434]. Speaking broadly, the presence of a low field effect is the hallmark that an otherwise spin-forbidden chemical process has been "turned on" and is now allowed by a magnetic field-mediated radical-coupling mechanism. Typically, this occurs by the lifting of a degeneracy between reactant and product energies that would otherwise render the effect negligible. The effect is to modulate the rate of an ISC process that is symmetry-forbidden in the absence of radical coupling or strong magnetic fields [906]. Although the RPM has become a widely accepted mechanism to rationalize the influence of WMFs on biological systems [379], the explanatory model offered by the RPM is not free from contradictions [429], and other models of biological magnetoreception have been proposed which include proposals for a magnetite-based sense [907] and the magneto-hydrodynamic effect [908].

Curiously, many examples of the influence of magnetic fields on physiology also rely on illumination. This reveals how collective cellular behavior, and in fact as we see in many cases of animal behavior, relies on the complex interplay between light and magnetic fields. Beyond assisting navigation in migratory species, cellular processes involving oxidative metabolism and redox signaling often exhibit light-dependent magnetic sensitivity.

Photochemical pathways that mediate these MFEs are generally linked to ROS modulation, cell growth, and tissue repair. Magneto-optic interactions of this kind form the basis for hypotheses that connect electromagnetic fields to mechanisms of carcinogenesis, pulsed magnetic field therapies, photobiomodulation, and multiple other areas of regenerative medicine and biology.

Static magnetic fields can have an impact on human physiology and sensation [909]. Blood cells have been shown to orient diamagnetically in strong magnetic fields, where a significant contribution to the total cellular anisotropic diamagnetic susceptibility was attributed to the microtubule cytoskeletal network [910]. Static MFEs on cellular ROS levels have attracted particular interest in studies because of their apparent ubiquity of cell physiology and presumed connection with the RPM [911]. Spin chemistry research has long been concerned with radical electron pairs, and also more recently the effects of free scavenger radicals such as dangerous ROS [441]. As far as ROS regulation is concerned, singlet oxygen has attracted considerable attention for its toxic and carcinogenic effect in the oxidation of lipids, proteins and nucleic acids [912]. Paradoxically, singlet oxygen has also attracted interest for its anti-tumor effect [913], and the production of mitochondrial singlet oxygen has offered a rationale to explain the effect of metabolic therapies in treating cancer [914].

Whilst a wealth of evidence has demonstrated that magnetic fields influence ROS levels in living cells, there is no established consensus on the nature of the effect, likely due to a lack of mechanistic understanding [775]. ROS are implicated in the hypomagnetic field effect on neurogenesis [915], as well as stem-cell mediated cell regeneration [478]. These effects were shown to be mediated by control of a weak magnetostatic field effect on endogenous ROS concentrations, indicating that weak magnetostatic fields can influence physiological redox homeostasis.

Exposure to GHz frequency electromagnetic waves were shown to modulate ROS in human HEK293 cells as a function of the signal amplitude [916]. Oscillating fields have been used to alter relative yields of ROS products in living cells, indicating coherent singlet-triplet mixing at the point of ROS production and concurrent metabolic changes [242]. Likewise, results in optogenetics are promising to circumvent conventional limitations on ROS control (through system-wide applications of ROS generating-or-quenching molecules) by enabling light to modulate activities of genetically-encoded effector proteins [917]. Quantum sensing techniques, such as magnetic relaxometry, have emerged to enable the unprecedented detection of radicals with subcellular resolution [918, 919].

Quantum mechanical methods are well-developed in biochemical and molecular electronics [920]. Despite the importance of ROS in health and disease, ROS physiology remains poorly understood [921]. Redox stress due to increased ROS in tumor cells is essential for cancer progression and recurrence [922], and cancer is being increasingly recognized as a metabolic disease [914]. Alterations to cellular free radical activities were also reported to affect DNA integrity, immune and inflammatory responses, cell proliferation and differentiation, wound healing, neural activities and behavior [853, 923]. Biological processes are known to respond to WMFs [436, 924], but there is still no viable hypothesis to rationalize all observed effects of electromagnetic fields on cellular radicals.

# Chapter 12 – Molecular Forces: Solvent Effects & Dispersion

Collective synchronizations are found everywhere throughout the natural world [821]. Living cells coordinate cooperation between thousands of active molecules that collectively carry out tasks which range from cell differentiation and mobilization to motility and replication [925]. Collective spontaneous movements exemplify emergent phenomena on the molecular scale where self-organized dynamics generate active responses in biology [925–927]. Yet manifestations of spontaneous synchronization challenge our understanding in many areas of biology [126]. Such is the problem of protein folding where it is unknown how a newly-translated protein strand can relax into its optimal folded conformation in fixed time when the number of possible conformational states is exponentially large with respect to the protein sequence length, and explorations of the funnel-like conformational "landscape" are obstructed by numerous topological restrictions on folding [928]. Nearest neighbor interactions in the protein strand cannot facilitate a timely reduction to a small number of conformation states, precluding a resolution to the problem based on relaxation by nearest-neighbor interactions. This contradiction has prompted a hypothesis that life relies on quantum-mechanical interactions and entanglement for development and survival [929, 930].

Although the "mean field" approach to solving a many-body problem by considering an equivalent one-body system draws from a rich history in physics [931], it did not develop as a formal method in quantum chemistry until it was adopted by Kohn and Sham based on previous work by Hohenberg and Kohn [278]. In the method of Hohenberg and Kohn, electronic structure of a system of many interacting electrons is represented as a function of its density alone, where it is then proven that the ground state of that system and all of its properties are uniquely determined by its density [280]. Kohn and Sham extended the work of Hohenberg and Kohn by substituting the many interacting electrons with a fictitious system of non-interacting ones that nevertheless give the same density as the true system of interacting electrons, resulting in a set of effective one-electron solutions to the many electron density problem (known as the Kohn-Sham equations) [281].

In principle, the Kohn-Sham equations provide an exact set of one-electron solutions to a many electron problem, suggesting (as Hopfield has [271]) that all of quantum chemistry is trivial in the formal quantum mechanical sense described. But there is a catch! Kohn and Sham's derivation depends on the assumption that the electron density changes "sufficiently slowly" in space [281]. Moreover, even though Kohn and Sham's effective one-electron solutions replicate the exact density of the equivalent many-electron system (and thus the solution to the many-electron problem) in principle, in practice they rely on the construction of a fictitious classical potential function that exactly reproduces all of the quantum mechanical correlations of the true many-electron system. As a consequence, the problem of solving a complicated quantum mechanical system of many interacting electrons is replaced by the problem of finding a fictitious classical potential that emulates the effect of introducing quantum mechanical correlations into an otherwise-classical system of non-interacting electrons.

In this respect, density functional theory (DFT) is similar to several other mean field methods that do not account for long-range electron correlations, despite the fact that these effects must necessarily be included in realistic simulations of condensed matter systems [282]. Countless efforts have been made to construct effective classical potentials to reproduce quantum mechanical exchange and correlation effects in quantum mechanical systems, with varying levels of limited success. The study of these so-called "exchange-correlation" functionals have come to represent a central problem in DFT because there is no general solution to the problem of finding an effective classical potential that reproduces all the consequences of quantum physics, or even a formal method for obtaining increasingly-accurate approximations to the results of quantum mechanics. In this respect, the formulation of DFT may seem like "robbing Peter to pay Paul," trading an intractable quantum mechanical problem for an impossible classical one, but the whole exercise entails one distinct advantage: It allows for the exact quantification of quantum mechanical effects in a chemical system, which are all included in the construction of the exchange-correlation functional. A system that is only "trivially" quantum mechanical will be accurately described by a simple classical approximation to its exchange-correlation effects, whereas a nontrivial quantum system may afford no effectively-classical solution to the exchange correlation problem at all. In either case, the solution to the question of whether a system is only trivially quantum is formally determined by whether it can be solved efficiently (using DFT) using an effectively-classical exchange-correlation functional.

Historically, accurate quantum chemical studies of biological systems have not been accessible to DFT calculations because the large size of biological molecules prohibit the elucidation of biological processes using quantum-mechanical calculations [257]. Biological systems present a challenge for DFT because of the ubiquitous presence of weak binding and charge transfer effects which are properly described by conventional DFT (*i.e.*, using local density or generalized gradient approximations [289]). To address this, approximate methods were developed for biological systems using self-consistent charge density-functional tight-binding (SCC-DFTB) [289], which have been modified with empirical dispersion corrections to successfully model the structure and dynamics of peptides, DNA, and ligands [290], as well as proton transfer steps in dehydrogenase and isomerase enzymes [932].

Nevertheless, conventional DFT methods do not intrinsically account for long-range dispersion effects in biomolecules [290]. Although many-body dispersion forces may only have a minor influence on the low-energy geometries of molecular van der Waals clusters (where vibrational frequencies of breathing modes can decrease by as much as 10% [933]), non-covalent van der Waals interactions constitute key aspects of many important biological processes such as protein folding and solvation, enzyme allosteric pathways, and electric dipole oscillations in biomolecules [291, 934]. Whereas covalent bonds are primarily responsible for the formation of individual molecules, noncovalent interactions can act over long ranges, reaching from several Ångstroms (in the case of van der Waals, electrostatic, hydrogen-bonding, and $\pi$–$\pi$ interactions) to hundreds of nanometers (for hydrophobic interactions) [935]. The participation of van der Waals interactions in the formation of molecular clusters, the assembly of macromolecular structures, and other complex chemical effects has motivated a proliferation of atom- and molecule-based approaches to modeling many-body dispersion effects in chemistry, biology, and material science [936, 937].

Hydrogen-bonds arise from electrostatic interactions which are also responsible for the primary structure of biomolecules, whereas stacking interactions are due to dispersion forces that arise from the quantum mechanical interaction between fluctuating electron densities [938]. Molecular solvation effects can present even greater complications, as the bulk hydrogen-bonding network of the aqueous environment must adapt to accommodate the presence of solvent molecules through hydrophobic effects. Although the dispersion interactions arise from spontaneous correlated electron-density fluctuations [939], current solvent models and molecular mechanics (MM) force fields include them only phenomenologically as a set of classical pairwise potentials that cannot capture the collective quantum nature of dispersion effects [937]. Accounting for long-range effects of correlated electronic fluctuations poses a substantial challenge due to their collective and ubiquitous nature, which scales nonlinearly with the system size [940].

Beginning with investigations of the anomalous properties of liquid water [941, 942], a growing number of landmark investigations have revealed the failures of the classical pairwise approximations to account for dispersion effects in materials [943]. Solute/solvent polarization effects are fundamentally quantum mechanical [944], and as such the van der Waals radius is directly related to the quantum-mechanical atomic dipole polarizability [945]. The complex relaxation dynamics of the hydration layer surrounding DNA plays a key role by influencing its flexibility and stability [946], where dispersion forces allow the DNA double helix to form by balancing the electrostatic repulsion between protons bound to the base pairs [947, 948]. Dispersion forces similar to those found in stacked DNA base pairs are also found in ribonucleic acid (RNA)–protein complexes where the stacking of aromatic molecular planes is achieved using a balance of classical electrostatic and nonlocal electronic interactions [949], as well as in the stacking interactions between ribophosphate groups and nucleotide bases found in RNA molecules [950]. Although they are often mistakenly attributed to classical electrostatic forces in standard texts, the $\pi$–$\pi$ interactions that enable stacking effects between aromatic molecules—such as the base pairs that make up the "rungs" of the DNA double helix [951]—define a fundamentally quantum-mechanical form of non-bonded force [952–958].

More broadly, non-covalent $\pi$–$\pi$ interactions between aromatic rings influence a wide variety of biochemical processes that extends beyond RNA and DNA base-pair stacking to include protein folding, molecular self-assembly and recognition, and the formation of supramolecular structures [959]. Dispersion forces have been implicated in water-mediated correlations in DNA-enzyme interactions [960, 961] and proposed to play a key role in some drug interactions [962]. Likewise, water molecules have been found to play a key role determining the specificity, kinetics, and thermodynamics of molecular DNA–drug complexes [963]. Dispersion forces are also predicted to play a key role in the dynamics of intrinsically disordered proteins (IDPs) [964] that transcend the classical biomolecular structural-functional paradigm (*i.e.*, the lock-and-key model [965]). Unlike folded proteins that have definite, compact three-dimensional structures, IDPs do not naturally exhibit distinct folded states [964]. Instead, IDPs exhibit glassy dynamics with a broad distribution of relaxation times under protein extension [966], making them suitable testbeds to investigate dispersion models (similar to studies of glassy dynamics in other media [967, 968]).

Although pairwise corrections to conventional molecular models can account for dispersion effects in biological systems of interest such as IDPs [964], those schemes are not always sufficient to address energetic aspects of van der Waals interactions that arise directly from the underlying quantum mechanics [939, 969], such as collective wavelike fluctuations that result from the nonlocality of the charge density in a wide range of biologically-important nanostructures [970]. Experiments have shown that the $\pi$-stacking of base pairs facilitates long-range charge transfer along DNA strands [971]. This has motivated broad interest in quantum transport in DNA assemblies [972, 973], coherent charge transfer along sequences of DNA base pairs [120, 974], and electronic properties of DNA assemblies [975, 976]. As a result, developing DFT-based methods that accurately account for collective dispersion effects has become a major goal in computational chemistry and materials science [943, 947], particularly for the study of proteins and DNA in an aqueous environment [939, 956, 977].

In contrast with IDPs, most globular, fibrous, and membrane-bound proteins are well-folded, with functions that are determined by a definite hierarchy of primary, secondary, tertiary, and quarternary structural motifs. Protein folding is primarily governed by hydrophobic forces which drive the protein into its stable folded state, where dispersion forces are enhanced by the tightly-packed protein interior [978]. Protein folding is a central problem in computational biology [979], where it is still unclear how the newly-synthesized peptide sequence relaxes into its folded state when the number of accessible states scales exponentially with respect to the sequence length [928]. Anfinsen's dogma dictates that the stable folded state of a protein must correspond to its thermodynamic free energy minimum [980], but the question of how a protein folds into this state in fixed time remains unanswered because the descent to the optimal folded state through conformational "landscape" of the protein is hindered by a plethora of topological folding restrictions [928]. Efforts to solve the protein-folding problem have circumvented this issue (known as Levinthal's paradox [981]) by employing machine learning (ML) techniques to predict protein structures empirically [982, 983], for which Baker, Hassabis, and Jumper received the 2024 Nobel Prize in Chemistry [984]. The problem of finding the protein fold that minimizes the free energy in an exponentially large search space has motivated interest in the use of quantum computing to solve hard problems in molecular biology [979, 985–987].

Unlike proteins, for which folding is primarily governed by hydrophobic interactions, single-stranded RNA folds in a hierarchical process controlled by the hydrogen bonds formed between nucleotides [988]. As a result, relaxation to the folded RNA tertiary structure is relatively slow, but results in definite, thermally-stable structures with relatively little flexibility [989]. In the absence of accurate classical force fields to use for structure prediction, intuitive physical arguments have been combined with phenomenological considerations to develop coarse-grained models with simplified effective potentials to simulate folding in DNA, RNA, and proteins [990]. Conformational changes can also have a dramatic effect on the electronic properties of DNA, necessitating the use of multi-scale modeling techniques and quantum calculations to predict electrical transport properties of stretched DNA [991].

Charge transport mechanisms in nucleic acids can differ markedly from those found in proteins [992], and DNA repair by the enzyme photolyase has been shown to depend on the intensity and direction of an external magnetic field [993]. DNA polymerase enzymes contain FeS clusters that enable the formation of active enzyme–nucleotide complexes [994] which can exploit the remarkable sensitivity of charge transport to DNA integrity [995]. Charge transport plays an essential role in DNA synthesis, where it is critical to detect and repair DNA mismatches [994]. These occur as a result of quantum tunneling effects which can produce mismatched base pairs via mutagenic tautomerism [996–998]. Studies elucidating the role of quantum mechanics in genetic mutations have motivated the theory that random point mutations may be directed adaptively by organisms according to the principles of open quantum systems theory [999], in line with proposals by McFadden, Al-Khalili, and Ogryzko [1000–1002]. Quantum mechanics has also been proposed to play a role in epigenetic mechanisms based on the charge transfer properties and CISS of DNA [1003].

Photoexcited electronic states of DNA have attracted scientific interest because of their role inducing genetic damage in the form of DNA lesions [1004], as well as the possibility that photogenerated radical pairs could be used in quantum information science applications [1005]. Infrared light was also proposed to guide the seemingly random motion of DNA chromatin in collective dynamics that provide context to the cell about how to translate cellular chemical and electrical gradients into the activation of specific genes. In that model, infrared light would mediate collective interactions amongst the chromatin segments of DNA in the cellular nucleus via a superradiance effect in a form of quantum optical information processing, enabling the cell to engage its entire genome in electrochemical context-based decision making and regulation [1006]. That idea was motivated by Babcock and coworkers's findings that the enhanced quantum optical effects observed in microtubules using fluorescence spectroscopy could be attributed to cooperative superradiant interactions exhibited by arrangements of Trp residues [70].

"Superradiance" is a quantum mechanical effect in which spontaneous light emissions are enhanced by entanglement between light emitters, whereas "subradiance" is the entanglement-mediated inhibition of spontaneous emissions [1007]. Superradiance has been called one of the most enigmatic phenomena in the history of quantum optics [1008], owing in part to its character as a true quantum effect which cannot be accounted for by semiclassical approximations. Superradiance has been linked to improved computation times in quantum computing algorithms based on quantum annealing (QA) [258], where the use of dissipation may be important for accelerating quantum algorithms that solve hard computational problems using quantum mechanics [1009, 1010]. Beyond research on superradiance in microtubule Trp networks [70, 659], studies have indicated roles for superradiance effects in actin filament bundles, amyloid fibrils [1011], and photosynthetic light-harvesting complexes [354, 1012–1014].

Cytoskeletal microtubules became a major topic of interest in quantum biology after nonlinear electrodynamic phenomena in microtubules were proposed to provide a theoretical explanation for the concerted organization of biological activities in the pioneering work of Hameroff, Smith, and Watt [1015]. Since that time, microtubules have been proposed as a potential source of quantum effects in biology [733, 1016]. Microtubules have been linked to a wide range of biological information processing functions including ciliary and flagellar control, axonal transport, and cognition [1017, 1018], with the understanding that "cognition" encompasses the whole set of mechanisms underlying the acquisition, storage, and processing of information at all levels of cellular organization [1019].

Though noise is often considered an "enemy" of signal detection, random fluctuations play a beneficial role in enhancing detection of weak signals in nonlinear systems that possess delicate detection thresholds [1020, 1021], such as precision photodetector systems [1022]. Microtubules are key developmental regulators that function at the intersection between biochemical and mechanical growth control [1023] where they respond to various forms of chemical, physical, and mechanical stresses [1024, 1025]. As such, microtubule fluctuations are crucial for sensing, learning, and memory [1026, 1027]. Indeed, label-free microscopy has revealed that switching between disassembly and assembly phases of microtubule growth is stochastic [1028]. Microtubules were also identified as effective light harvesters with energy transport properties that are not explained by conventional resonance energy transfer (RET) theory [78] and fluorescence properties that are not explained by the conventional theory of fluorescence [70].

Far from being counterproductive, noise is strongly implicated in self-organized decision-making processes, where it provides the drive to adapt flexibly in a dynamically changing environment [1029]. Fundamental noise—*true randomness*—is an essential resource for computing tasks in many scientific fields including cryptography and optimization, machine learning, chemistry, and materials science [1030–1032]. Noise-driven signal amplification may be achieved by stochastic resonance in nonlinear response systems [1033], where a combination of spatially coherent and incoherent noise can effectively sculpt constructive and destructive interferences into an amplified signal as desired [1034]. Quantum uncertainty can provide a source of fundamental noise, and oscillator-induced synchronization has been demonstrated to amplify spin state fluctuations and thus measure the spin dynamics of individual atoms in real time using quantum stochastic resonance (SR) [815]. As necessary conditions for amplification, chirping, damping, and stationary solitonic modes have been devised for the nonlinear Schrödinger equation [1035, 1036], making SR a good candidate mechanism for nonlinear amplification effects in biology [1037].

SR has been investigated as a noise-driven signal amplification mechanism in tubulin, where the amplified signal-to-noise ratio was derived as a function of temperature for an effective potential governing delocalized electron dynamics in a tubulin dimer modeled as a bimodal well [1038]. That bimodal potential model was further used to derive the equilibrium condition for a solitonic microtubule state subject to a perturbation [1039]. Microtubules provide the structural framework for a cylindrical cytoskeletal structure known as the axoneme in cellular appendages such as motile cilia and flagella, where a dynamic influx of $Ca^{2+}$ ions initiates and synchronizes the action of the dynein motor proteins which drive the motile power stroke to propel the cell or surrounding fluid [1040, 1041]. $Ca^{2+}$ ions are crucial to many signaling pathways in cilia and flagella, even though very low concentrations of $Ca^{2+}$ ions will disassemble microtubules (unlike $Mg^{2+}$ ions which were found necessary for microtubule polymerization [1042]). Calcium signals also control bundle motility in actin filaments [1043],

Nonlinear amplification is an essential ingredient in the formation of solitons, self-reinforcing "solitary" waves which move with constant velocity and shape [1044]. Microtubule networks [797, 799] have be proposed to support the propagation of charge-transfer solitons [692] and soliton-like "clouds" of $Ca^{2+}$ ions [1045], similar to models of solitonic electrical impulses in actin filaments [1046]. Electrical solitons have been observed in neuronal dendrites, where they were proposed to influence the generation of ionic flows [1047]. More importantly, solitonic "wetware" of this kind has been proposed to work as a liquid core waveguide for use as a polariton transmission line in a polaritonic network for applications of strong light–matter coupling [1048, 1049]. Coherenct interactions of this nature are proposed to have central importance to the spatiatemporal coordination in the cell [665, 1050], which is in turn a determining factor controlling tissue maintenance and processes such as embryological development, angiogenesis, bone formation, collagen synthesis, immunity, inflammation, and metastasis [1051–1055].

In contrast with motile cilia, primary cilia are solitary, non-motile extensions of the centriole found in most eukaryotic cells, where they operate as specialized calcium-signaling organelles [1056]. Also known as sensory cilia, primary cilia transduce optical, chemical, and mechanical signals that enable cell signaling pathways [1057, 1058]. Its essential role as the "antenna" of the cell may explain why many primary ciliary protein mutations are linked to severe developmental defects [1056, 1058]. A growing body of research is beginning to uncover the complex relationship between cellular mechanical force sensing and chemical signaling via $Ca^{2+}$ ions, hormones, ROS, and the immune system [1059]. Synchronized interstitial calcium oscillations, in particular, are crucial to enable tissue functions which have synchronization properties determined by the dissipation of coupled cellular oscillators [1060].

Calcium ions are essential to maintain homeostasis and coordinate function in mitochondria [1061], which in turn produce most cellular energy, control oxidative metabolism, maintain cellular $Ca^{2+}$ levels, regulate high-energy metabolites, and mediate apoptosis [845, 1062]. All this relies on a complex system of $Ca^{2+}$ sensors which mitochondria use regulate their internal and surrounding $Ca^{2+}$ concentrations [1063–1065]. Calcium regulation also play a key role in gene expression [1066]. Although calcium homeostasis is usually modeled in terms of an intricate system of chemical machinery made up of pumps, ion channels, and auxiliary proteins [1067, 1068], some recent proofs-of-principle have demonstrated photosensitizer-based mechanisms that use light to control cellular calcium regulation [1061, 1069–1071]. Light-based $Ca^{2+}$ signaling methods involving endogenous calcium reserves and native photosensitizers (such as flavin-containing compounds [1070]) are particularly significant, as they suggest the possibility of light-based homeostatic regulation in-and-by the cell. Indeed, mitochondria have been shown to demonstrate light-based interactions [1072], consistent with a wide body of documented examples of nonlinear cellular responses to low-intensity radiation [1073, 1074]. Mitochondria are critical integrators of calcium, ATP, and ROS signals in the cell [1062, 1075, 1076], but the mechanisms by which $Ca^{2+}$ ions elicit mitochondrial dysfunction have been elusive [1077]. Advances in our understanding of the integration cellular signals between the nucleus, mitochondria, and the cytoskeleton are inspiring a new vision of the highly multifunctional and responsive nature of these organelles and the cellular microenvironment that they control [1078–1080].

# Chapter 13 – Multiscale Modeling of Biomolecular Systems

The central dogma of molecular biology—that DNA provides the code-script for the synthesis of proteins that make of the body of life [1081]—does not solve the problem of how the greater structure of living organisms can emerge from the warm and wet environment of the cell [1082]. In fact, Schrödinger had already posed the question of how life should overcome the statistical tendency to devolve from order to disorder [39] a full decade before Watson and Crick identified the structure of the double helix (but still more than 70 years after Mieschner's original discovery of DNA). To the classical physicist, the noisy statistical nature of the biological microenviroment can only be overcome in the limit of large numbers, in which the physical processes affecting the gross structure and function of the organism are determined by the law of large numbers. To the classical physicist, ever-present statistical noise would preclude the existence of any significant biological structures on the molecular scale, in order for the organism to reap the benefit of accurate and reproducible processes governing its overall function.

It is crucial to recognize that there are many quantum effects which are not captured by quantum mechanics in its conventional form, including effects that have no analogy in standard quantum mechanics [1083]. As Schrödinger realized even in his time, distinct biological features are not limited to the gross structure of the organism and in fact can be identified all the way down to the molecular (and even atomic) scales [39]. Students of the history of science will be aware that the real genius behind the theory of the warm wet and noisy systems was Einstein, who laid the foundation for it in the third paper of his *annulus mirabilis* on Brownian motion. Jean-Baptiste Perrin later verified Einstein's theory of Brownian motion and in turn confirmed the atomic nature of matter, for which he was duly awarded the Nobel Prize in 1926 "for his work on the discontinuous structure of matter, and especially for his discovery of sedimentation equilibrium."

This point of history was not lost on Schrodinger who duly recognized it in his treatment of the warm wet noisy micro environment of the cell. Authors who quote the warm wet and noisy counter-argument typically fail to cite either Einstein or Schrodinger as a progenitor of the idea, because the actual theory of warm wet and noisy systems is definitively quantum mechanical. Rather than supporting a semi-classical picture of biology in which quantum effects are trivial, Einstein's theory of warm wet and noisy Brownian motion provided one of the first tests of quantum mechanics and demonstrated the reality of quantum theory as the basis for atomic physics.

To overcome the irreversible degradation from order to disorder imposed by the second law of thermodynamics, Schrödinger proposed that living organisms must take in a constant influx of thermodynamic free energy [39], which he described in terms of a need harvest orderliness (or 'negative entropy') to stave off one's inevitable decay into an inert mass in equilibrium with its surroundings. Schrödinger was not the first to characterize the struggle to survive in this light. It was rather Boltzmann, who's fundamental definition of entropy remains engraved on his tombstone, who wrote in 1875 that, "The general struggle for existence of animate beings is not a struggle for raw materials ... nor for energy which exists in plenty in any body in the form of heat, but a struggle for *entropy*" [615]. In this sense, the most essential question of Boltzmann's career was why any material excitation should relax to equilibrium at all.

In the 1870s, Boltzmann published his celebrated "H-theorem" [1084] to provide a statistical foundation of the second law of thermodynamics, but that theorem relied on the unsubstantiated hypothesis that the velocities of colliding particles should be uncorrelated and position independent [1085]. In the 1920s, von Neumann set forth to correct the circular reasoning in Boltzmann's arguments by developing his own quantum mechanical "ergodic theorem" [1086], but in the same vein as Boltzmann's theorem, von Neumman's relied on the statistically-independent action of a classical measurement device to provide the source of randomness necessary to bring about thermodynamic decay [1085]. Although much work was carried out over the ensuing century to find a theoretical proof of the second law, the problem remains as intractable today as it was in Boltzmann's time.

The *a priori* assumption of inherent randomness is necessary to both Boltzmann's and von Neumann's theorems to recover the principles of equilibrium thermodynamics [1087]. Although the assumption that the motions of component particles are essentially uncorrelated may be reasonable for large material bodies, it is not true in general, and it certainly falls into question in the highly coordinated microenvironment of the living cell. Nevertheless, great care is taken to recover the approximations of statistical mechanics in numerical models of molecular biology, where physical and chemical processes occur in thermal equilibrium with the surrounding biological milieu [1088]. This is a natural approach both for living organisms which adapt to the temperature of their environment as well as warm-blooded animals which maintain control of their body temperatures.

Classical molecular mechanics (MM) simulations have been used successfully to model molecular conformational changes and protein folding in countless biological structures which range from small peptides to large macromolecules in solution, membranes and membrane-bound proteins, and even macromolecular complexes such as ribosomes and nucleosomes [1089, 1090]. A wide range of physical properties can also be calculated, including thermodynamic quantities such as free energies and binding affinities [1091, 1092], as well as material properties such as densities, viscosities, and thermal conductivities [1093]. Physical interactions between atoms and molecules in these molecular simulations must account for electrostatic forces between both non-bonded atoms and bonded atoms (with the necessary equilibrium bond lengths, angles, and elastic force constants) using effective force field parameters obtained from quantum mechanical calculations. Semiclassical force field parameters must also be supplemented by quantum mechanical corrections which typically take the form of a Lennard-Jones potential which approximately accounts for the van der Waals attractions and very short-range repulsions between atoms [1094]. Polarizable atomic models based on the classical Drude oscillator approximation may also be used for simulations that must account for the explicit polarization of charged particles in polar media [1095].

In addition to the MM force fields outlined above, molecular simulations typically replicate the equilibrium statical ensembles that reproduce relevant temperatures, free energies, and other thermodynamic properties of interest. As such, a variety of methods have been developed to control simulation temperatures using Langevin dynamics [1096], as well as Anderson [1097], Berendsen [1098], and Nosé-Hoover thermostats [1099]. Physical simulations of this kind obey Newton's classical laws of motion, so that even though molecules can bend, stretch, and twist, the molecular bonds themselves cannot be broken or formed. To account for quantum effects such as the formation and cleavage of chemical bonds, molecular simulations must include quantum mechanical models [1100].

Accounting for explicit quantum mechanical effects poses a dilemma to molecular modelers, because whole biological molecules are much too large to simulation quantum mechanically—even on present-day supercomputers—but classical simulations alone cannot account for the most critical steps that occur during biological catalysis and synthesis. For this reason, experts in molecular simulation have developed "multi-scale models" techniques to enable the simulation of molecular dynamics (MD) that occur over multiple scales in biology: modeling the quantum mechanics explicitly where it is absolutely necessary (*e.g.*, at the active sites of enzymes) and treating the quantum mechanics only implicitly in the surrounding molecules and solvent where the classical approximations described above are suitable. For their work developing the molecular simulation techniques needed to span multiple biological scales, Karplus, Levitt and Warshal were awarded the Nobel Prize in Chemistry in 2013 [1101].

This work began early in the 1970s as Warshal and Karplus developed some of the first calculations to study excited states of conjugated organic molecules where the electrons are quantum mechanically delocalized across multiple bonds [1102, 1103]. That research was developed in subsequent studies of protein folding [1104], culminating in 1976 with research on the mechanism of the lysozyme enzyme reaction that demonstrated the primary role of electrostatic effects in governing enzyme catalysis [1105]. In that research, Warshell and Levitt used a multi-scale model in which a semi-empirical quantum mechanical (QM) model of the lysozyme enzyme active site was coupled to a classical model of the surrounding ions, polarized atoms, and water molecules [1105]. Those calculations relied on self-consistent field (SCF) models of the active sites based on molecular orbital molecular orbital (MO) theory using the Hartree-Fock (HF) method with corrections due to configuration interaction (CI) effects [1102].

The integration of quantum mechanical (QM) methods into classical models continued over the ensuing decades with the use of continuum solvation approximations to predict properties of solvated molecules in tandem with the coupling of molecular mechanics (MM) to QM methods in hybrid approaches [1106]. Simulations based on mixed quantum-mechanical/molecular-mechanical quantum mechanics / molecular mechanics (QM/MM) models gradually became the "method of choice" for studying enzymatic reactions involving electronic excitations or charge transfer [1107]. Based on the enormous success of these techniques [1108–1111], numerical methods for simulating enzymatic reactions are now routinely taught to students in classrooms and laboratory tutorials [1112]. Despite this success, researchers in this area have yet to establish a consensus on the best way to implement QM/MM techniques for any given application. Rather, the performance of a particular QM/MM model is likely to depend on the details of the system and the physical property of interest [1113]. The accuracy of a QM/MM method generally reflects the choice of QM method, size of the QM region, and details of the embedding potential, where the inclusion of a sufficient amount of water at the QM region can be critical to reduce errors [1114].

Enzymes catalyze biological reactions, enabling biochemical reaction rates to far exceed their uncatalyzed counterparts. Enzymes can accelerate reaction rates by many orders of magnitude [1115], raising questions of how such enormous rate enhancements are achieved [1116]. A leading proposal that catalytic rate enhancements may be attributed to the free energy invested in the electrostatic organization of the enzyme structure [1116], where the electrostatic energy cost of enzyme pre-organization is paid by the protein folding energy [1117]. Whereas most enzymatic rate enhancements may be attributed to a decrease in a reaction's free energy of activation (due to electrostatic pre-organization), these enhancements are fundamentally different from dynamical effects (represented by the quantum electrodynamic transmission factor [1118]). Quantum dynamic effects (*e.g.*, dynamical coupling between the conformational and chemical coordinates) do not contribute significantly to the catalytic action of enzymes, except in reactions involving ultrafast dynamics such as photoisomerization and electron transfer [1119].

QM/MM methods are not only applicable to "trivial quantization" effects [141] such as ground-state structures and energies, but are also increasingly being applied to study photochemically-excited states, ultrafast dynamics, and spectroscopic properties of enzymes and other biological molecules [1107]. Quantum chemical dynamics of this kind have been treated in depth by Nitzan [72], and include many common biological processes such as electron transfer reactions which occur during photosynthesis and vision, solvent-controlled and bridge-mediated electron transfer, and other excited-state relaxation effects [1120].

For example, QM/MM simulations were employed to investigate the photodynamics of vitamin D photosynthesis, providing mechanistic insights into the photodynamic processes involved in synthesizing vitamin D and presenting a theoretical explanation for experimentally observed excited-state decay rates [165]. QM/MM simulations have also been used to simulate the complete photocycle of phototropin, a blue-light photoreceptor found in green algae [1121]. Those findings indicated that the key bond formation step during phototropic photoreception relies on radical pair dynamics initiated by the coordinated action of electron and proton transfer steps, suggesting that photo-induced signal transmission in phototropin may be mediated by the altered dynamical properties that result from covalent bond formation [1121].

Although fast nonadiabatic processes are ubiquitous in biology where they enable many photo-induced reactions [1120], the simultaneous use of classical and quantum methods in QM/MM techniques can result in incorrect descriptions of quantum processes such as wave packet branching and tunneling [1122]. In fact, traditional methods for simulating enzyme catalysis using QM/MM can fall short even in electrostatic systems if they contain strong quantum correlations [272, 1123]. Strongly-coupled quantum resonance processes have also recently been proposed to play a possible role in certain aspects of cellular growth and development [604]. Several recent proposals have been aimed at replacing the QM portion of QM/MM simulations with machine-learning (ML) models for ML-guided enzyme engineering [1124–1126], but biological systems pose a challenge for these proposals due to the ubiquitous presence of quantum mechanical correlations and long-range interactions in biomolecules [1127].

For strongly coupled systems, conventional QM/MM methods (based on SCF theories) can fail to capture non-local electronic correlations that are key to the fundamental electronic and photophysical properties of enzymes [1128–1130]. To overcome this issue, multi-level embedding schemes have been proposed to allow enzymatic reactions to be simulated on quantum computers [272, 1131]. Multi-level schemes of this kind generally incorporate a virtual embedding strategy comprising an inner level where the strongly-correlated quantum subsystem may be simulated using a quantum computer, surrounded by a weakly-correlated quantum region that is represented using conventional (mean field) QM methods on a classical computer, which is finally surrounded by an outermost level of simulation that is performed using a classical continuum approximation and/or conventional MM methods [272, 1131]. To this effect, a number of hybrid methods have been proposed, such as density matrix embedding theory [1131] and dynamical mean field theory [272] following previous work using continuous-time quantum Monte Carlo (CTQMC) impurity solvers [1132]. These powerful computational methods may be particularly valuable for calculations of ligand screening used for drug discovery, which have often employed in parameterized energy functions that are computationally inexpensive but unsuitable for calculations of chemical properties [1128].

# Chapter 14 – Quantum Correlations in Biological Cofactors

Despite enormous gain in understanding that was made through characterizations of cellular metabolism during the 19$^{\text{th}}$ and 20$^{\text{th}}$ centuries [1133], there is still little known about how many small molecules and metabolites control gene expression [1134]. Proof-of-principle demonstrations have shown that pulsed electron paramagnetic resonance (EPR) can be used to coherently manipulate entangled spin states of chromophores incorporated in strands of DNA for quantum information processing applications [1005], and quantum mechanical effects have been proposed to enable synchronization in DNA–enzyme interactions [1135, 1136]. Long-distance charge transport depends intimately on the stacking arrangements of base pairs in DNA, making it sensitive to the presence of DNA lesions, mismatches, binding proteins, and chemical reactions [995]. DNA-repair enzymes that contain FeS clusters have been shown to exploit long-range charge transport chemistry to search the genome quickly and efficiently for errors [1137], but the innate quantum effects that govern these processes are only minimally understood.

Metallic active sites, like the FeS cluster contained in the DNA repair enzyme endonuclease III [1138], carry out key roles in many crucial biological functions including DNA repair, transport, and photosynthesis [1139]. Iron-containing enzymes, for example, play a central role in biochemistry where they remain poorly understood despite years of intensive study [1140–1142]. Heme proteins, known for the characteristic porphyrin ring which binds a single iron atom at the active site, are one of the most versatile classes of proteins with functional roles in electron transfer (ET), substrate oxidation, ion storage, oxygen transport, transcriptional regulation, chemical sensing, circadian reguation, and RNA processing [1143, 1144]. Their high versatility has made heme proteins a major topic of interest in areas of protein structure prediction and design for medical and clean energy applications [1144]. Yet in-depth studies of heme protein biophysics are hampered by the presence of quantum many-body effects at the heme active site, where the spin-state energetics of the system depend strongly on small changes in the active site geometry [1142]. These difficulties have led some authors to conclude that functional properties of heme proteins, such as the binding of molecular oxygen, may result from the presence of magnetic correlations, indicating that magnetic effects may be involved in the function of iron-containing proteins such as hemoglobin [1140, 1145].

Iron is also present in many non-heme proteins [1146, 1147], such as the alkylation B (AlkB) family of metalloenzymes known for repairing methylated DNA (as DNA repair enzyme AlkB [1148]) and converting straight chain alkanes into alcohols (as alkane monooxygenase AlkB [1149]). DNA repair AlkB enzymes play an essential role protecting the genome from alkylation damage, where they employ molecular oxygen to demethylate damaged DNA and RNA chains by hydroxylating the methylated nucleic acid base. QM/MM studies of the oxygen binding and hydroxylation steps of the reaction enabled a detailed characterization of the electronic and structural changes that occur during the catalysis, revealing important insights into the nature of the nucleotide- and oxygen-binding channels of the enzyme [1141]. Simplified quantum mechanical models based on HF and DFT have proven effective at modeling chemical reactions as electrostatic phenomena involving short-range correlations [1107, 1117], with examples that include simple proton transfer events and cleavage reactions such as phosphate hydrolysis [1150]. However, functional aspects of biomolecules often depend on complex nonlocal electronic correlations [1151, 1152], particularly in circumstances where there is a narrowing of the gap between the highest occupied and lowest unoccupied molecular orbitals [1153–1155].

Quantum mechanical correlations are a common feature of metalloproteins [1156], particularly those containing transition or rare-earth metal atoms configured with partially-filled $3d$ or $4f$ electron shells [1139]. Conventional methods for simulating these systems using DFT with LDA or GGA techniques fail to describe the intrinsic electronic correlations, resulting in predictions which can differ dramatically from experiments. For metalloproteins that contain transition metals such as iron (in hemoglobin), copper (in hemocyanin), or manganese (in oxygen-evolving complex), computational studies must incorporate more advanced approaches. Semilocal methods involving semiempirical "hybrid" orbitals or the Hubbart $U$ correction provide only limited improvements over standard quantum chemical models because they do not account for complex electronic correlations caused by quantum fluctuations [1139, 1157] or subtle quantum mechanical effects which occur near energy level splittings [1158].

Single electronic-configuration methods can significantly miscalculate estimates of reaction barriers even when perturbative quantum mechanical corrections are applied [1153], and the accuracies of DFT-based techniques are notoriously hard to predict [1159]. These issues arise primarily as a result of the mean field approximations used in HF and DFT which assume that the quantum system of interest can initially be approximated by single (exchange-correlated) non-interacting electronic configurations. Methods based on HF and DFT are most appropriate where the single configuration model is exact or where the interactions between multiple electronic configurations are small enough to be treated using perturbation theory [1154]. Quantum mechanical projection-based techniques have been developed to allow exact quantum mechanical simulations of enzyme active sites to be embedded in simplified DFT-based models [1160], which is itself embedded in a MM representation of the extended protein environment [1153]. Simulations of the correlated electronic effects in the exactly-embedded quantum subsystem can then be treated using higher-level correction techniques such as coupled cluster methods [1161] or spin-component scaled second-order Møller–Plesset perturbation theory [1162]. As the CI between multiple electronic reference states becomes large, methods based on correlation-corrected forms of HF and DFT may fail as the "multi-reference" character of the quantum mechanical system becomes dominant [1163].

Nevertheless, single-reference methods based on HF and DFT have demonstrated unparalleled success in computational chemistry research [1164], where they have become workhorses of molecular simulation [1165, 1166]. A vast number of techniques have been developed at varying levels of approximation, and the resulting menagerie of computational chemistry methods are arranged in increasing order of complexity [266, 1156, 1167]. These begin with single-reference approximations and the perturbative corrections thereof, and ascend to a wide range of methods to account for strong multi-reference correlations which include multi-reference CI, complete active space self-consistent field (CASSCF) theories, multiconfigurational perturbation theories, multi-reference coupled-cluster theories, and density matrix renormalization group (DMRG) methods [1168, 1169]. Although some of these techniques have even been extended to account for quantum electrodynamic effects inherent to polariton chemistry [1170], all approximate methods often fall fundamentally short of exact *ab initio* theories [1171–1173]—necessitating the use of experimental reference data against which to benchmark the quality and applicability of the many different quantum mechanical approximations [1174].

Iron-sulfur clusters perform a diversity of biological functions related to redox chemistry, electron transfer, and oxygen sensing which rely on electronic configurations with strongly-interacting entangled quantum states which are not described by HF or DFT [1175, 1176]. Symmetry-breaking introduces nontrivial physical effects [1177], and broken-symmetry DFT has been used to produce detailed and accurate computational analyses of structural models of FeS clusters in rubredoxin and nitrogenase enzymes by incorporating weighted sums of spin multiplets in calculations [1178, 1179], but this method only provides an average of the system energy spectra and does not allow for direct simulations of individual electronic states [1175]. The problem is exacerbated by the large exchange-correlation functional dependence in calculations of the open-shell electronic structure of FeS clusters that is not well understood [1180]. Extending DFT beyond descriptions of static magnetic moments associated with spin symmetry-broken states requires the inclusion of spin-orbit interactions to account for dynamical correlations involved in magnetic coupling [1181] and altered spin states due to quantum mechanical and thermal fluctuations [1142, 1182].

A sophisticated approach to the problem of simulating strongly-correlated electronic behaviour is provided by dynamical mean-field theory (DMFT), which can describe electronic interactions by introducing local quantum fluctuations the extend beyond the single-electron picture assumed in conventional quantum chemistry methods (HF and DFT) [1183, 1184]. Combining DFT and DMFT, the DFT+DMFT approach is suitable for simulating magnetic nanostructures and metal clusters similar to those found ubiquitously in biological systems [1185–1187]. Combined DFT+DMFT has been used successfully to model entangled spin state transitions at the active site of hemoglobin, consistent with experimental observations of the transition between high and low spin states of the hemoglobin molecule [1188]. In principle the use of DMFT is limited to high-dimensional systems as the system dimension scales to infinity, but in practice it is an effective approximation for crystal lattices and ligand systems with large coordination numbers [1185, 1186]. Use of DMFT in these systems was developed in pioneering work by Metzner and Vollhardt [1189, 1190] who showed that the method could be used to simplify studies of strongly-correlated quantum mechanical systems by taking the average of the spatial quantum correlations in a scenario where the quantum mechanical time correlations remain non-trivial [1191, 1192]. This unexpected result revealed that quantum mechanical systems could exhibit strong local electronic correlations even in the absence of nonlocal spatial correlations, resulting in complex phase diagrams and exotic physical effects that make their study challenging and their applications exciting [1193].

Quantum embedding theory provides a natural framework in which to consider the interaction of a strongly-correlated quantum system with its molecular environment [1194]. Initially developed to address the infinite limit of large coordination-number lattice systems [1195], DMFT in its original form was well-suited for the study of local dynamic correlations in quantum mechanical systems (as its name implies [1191]), and was developed particularly to model strongly-correlated materials where DFT fails to predict even basic material properties [1196]. Although DMFT offers tremendous predictive power by mapping the dynamics of a strongly-correlated, infinite-dimensional quantum system onto a frequency-dependent one-particle Green's function [1192], DMFT in itself cannot fundamentally account for the nonlocal static correlations which lie at the heart of quantum mechanics. Practically since its inception [1192], DMFT has been refined and extended to address finite-coordination systems where static and dynamic correlation effects coexist [1191, 1193]. Cluster-based extensions of DMFT are important to bridge the gap between the single-site mean-field approach and the exact nonlocal many-body expansion, describing a wide range of symmetry-breaking effects which include magnetism, superconductivity, and charge and orbital order [1192].

Cluster DMFT has been used to show the importance of many-body effects to ligand binding at the active sties of hemoglobin and myoglobin [1197]. Those studies identified the metal-to-ligand bond length as a key control parameter for spin transitions in metalloporphyrin systems [1142], consistent with previous reports of entanglement in the ground state of the oxygen–myoglobin complex [1142, 1198, 1199]. Although Pauling and Coryell proposed that magnetic effects could play an important role in enzymatic systems like hemoglobin in the 1930s [1200], the role of magnetism in the functioning of heme proteins has only recently been defined as a research topic in quantum biology [1140]. Many-body DMFT has proven itself as a powerful method in studies of biomagnetism, where it simultaneously accounts for both electron valence and spin fluctuations when combined with large-scale DFT calculations [1198]. Indeed, large-scale DFT simulations have had a major impact on biological studies where it is common for large simulations of 500 atoms or more to be required to predict biological properties of interest [1151].

Large scale quantum mechanical simulations are likely to continue to play an important role in investigations of complex organometallic structures found in countless biological catalysts, such as the tetramanganese–calcium ($Mn_4Ca$) cluster at the core of the oxygen-evolving complex of photosystem II (PSII) [1201]. Metal-ligand systems with multiple metal centers [1202] are likely candidates for investigations of magnetic effects in systems with properties that are controlled or enhanced by quantum magnetic effects such as superexchange [1203, 1204]. In spite of their wide success in modeling metabolically-important biochemical processes, the accuracies of conventional QM/MM methods are still not well understood [1113]. Complex biochemical systems present a challenge to investigate using electronic structure methods because of these systems' large sizes, numerous subtle interactions, coupled dynamical timescales, and nonadiabatic electronic effects [1160]. Strong quantum mechanical correlations are present in many biological photosystems and catalysts which make them inherently different from many systems which have been successfully modeled using conventional quantum chemical theories [1205]. This makes it necessary to employ comprehensive models of static and dynamic electron correlations involving the use of sophisticated multi-reference (MR) quantum mechanical calculation methods to obtain efficient and quantitatively accurate descriptions of many biochemical processes [1206–1210]. Noteworthy examples of methods which extend the conventional QM/MM formalism include quantum embedding methods based on projector techniques, as well as hybrid, multilayered and fragment-based methods, not to mention density functional- and density matrix-based embedding schemes [1131, 1160, 1211, 1212].

Sophisticated computational methods are accompanied by the need for increasingly-advanced numerical techniques to implement them, including high-level parallel computational packages designed to carry out DMFT and continuous time quantum Monte Carlo (CT-QMC) algorithms on classical supercomputing architectures [272, 1213, 1214]. The vast computational resources required to perform these simulations have motivated interest in the use of quantum computing to study biochemical systems [272]. Although the computational armory of numerical quantum chemistry is well-stocked with methods that range from primitive Hartree–Fock techniques to highly ac-curate coupled-cluster and CI methods, exact treatments of quantum systems with more than a few dozen electronic orbitals remain intractable on classical computers [1131]. Fortunately, embedding methods developed to simulate complex quantum mechanical systems on classical computers are readily translated to hybrid quantum/classical computing schemes where some parts of the computation are executed on classical computers while other parts are executed on quantum computers [272, 1131]. These methods suggest a practical and systematic path to multiscale descriptions of quantum systems for use in quantum simulations of large, realistic molecules [1131, 1215, 1216].

Quantum dynamics play a critical role in the excited state properties of magnetic molecules and materials [1217] like the FeS clusters found in the enzyme nitrogenase which exhibits strong correlations amongst its eight transition metal atoms and has been called a holy grail of quantum chemistry [1131, 1218]. Quantum information processing will be critical to predict the dynamics of complex quantum systems such as these, which are beyond the reach of simulations that employ even the most advanced classical algorithms [1219]. For example, the molecular fragmentation pathways of the photo-dissociating sulfonium ion $H_3S^+$ have been simulated on a superconducting quantum processor, demonstrating the usefulness of hybrid quantum/classical algorithms at simulating molecular decay spectra [1220]. The ubiquity of ultrafast nonadiabatic processes in biological systems have made investigations of quantum nonadiabatic molecular dynamics an important topic of study for small molecular systems using high-level electronic structure calculations which account for the quantization of nuclear motion [1122, 1221]. Whereas a decisive demonstration of an advantage at simulating quantum dynamics would appear to require achieving the elusive goal of developing a fault-tolerant quantum computer, present-day and noisy intermediate-scale quantum (NISQ) devices may provide as-of-yet unclear benefits in efforts to simulate molecular dynamics [1222].

Quantum computers offer a diverse range of potential advantages in biochemical and biophysical computations, where even the preparation of a suitable initial state (such as a molecular ground state or a thermal Gibbs state [1219, 1223, 1224]) may constitute a hard computational task [1152, 1225]. Thus, implications of quantum computing for computational biology have broad applications which range from quantum mechanically embedded simulations of enzyme catalysts and reaction dynamics to protein folding, molecular recognition, nucleic acid research, and functional genomics [269, 979, 986, 987, 1226]. The development of quantum computing algorithms for biochemistry are likely to have major applications in pharmaceutical discovery, material design, and catalyst optimization [1131, 1227–1230].

Nuclear quantum dynamics beyond the (quasi-static) Born–Oppenheimer approximation are also becoming recognized in a growing number of biochemical processes [68]. Quantum simulations of nonadiabatic dynamics involved in photophysical internal conversion and intersystem crossings, the Jahn-Teller effect, and vibrationally-assisted energy and/or electron transfer reactions provide natural starting points for investigations of this kind [1122]. Simulations of the conical intersections which couple the diabatic energy surfaces during nonadiabatic reactions have been carried out on trapped ion quantum computers, providing insights into the entangled nuclear and electronic (spin) dynamics which occur around quantum singularities as the geometric phases accrued there become non-trivial [1231]. Nonadiabatic dynamics can also become significant in photoinduced PCET which are crucial to long-range signaling in cells [234]. Quantum dynamics are particularly relevant in photo-activated phenomena which are key to essential aspects of biological systems including photosynthesis and light-harvesting, the photostability of DNA, and various forms of photocatalysis [1222, 1232–1234].

# Part III: Nanomedicine & Biotechnology

> Physics constitutes a logical system of thought which is in a state of evolution, whose basis cannot be distilled, as it were, from experience by an inductive method, but can only be arrived at by free invention.
>
> — Albert Einstein, 1936 [1235]

## Chapter 15 – Photobiomodulation & Electromagnetic Therapies

While the healing power of sunlight has been known since antiquity [1236–1238], light-based therapy was established for the modern world when the 1903 Nobel Prize in Physiology or Medicine was awarded to Niels Ryberg Finsen "in recognition of his contribution to the treatment of diseases, especially lupus vulgaris, with concentrated light radiation, whereby he has opened a new avenue for medical science" [1239]. In addition to solar (UV) treatment for dermal tuberculosis, Finsen also developed phototherapy using red light to treat smallpox [1240, 1241]. However, subsequent work on phototherapy lapsed until the mid-1960s when the regenerative powers of red light were rediscovered [1242].

The first research demonstrating that ruby laser light could accelerate wound healing and hair growth was carried out in 1967 by Mester at Semmelweis Medical University in Hungary [1238, 1243] in an attempt to reproduce research by McGuff *et al.* ablasing malignancies in rats with the (recently invented) ruby laser [1243, 1244]. Cancer mitigation was unsuccessful because of the low power of the laser he used, which produced enhanced tissue generation instead. The resulting treatment was originally known as low level laser/light therapy (LLLT) for this reason, garnering interest from National Aeronautics and Space Administration (NASA) scientists for the prospect of accelerating wound healing in space [1245]. The therapeutic application of light to modulate tissue function (by activation or inhibition) is potentially important for many physiological processes such as angiogenesis, collagen and cartilage production, muscle and nerve regeneration, and neovascularization, as well as decreased inflammation, edema, and pain [79]. More than 4000 laboratory studies have been performed on LLLT, as well as more than 400 Phase III randomized, double-blind, placebo-controlled trials [1246].

It is now evident that photophysical forces applied at low energies and specific frequencies can have profound effects on physiological processes [1247]. The scope of these photobiomodulation (PBM) effects have been broadened to include other light sources such as light-emitting diode (LED)s, mainly in the red to near infrared spectrum [161]. Moreover, use of the term LLLT to refer to PBM techniques has become discouraged, because the broad connotations of "low level light" could also be applied to implementations of photodynamic therapy (PDT) and optogenetic techniques wherein low-doses of laser or LED light can be used to activate photochemical switches, rendering the term LLLT overly vague and misleading for precise usage in medical physics [1237]. The need for a more specific term to characterize the use of red through infrared light for the (nonthermal) biological stimulation of tissue repair and vascularization, as well as pain reduction and mitochondrial ATP production, has lead to the wide adoption of the term PBM [79, 1244], which is formally defined as any "nonthermal process involving endogenous chromophores eliciting photophysical (*i.e.*, linear and nonlinear) and photochemical events at various biological scales" [1248].

Hence, PBM therapy encompasses both conventional low intensity laser and LED-based treatments [79, 1248], as well as the growing potential for high-intensity LED-based irradiation in hard tissue regeneration therapy [1249]. PBM effects depend on the dose (intensity, duration, and wavelength) of light. During PBM therapy, light is usually applied from an emitter in the 10 mW to 500 mW power range [1246] using monochromatic light at a wavelength in the 600 nm to 1000 nm region of the optical spectrum (*i.e.*, red to near infrared) [1250], affording the added clinical advantage of treatment within the 650–1000 nm tissue-transparency window [1251]. These power levels and wavelengths have the ability to penetrate skin as well as other soft and hard tissue structures, and were demonstrated to produce good effects on pain, inflammation and tissue healing in various clinical studies [1246].

For the treatment of musculoskeletal injury and pain, an optical power density of $\sim$ 5W/cm$^2$ in red to nearinfrared spectrum (660–905 nm) may be applied to the injured or painful site for up to a minute a few times a week for a period of weeks to reduce inflammation and pain while accelerating tissue regeneration [1246]. PBM using far red and NIR light was found to reverse effects of neurotoxins on neuronal cell cultures, supporting the hypothesis that PBM up-regulates cytochrome *c* oxidase to increase energy metabolism in a signaling cascade that prevents poisoned neurons from undergoing cell death [1251]. PBM using red and NIR light has also been shown to have therapeutic effects on neurological diseases and nerve injuries including stroke, traumatic brain injury, Parkinson's disease, Alzheimer's disease, and major depressive disorder [1252–1256]. Therapeutic benefits are also reported for PBM using blue light [1257], particularly for use in dermatology and would healing [1258], and PBM therapy is also being studied for its capacity to improve skin health, joint mobility, and athletic performance [1259]

PBM mechanisms are primarily believed to operate via photoactivated retrograde signaling pathways whereby mitochondria relay their functional status to the cell nucleus [1257, 1260]. This was proposed by Karu as a result of the appearance of a 300–860 nm action spectrum for DNA and RNA synthesis of cell cultures, even though cell nuclei do not have dedicated chromophores to absorb in this region [1250]. Instead, those action spectra data revealed PBM-active chromophore residing in the respiratory chain, implying the existence of a "retrograde" signaling mechanism from the mitochondria to the nucleus. Retrograde signaling is also a key factor in cellular homeostasis and regulation of neuronal function [1261]. Cytochrome *c* oxidase, hemoglobin, and myoglobin are the three main photo-acceptor molecules known to absorb NIR light in mammalian tissues, but of these three only cytochrome *c* oxidase is directly associated with energy production [1251]. Thus, PBM is believed to improve cell viability by stimulating mitochondrial function [1245] via light–matter interactions with mitochondrial chromophores [1243].

Although there is ample evidence that PBM is mediated by mitochondrial retrograde signals in irradiated cells, cellular responses to PBM initiated in the nucleus have also been documented, and a number of PBM mechanisms have been proposed that do not rely on cytochrome *c* oxidase [1250, 1262]. In particular, new evidence has shown that PBM at 660 nm can stimulate cell proliferation without the involvement of cytochrome *c* oxidase [1263]. This indicates that a number of other candidate molecules such as ROS, genetic transcription factors, and hypoxia-inducible factors are also involved in PBM-induced changes that enhance cell metabolism and growth. PBM using NIR light has also been shown to stimulate assembly and disassembly in tubulin and microtubules, respectively [1244]. Even though PBM is used to treat a large number of ailments, it has been slow to achieve wide acceptance because its underlying biochemical mechanisms remain unlear. Partly for this reason, optimal PBM treatment parameters (such as wavelength, power density and fluence, pulse shape, and treatment timing) are not known definitively [1264, 1265], although suboptimal parameter choices can reduce treatment effectiveness or result in negative outcomes [161].

Despite its broad range of applicability, widespread adoption of PBM treatments have been limited by sparing knowledge of its physiological basis [161, 1245, 1258, 1266]. Full characterizations of the effect of PBM on living cells have remained elusive because of the uncontrolled variations that distinguish different PBM protocols, although progress is being made to establish more favorable methods for promoting the proliferation and differentiation in stem cells *in vitro* [1267, 1268]. Mitochondrial function is linked to cytosolic $Ca^{2+}$ ions, ROS, and adenosine monophosphate (AMP) levels, and changes in these levels can activate a diversity of retrograde signaling pathways [1269]. PBM can initiate mitochondrial retrograde signaling by way of a $Ca^{2+}$ ion and cyclic AMP-related cascades which modulate cellular functions (such as proliferation and migration) and can result in tissue regeneration [1270]. Under normal circumstances, retrograde signals are linked to mitochondrial biogenesis, cell phenotype, and metabolism. However, the exact role of retrograde signaling in mitochondrial dysfunction is complex, and the mechanism by which mitochondria influence gene expression is still unclear [1271, 1272].

PBM has been shown to stimulate growth and metabolic activity in mammalian cells, yeast, and bacteria [1250, 1260, 1273]. PBM is proposed to influence physiological (*i.e.*, neuronal and circadian) rhythmicity [1247], consistent with findings that PBM effects exhibit a biphasic dose–response curve and that biological stimulation for a given cell or tissue type occurs only through a therapeutic dose window [1268, 1274]. There is growing evidence that PBM may be indicated in cancer therapy, both as a primary treatment method [1275] and for chemotherapy-related neuropathies [1255, 1256]. Furthermore, preliminary studies show that PBM may be combined with other therapeutic applications including hyaluronic acid and gold nanoparticles to significantly accelerate tissue healing [1276].

As with PBM, electromagnetic effects have been ascribed to a plethora of biological processes ranging from wound healing and cell migration to gene expression and DNA synthesis [1277–1279]. Although magnetic fields have been known to interact with living organisms for centuries [1280], interest in the influence of electromagnetic fields on biology and ecology grew over the 20th century [1281] as electronic devices came to influence practically every aspect of culture—from health and medicine to science, engineering, and communications. Public interest in the health effects of ambient electromagnetic fields has grown [1282], with studies finding that these fields can exert stimulating or inhibiting effects on immune function [1283].

Pulsed electromagnetic field (PEMF) therapy influences biological systems through the application of short bursts of electromagnetic fields which are both tissue and frequency-specific in their effects [1284, 1285]. It has emerged as a promising method for treating chronic inflammatatory and immunological conditions such as rheumatoid arthritis by targeting mesenchymal stromal cells through modulation of differentiation, proliferation, cell cycle regulation, and metabolism, as well as cytokine and growth factor secretions [1284, 1286]. PEMF is proposed to affect cellular function and differentiation by altering intracellular calcium ion concentrations and restoring potassium channel activity, down-regulating inflammatory cytokines, heat-shock proteins, and proangiogenic molecules to make it possible for cells to commence regeneration of healthy cartilage. The use of PEMF as an inflammation and immune system modulator is considered relatively safe in comparison with other broad immunosuppression techniques currently used to treat arthritic conditions [1284]. Likewise, extremely low frequency (ELF) electromagnetic fields have been found to accelerate wound healing [894] and have been used to treat pain and inflammation without causing cytotoxic or genotoxic effects in human mesenchymal stromal cells [1287].

Oscillating electromagnetic fields have been shown to induce cell autophagy *in vitro*, promoting homeostasis in a cytoprotective effect [1288]. Furthermore, combinations of $\mu$T-scale static and alternating magnetic fields were found to reduce amyloid plaques in the cortex and improve memory in transgenic mice used to model Alzheimer's disease, suggesting that magnetic fields may combat neuronal degeneration early on in brain diseases with Alzheimer's-like symptoms of amyloid deposition [1289]. Pulsed electromagnetic fields have been demonstrated to have beneficial effects on an intracerebral hemorrhage (ICH) model system *in vitro*, protecting against brain injury by alleviating inflammation and promoting hematoma clearance [1290]. As such, transcranial magnetic stimulation is being clinically investigated as a potential non-invasive therapy for neurological and psychiatric disorders [1291].

While the exact physiological basis for these effects remain to be fully understood, it is clear that practically all living cells are diamagnetic [1292]. In fact, electromagnetic fields have been used clinically to aid in the healing of delayed union and non-union bone fractures for several decades [777, 1293], and magnetic fields have been shown to induce deformations in bone-forming osteoblast cells *in vitro* [1294, 1295]. PEMFs have also been shown to induce ROS production levels in a cryptochrome-dependent effect in mammalian cells [873]. Those findings prompted interest in the potential use of magnetic field therapies to influence circadian clock activity to control such factors as glycolysis, lactate production, extracellular acidification, the ADP/ATP ratio, and cytosolic ROS levels—with substantial implications for progressive conditions such as inflammation, infarction, cancer and stroke [1296].

Cancer cell proliferation can be inhibited by amplitude-modulated electromagnetic fields using tumor-specific frequencies [1297–1299], and PEMF stimulation has demonstrated promise in treating a variety of cancers (including bladder, breast, colon, liver, lung, ovarian, pancreatic, prostate, skin, and thyroid), although only a few limited examples of PEMF cancer treatments have been demonstrated on humans [1300, 1301]. This is consistent with the picture of a cancer-like phenotype in which cell metabolism is upregulated in the absence of a magnetic field, accelerating anaerobic glycolysis [771] and in turn reducing ROS production [875, 876]. Although relatively little is known about the potential benefits of combining PEMF and radiation therapies in cancer treatments, the radiosensitizing effect of electromagnetic field treatments hold promise to increase the efficacy of established radiotherapies by enhancing the effects of ionizing radiation effect on cancerous tumor cells [1302].

Studies have shown that specific PEMF protocols can induce membrane permeabilization and apoptosis in medulloblastoma cancer stem cells *in vitro*, where ROS generation was observed in the surviving cells together with cell-cycle arrest associated with a senescence-associated phenotype [1303]. Those findings reveal a promising new therapeutic option to target medulloblastoma cancer stem cells which may provide an effective means to overcome radiotherapy resistance in cancer stem cells. This finding is particularly important because medulloblastoma is a highly malignant tumor of the cerebellum that constitutes about 20% of all primary childhood central nervous system cancers—which have been characterized as the leading cause of death in children and adolescents [1303].

Transduction of electromagnetic signals to implement control of cell processes—ranging from nerve impulses to gene expression and apoptosis—have been demonstrated both *in vitro* and *in vivo* [1291]. Chronic ROS levels can promote cell proliferation and tumor metastasis, whereas a sudden burst of ROS can initiate apoptosis immediately [1304, 1305]. In fact, cellular ROS levels are increased by most chemotherapeutics today, which often alter redox homeostasis in cancer cells as part of their drug action [1306]. In contrast, iron oxide ($Fe_3O_4$) magnetic nanoparticles delivered to the tumor can generate ROS under the influence of magnetic fields, modulating the processes governing mitochondrial catalysis, DNA damage, and (abnormal) cell growth—and consequently altering cell morphology and ROS production while *promoting* tumor growth [1307].

Blocking cellular ROS production can also reduce cancerous cell populations in cancers that are characterized by ROS overproduction, effectively inhibiting oncogenic signaling [1308]. Extremely low frequency electric fields are widely reported to contribute to the risk of some types of cancer [678, 1309], and non-ionizing wireless radiation from wireless devices is increasingly being considered as a possible carcinogen [1310]. Under certain conditions, PEMFs have been found to exacerbate the cytotoxic action of DNA-damaging agents bleomycin and methyl methanesulfonate [1311] Nevertheless, cancer-suppressing magnetic fields have been used to *inhibit* tumor growth over a broad range of tumor cell lines representing lung, kidney, gastric, and pancreatic cancers [1312]. Thus, the potential of magnetic field-based treatments for cancer and other ischemic disease is diminished by the lack of clear biological mechanisms underlying the inconsistent effects of WMFs and ROS on cancer metabolism [1306], presenting an urgent need for fundamental advances in this area.

Healthy, neoplastic, and cancerous tissue differ substantially in terms of electromagnetic and mechanical properties [1313]. While differences in tissue stiffness may be well-known (*i.e.*, thanks to conventional medical guidance to check manually for tumorous lumps), electromagnetic methods of imaging cancer and other pathologies have become a mainstay of medical science, subject to ongoing development and improvements [1314–1316]. More recently, advanced imaging methods been developed using non-linear resonance interactions with weak electromagnetic fields to improve the discrimination of cancer, bone disease, inflammation, and vascular disorders [1317]. The tissue resonance interaction method (TRIM) has been shown to detect malignant tumors in the bladder [1318], colon [1319], and liver [1320], holding promise for cancer diagnostics without the need for invasive procedures.

Nuclear magnetic resonance (NMR) imaging, notably, has been used to visualize metabolic and physiological signs of health, and for the prediction and early detection of responses to cancer treatment, allowing for the possibility of NMR-guided individualized cancer therapies [1321]. Electromagnetic fields have been found to affect many biological functions such as gene expression, cell fate, and cell differentiation [1285]. Studies on the effect of NMR modulation of cell proliferation and gene expression have motivated interest in the therapeutic use of NMR technologies to treat a variety of medical conditions such as nerve injuries [1322]. Therapeutic NMR has been found to interact with cell hypoxic signaling pathways through circadian clock mechanisms via cryptochrome and clock protein expression levels according to a nonlinear dose–response curve [1323]. This has prompted a hypothesis that quantum mechanical effects based on the RPM underlie the effect of therapeutic NMR on cell metabolism [1296].

While there is some broad understanding of the physiological mechanisms that are likely to be involved in mediated cellular interactions with electromagnetic fields via PBM and PEMF treatments, the physiological action mechanisms of static magnetic fields are more enigmatic [1279]. As previously outlined, radical-pair recombination reactions provide established mechanisms by which magnetic fields can interact with chemical dynamics in biological systems [1280], altering the activities, concentrations, and lifetimes of paramagnetic free radical species [1324]. Low-intensity static magnetic fields are known to influence a wide range of cellular photochemical reactions [436]. Thus, as an uncontrolled factor in biological experiments, variations in ambient magnetic fields may alter the production of free radicals and heat shock proteins, producing significant changes in rates of cell replication, differentiation, and survival [1325]. Biological electron transfer is implicated in cellular responses to WMFs because the sudden transfer of an electron can result in the formation of a magnetically-sensitive radical electron pair. Low field effects (LFEs) of this kind have been implicated in a wide range of physiological processes ranging from the development of cancer metabolism (*i.e.*, the Warburg effect) to beneficial cellular regeneration effects. Likewise, electromagnetic fields have been found to alter redox metabolism, biological ROS signaling, and cell growth *in vitro* [1326]. Findings of this nature have inspired proposals to assess and limit the environmental risks associated with long-term human exposure to ambient electromagnetic fields [1288, 1327].

Cancer cells display distinctive bioelectric, biochemical, and biomechanical signatures during tumorogenesis [1328, 1329], which is known to coincide with the bioelectric decoupling of those cells from their surrounding somatic environment, where the cell's bioelectric state can either induce metastatic behavior or suppress tumorigenesis [1330]. In spite of these research efforts, the biophysical mechanisms that stem cells use to sense and interpret electromagnetic cues in their environment remain unclear [1331]. As a result, cancer patients are often faced with dismal prognoses and marginal recovery outcomes amidst a proliferation of treatment-resistance and recurrent cancer phenotypes [762, 1332, 1333]. Given its high mortality rate, severe side effects, and low prospects for complete remission, there is great need to understand the root causes of cancer and to develop new therapeutic modalities to combat it [1300]. These prospects are brighted by a growing body of evidence that is revealing the importance of electromagnetic fields in regulating cellular processes, particularly in maintaining tissue homeostasis, cell differentiation, and organ morphogenesis [1334].

# Chapter 16 – Photodynamic Therapy & Nanotheranostics

In the 1960s, a general method of carrying out photochemical transformations "without light" was proposed by White, Wiecko, and Roswell [1335, 1336]. In this method, the need for direct interaction with light in a photochemical transformation is circumvented by the use of a primary "photosensitizer" molecule which is prepared in a photophysically-excited spin triplet (or singlet) state [1336]. The resulting photoproducts can then promote high-energy biochemical transformations without light. These have immediate importance for nucleic acid photodamage and repair, flavin photochemistry, vitamin D production, lipid peroxidation, urocanase activation, chromophore excitation, and aerobic metabolism [607]. Energy transfer from photo-excited species to cellular targets will often result in luminescent products [610, 1337], and demonstrations of biological ultraweak photon emissions reveal that the phenomena of "photochemistry without light" is in fact a universal characteristic of living organisms.

Flavin and porphyrin coenzymes are natural photosensitizers which act as strong oxidants under photo-excitation [1257, 1338–1340]. The extraction of hematoporphyrin from dried blood in 1841 marked the first isolation of a photosensitizer molecule from its biological milieu [1340]. Its structure took the better part of a century to characterize, culminating with the 1930 Nobel Prize in Chemistry to Fischer for his research on the structure and synthesis of porphyrins. Building on the work of 1915 Chemistry Nobel Laureate Willstätter, Fischer showed that chlorophyll and heme are structurally related, being formed from the same pyrrole constituents, where heme carries a single iron atom at its center in the same way that chlorophyll holds an atom of magnesium [1341]. The structural similarities shared by chlorophyll and heme make them essential for energy harvesting and transduction in plants and animals [1342, 1343]. For this reason, heme and chlorophyll have been called "the two most important pigments for life on earth" [1344].

The effects of combining specific dyes with certain frequencies of light has been known since the beginning of the 20$^{th}$ century, when Raab observed that acridine red dyes would kill paramecia when exposed to light but not in darkness [1345]. The term "photodynamic reaction" was later coined by von Tappeiner in his work on studying the role of oxygen in the photo-sensitization of fluorescent substances [1346]. With his colleagues, von Tappeiner carried out the first tests to explore the therapeutic benefits of the photodynamic effect in a trial on patients with facial basal cell carcinoma using the photosensitizer eosin, treating the affected areas with a 1% eosin solution in sunlight or arc-lamp light, and finding that two thirds of the patients treated experienced complete tumour resolution and a relapse-free period of twelve months [1347, 1348]. Systematic studies on tumor localization and the phototoxicity of porphyrins were carried out throughout the mid-20th century, when Lipson *et al.* developed hematoporphyrin derivative (HPD) by chemically modifying hematoporphyrin [1349, 1350]. Injections of hematoporphyrin preparations and derivatives led to selective tumor fluorescence into oncology patients, establishing the basis for modern day photodynamic therapy as HPD became the precursor the first photosensitizer to be approved by the Food and Drug Administration (FDA) [1351].

PDT has emerged as a modern, non-invasive form of therapy which can be used to treat cancers as well as many non-oncological conditions. It has been applied in areas of dermatology, oncology, gynecology and urology, and in treatments of chronic inflammation and drug-resistant bacterial infections [1352, 1353]. Treatments entail the local or systemic application of the photosensitizer, a photosensitive compound which accumulates in the affected area(s) where it is activated by light to initiate the targeted destruction of pathological tissue or cells [1337, 1352]. Light-initiated ROS production constitutes the primary action mechanism of PDT treatment, so effects of PDT are abolished in the absence of oxygen [1354]. PDT protocols are typically painless, efficient, simple to use, free from side effects, and well-tolerated by patients because of PDT's tissue-selective action [1352, 1355].

During PDT, the absorption of a photon transforms the prescribed photosensitizer molecule from its ground (singlet) state into a relatively long-lived electronically excited (triplet) state via ISC from a short-lived photoexcited singlet state [1356]. The excited triplet can then react in one of two ways, either by reacting with a substrate (*e.g.*, a membrane or another molecule) by transferring a hydrogen atom to produce a radical which ultimately combines with oxygen to form an oxygenated product, or by direct energy transfer to oxygen in a quantum mechanical "triplet-triplet annihilation" reaction to form a highly-reactive singlet oxygen molecule ($^1O_2$) [1337, 1354]. Both reaction pathways can occur simultaneously, and the ratio between them depends on the sensitizer type, the oxygen and substrate concentrations, and the sensitizer–substrate binding affinity [1354]. Subsequent reactions of $^1O_2$ with organic molecules invariably produce a peroxides in chain reactions which readily yield a wide range of other oxygen-based radicals, reactive intermediates, and final products [1357]. This abundance of action has made maximizing the $^1O_2$ quantum yield a priority in photosensitizer research, where molecules can be engineered to harness radical pair dynamics and symmetry-breaking effects that enhance the $^1O_2$ and fluorescence quantum yields of photosensitizers when designing high-performance applications for PDT and diagnostics [1356].

PDT, also known as photodynamic inactivation, combines light with a biological substrate and photosensitizer to generate ROS that eradicate undesired cells involved in cancers and infections [1236, 1358, 1359]. Following Photofrin's success and developing on its limitations [1354, 1360, 1361], many photosensitizer molecules have been proposed for similar medical applications [1362–1365]. However, PDT is just one of many tools being developed that use nanomaterial platforms to combine therapy and diagnostics in the emerging field of nanomedicine [1352, 1366, 1367]. In principle, the use of targeted nanoparticles can improve the effectiveness of anti-cancer and other therapies while eliminating harmful side effects by leveraging the non-invasiveness and high sensitivity of these systems [1366, 1368–1371]. A growing number of nanodynamic therapies employ sensitizer molecules to generate cancer-killing ROS upon the application of an external energy flow such as an electric field during electrodynamic therapy (EDT) [1372, 1373], ultrasound during sonodynamic therapy (SDT) [1374–1378], or X-rays during radiodynamic therapy (RDT) [1379–1383].

Proposed in the modern context by Feynman [1384, 1385], nanomedicine encompasses the visionary idea that nanoscale structures, devices, machines and robots may be be designed and manufactured to perform comprehensive surveillance, control, repair, and defense of biological systems. The many diverse applications of nanoparticle systems include diagnostic instruments, imaging methods, targeted medicines and pharmaceuticals, biomedical implants, and tissue engineering [1384]. Nanomedicine is a growing and interdisciplinary field that integrates nano engineering and technology with the life sciences where it may eventually be integrated into all aspects of biomedical science [1386–1388]. Since Feynman put forth the revolutionary proposal that the future of technology lay at the nanoscale [1389], advances in nanotechnology have revolutionized practically all aspects of human life [1390].

Nanomedicine itself has branched off in hundreds of directions that encompass all manner of materials and devices structured to carry out medical diagnostics and clinical therapeutics at the nanoscale [1385, 1391]. Similar to PDT, chemodynamic therapy (CDT) defines a therapeutic strategy to eliminate cancer using Fenton or Fenton-like reactions to produce hydroxyl radicals directly at the tumor site [1392]. Originally described by Fenton in 1894 [1393], the Fenton reaction is an iron-mediated chemical process in which hydrogen peroxide ($H_2O_2$) is converted to a hydroxide ion ($OH^-$) and a hydroxyl radical ($OH^\bullet$) by way of iron oxidation ($Fe^{2+} \rightarrow Fe^{2+}$) [1394]. Proposed by Zhang *et al.* [1395], CDT exploits the mildy-acidic, peroxide-rich tumor microenvironment to ionize amorphous iron nanoparticles, generating cytotoxic $OH^\bullet$ in cancer cells without nanoparticle photoactivation. Compared to traditional cancer treatments, CDT offers higher tumor specificity and selectivity, lower systemic toxicity, and fewer side effects [1396, 1397]. Also—unlike PDT, SDT, and EDT—applications of CDT operate independently of external stimuli (such as light, sound, or electricity) and do not depend on a source of oxygen [1398].

Although CDT has been scientifically validated and is widely recognized, clinical applications of CDT remain challenging because its is still poorly understood [1394, 1396]. Clinical performance of CDT is limited by certain key mechanistic issues, primarily the insufficient endogenous $H_2O_2$ content of tumors and the low catalytic efficiency of existing chemodynamic agents *in vivo*, in spite of expectations of high catalytic efficiencies for iron-based catalysts and copper-based nanomaterials used in CDT [1397]. These limitations have made CDT insufficient to completely eliminate malignant tumors, motivating the development of synergistic treatment strategies that combine CDT with methods to increase $H_2O_2$ levels, reduce pH levels, and eliminate native reductants that counteract $OH^\bullet$ production in many tumors [1397, 1399–1401]. More fundamentally, the current limitations of CDT raise questions about the intrinsic cell interactions that influence the Fenton reactions, biochemistry, and biological signaling pathways that govern the CDT mechanism(s) [1396, 1402]. Precision control of quantum mechanical properties of the Fenton reaction active site include its electron density, configuration, and spin state and are significant for designing optimal nanoparticles for CDT [1396]. Recognition of these issues has inspired recent efforts to comprehensively optimize the CDT Fenton reaction centers to systematically overcome the current drawbacks of CDT [1387, 1403, 1404].

Photothermal therapy (PTT) designates another emerging method of combating cancer by using photosensitizer molecules by convert light energy into heat to carry out thermal ablation of tumor cells [1405, 1406]. PTT is a short, simple procedure with a rapid recovery time, which has been tested intensively in clinical trials [1407]. Its drawbacks are primarily associated with the systemic uptake of photosensitizers into the body during treatment and limits on the precision of lasers used for photoactivation [1405]. To improve its selectivity, PTT can be administered with microneedles to deliver photosensitizers directly into an affected area [1408]. A wide range of nanomaterials have been proposed for use in photothermal nanotherapeutics [1409, 1410], and heat produced during PTT can be used to initiate the release of therapeutic compounds on demand, to regulate gene transcription and enzyme activity, and to enhance other chemical processes in pathological tissues [1411, 1412]. PTT can be combined with other treatment modalities [1411, 1413], such as PDT which can have a synergistic effect thanks to the increased action of ROS at elevated tissue temperatures [1406]. Nanomaterials that combine efficient ROS production with antioxidant consumption can facilitate apoptosis-inducing ROS bursts while eliminating cancer cells' defenses against ROS overproduction [1304]. Biomimetic nanoparticles can also be functionalized to combine immunosuppresive signal blocking with photothermal tumor ablation to enhance anti-cancer treatments [1412, 1414].

Although the study of photosensitive processes dates back at least to the 19th century and heliotherapy has been practiced for millenia [1236], characterizations of these phenomena in quantum biological terms are a recent development [918]. This may be due to the renewed interest in medical nanoparticle systems that has accompanied recent advances in quantum dot technologies for medical applications [1415–1418]. For example, iron selenide (FeSe) quantum dots have been functionalized with a coating of glutathione, polyethylene glycol (PEG), and human epidermal growth factor receptor 2 (HER2) antibodies to enable multiphoton imaging of tumor cells [1419]. It is common for HER2 to be overexpressed in several types of cancer including breast, ovarian, bladder, stomach, ovary, colon, bladder, lung, uterine cervix, head and neck, and esophageal cancers, as well as uterine serous endometrial carcinomas [1420]. As a consequence, HER2 has become a predictive biomarker for certain forms of cancer, particularly in breast and gastric cancers where therapies targeting HER2 have enabled revolutionary improvements in clinical outcomes [1420]. Nanoparticles conjugated with HER2 antibodies will bind preferentially in cancer cells where HER2 is overexpressed, allowing for targeted imaging, diagnostics, and therapeutics.

Quantum dots have been called "designer atoms" [1421] because of their unique, tunable optoelectronic properties, which can differ substantially from their properties in bulk. They are typically semiconductor nanocrystals with 2–10 nm diameters that contain between 10 and 50 atoms [1417]. They are known for their vibrant optical properties which include high brightness and resistance to photobleaching [1416], as well as excellent surface-to-volume ratios which enhance their optical, mechanical, magnetic, and catalytic qualities, and afford them with high absorbencies for pharmaceutical applications [1384]. They are increasingly being developed for photochemical pharmaceutical applications that require photoresponse patterns to interfere minimally with imaging while maximizing responses to the wavelengths used for photo-activated drug delivery [1422, 1423]. Photopharmaceuticals have demonstrated progress in mammalian disease model systems, but the general use of photopharmacology requires precise drug targeting following system-wide administration. This necessitates a combination of high bioactivity, photoswitchability, and potency, while maintaining metabolic robustness, solubility, and distinguishability for imaging [1424].

Similar to other nanoscale drug delivery systems [1425], "nanotheranostic" techniques combine targeted nanotherapies with diagnostic imaging [1426–1428]. They have been developed steadily over recent decades using an extensive range of nanoparticles and strutures [1429]. In principle, theranostic treatments enable specific diagnoses, time-resolved drug monitoring, and enhanced drug action when treating a wide range of pathological phenotypes. Nano-sized magnetic particles are increasingly being investigated across a wide spectrum of biomedical fields, as functionalized magnetic nanoparticles may be situated in living cells using magnetic fields, where they can be activated to enable applications like DNA transfer, gene therapy, thermal induction, or tissue engineering [1430, 1431].

Photo-driven nanotheranostics, or phototheranostics [1432, 1433], rely on diverse excited-state energy conversion properties of theranostic elements that integrate diagnostic imaging and therapeutic treatments and employ particles with favorable biocompatibility, tuneable photophysics, and photochemical characteristics upon aggregation [1434]. These systems pave the way for the development of tailored theranostic treatments with ideal activa-tion characteristics, that may be readily prepared as all-in-one multimodal interventions that combine specialized photothermal and photodynamic properties with diverse modalities. These currently include fluorescence-guided PDT [1435], photoacoustic-guided PTT [1436, 1437], and more general multi-modal design strategies that combine high-performance photosensitizers with high photothermal conversion or other "multimodal" functional capabilities that tune the relevant excited-state energy dissipation pathways as needed on a patient-centered basis [1428, 1438].

Precision treatments are being developed to target, not only individual (cancerous or infectious) cells, but specific proteins within cells. Chromophore-assisted light-inactivation (CALI) is a method to inactivate (*i.e.*, damage or denature) specific proteins with high spatiotemporal accuracy using photochemically-generated ROS [1439]. CALI can be used in cancer therapies to target cancer-associated proteins (such as integrins which are associated with tumor development, angiogenesis, invasion, metastasis, and treatment resistance [1440]), or in various protein-targeted research applications (such as studies of actin polymerization [1441]) [1439]. Photoswitchable pharmaceuticals present a growing portfolio of biomedical applications, based on the development of photoswitchable designer molecules including oligonucleotides, saccharides, peptides, and proteins [1442]. Proposals for photo-active drugs run the pharmaceutical gamut of from analgesics and inhibitors to antidepressant, antidiabetic, antiestrogenic, anti-inflammatory, antihypertensive, antipsychotic, antimycobacterial, antioxidant, and antitumor drugs [1443, 1444].

This wide range of applicability presents an open challenge to photochemical pharmacology to identify applicable reagents for therapeutic investigations *in vivo*, where photoswitchable reagents present vast new possibilities for high-precision research on cell development, division, motility, and transport [1445]. One approach has been to substitute traditional anti-mitotics with photoswitchable ones, enabling light-mediated spatiotemporal control of microtubule dynamics [1446]. Anti-mitotic cancer treatments interfere with cell growth by suppressing microtubule dynamics with a systemic drug action that produces the well-known and often-devastating side effects of chemotherapy [1447]. Photoactive tubulin-polymerisation inhibitors developed for optical control of microtubule dynamics hold promise to enable methods for the quantitative photocontrol of the microtubule cytoskeleton [679, 1445, 1448]. Thus, photochemical approaches aim to revolutionize cancer chemotherapy by opening a new horizon of targeted treatments involving photoswitchable anticancer medications.

Direct control of cancer cell redox chemistry defines a grand challenge of oncology, where it would constitute a means of inducing apoptosis or modulating cancer metabolism, if not reversing carcinogenesis itself. In lieu of direct control of cancer metabolism [1329, 1449, 1450], cellular redox control is increasingly being achieved using ROS-generating photosensitized nanoparticles—even in the absence of a definitive picture of the photopharmacological mechanisms in these systems [1359, 1451, 1452]. Magnetically-driven ROS production provides another novel method for treating cancer, tissue damage, and cellular degeneration [1453]. Functionalized magnetic nanotransducers have been shown to enable magnetic control of certain cellular processes such as neuronal firing, gene expression, and apoptosis [1291]. Weak mT-range magnetic fields can also interact synergistically with drugs to potentiate effects involving cholinergic, dopaminergic, opiate, nitric oxide, and serotonergic signaling pathways [1454]. Cobalt ferrite ($CoFe_2O_4$) nanoparticles have been used to direct the alignment of microtubules, creating ordered microtubule ensembles which may have applications for the control of cellular structures and transport mechanisms [1455].

Transformative new schemes for implementing light-based control of biological systems are being developed in the emerging field of optogenetics [1456–1458]. In optogenetics, natural and bioengineered photoreceptors are introduced genetically into the cells of interest for the purpose of making previously light-insensitive biological mechanisms both photosensitive and optically addressable [234, 1456]. These include many mechanisms involved cell cycle transitions, embryonic development, metabolic homeostasis, and stress responses [1458]. For example, optogenetic tools provide an ideal method for probing the complex spatial and temporal dynamics of gene expression signals [1459], enabling a high degree of spatiotemporal control over many biological processes [1460]. Optogenetic engineering has been used to engage protein oligomerization, protein-target interactions, and conformational switching to enable the reversible control of light-induced calcium-channel gating, dynamic protein-microtubule interactions, and membrane contact-site assembly [1461]. These methods show promise for the precise control of physiological development by patterning cell ion currents and signaling cascades [1462]. Originally conceived as a tool for neuroscience by Crick in 1979 and developed formally by Deisseroth *et al.* in 2006, the growing range of biomedical applications for optogenetics now include microbiology, cardiology, endocrinology, neurobiology, and oncology [1457, 1463, 1464].

Light-activated molecules are involved in many cellular processes and have inspired a growing assortment of nanomedical and nanotheranostic mechanisms and methods for advanced healthcare solutions with the potential to improve treatment and patient outcomes on a global scale [1465]. Minimally invasive and often straightforward to administer, these nanomaterial platforms enable precision diagnostics and drug delivery, improving the effectiveness and reducing side-effects of treatments for many common and deadly diseases [1352]. Photosensitive chromophores, quantum materials, and other biomolecular cofactors function at the heart of these new biomedicines, with quantum mechanical properties that depend critically on the physics and chemistry of the cellular surroundings.

## Chapter 17 – Regenerative Processes: Cells, Tissues, & Organs

Physiology and behavior follow specific daily programs that are adapted to the alternating challenges and opportunities of day and night [1466, 1467]. Diurnal cycles of light and darkness govern rhythmic changes in the physiology and behavior of most living species [1468, 1469], along with other environmental and seasonal cues [1470, 1471]. Light plays an essential role in modulating countless physiological processes which are influenced by the intensity, duration, rhythmicity, and color(s) of light signals [1472, 1473]. Many essential processes need to be regulated over each 24-hour daily cycle, making circadian rhythms essential to practically every aspect of physiological regulation [1474–1480].

These considerations have drawn the attention of scholars to the idea that the repetition of low-energy stimuli can have profound effects on physiological function [1247], with broad implications for the mechanisms that govern biological timekeeping and the associated field of chronobiology [1468, 1481]. This is consistent with observations that the outcome of PBM therapy can depend on the wakefulness of the treatment subject and the time of day when treatment is applied [1482]. Naturally, these ideas have been linked to research on UPE and the hypothesis that cells can communicate via metabolic biophoton emissions [1482].

Observations that PBM treatments using red and NIR light can have a regenerative influence on cell function have inspired proposals for the strategic use of light therapy to regenerate human tissues and organs [1483]. Regeneration is necessary after an injury to restore damaged tissue and recover lost organ structure and function, making it the basis for healing and a key aspect of the preservation of life [1484]. Regeneration is not trivial, and substantial loss of life can be attributed to organ-system regenerative failures underscored by inflammation and tissue fibrosis [1485]. Identifying functional cues which direct tissue healing is a critical yet elusive goal. Hence, regenerative science has emerged as a key topic at the forefront of medical research [1486, 1487].

Stem cell therapies have shown recent promise to aid the restoration of tissue or organs compromised by injury or disease. These can be highly effective, especially when combined with adjunct treatments such as gene-editing or anti-inflammatory therapies among others, but can be cost and labor intensive to implement [1485]. In spite of continuing advances, the translation of regenerative science into standardized clinical applications has been limited by a mixture of outcomes across specialties which may be attributed to differences in clinical practice and implementation [1486]. The use of stem cells to repair, replace, and regenerate damaged tissues during stem cell therapy is itself associated with risks because it is not straightforward to isolate stem cells or to commit them to develop into a desired phenotype. Stem cells can differentiate into all cell types, including cancer cells, making tissue overgrowths (as tumors or teratomas) a main concern for stem cell treatments [1484, 1485].

These factors make PBM a promising choice as an adjunct treatment to stem cell therapy, because PBM has been developed as a potential treatment for controlling stem cell differentiation, proliferation, and migration *via* ROS modulation [1488, 1489]. Full characterizations of the effects of PBM on living cells have remained elusive because of the uncontrolled variations that distinguish different PBM protocols, although progress is being made to establish more favorable methods for promoting the proliferation and differentiation in stem cells *in vitro* [1267, 1268]. PBM has been found to enhance stem cell proliferation and promote differentiation *in vitro*, where specialized media can be used to induce cell differentiation into mature phenotypes that include adipocytes, chondrocytes, and osteoblasts [1490]. It has been shown to stimulate the expression of osteogenic genes in periodontal ligament stem cells, promoting enzyme activity related to bone growth and mineralization [1491]. Promising results of this kind support the idea that PBM can be used to systematically improve methods for proliferating and differentiating bone marrow stem cells *in vitro* prior to host transplantation during stem cell therapy [1492].

Although only a relatively small number of stem cell therapies are clinically available today, stem cell engin-eering has the potential to revolutionize clinical practice with the application tailored biomaterials with tuneable mechanical and biochemical properties by effectively recreating tissues *in vitro* [1493, 1494]. Lab grown tissues have tremendous implications for fundamental physiological studies for translational therapies intended to repair, preserve, and enhance tissue function using stem cell implants [1483, 1494, 1495]. These therapies work by enabling the controlled migration, stimulation, and differentiation of stem cells to materially reconstructing compromised tissues [1248, 1496, 1497]. However, practical implementations of stem cell engineering in the form of designer tissues and engineered cartilage implants remain elusive, leaving a persistent need for hydrogels that incorporate growth factors into functional three-dimensional (3D) cell cultures [1498–1501].

Reconstitution of damaged cartilage poses one of the greatest challenges to regenerative medicine because the regenerative capacity of cartilage is restricted by its avascular composition [1502–1507]. Repairs to the articular cartilage covering joint surfaces pose a significant difficulty for joint surgery due to its limited capacity for self repair [1503, 1508–1510]. For this reason, current therapeutic interventions to treat osteochondral defects remain primarily palliative (rather than curative) [1502]. Nevertheless, a variety of innovative approaches have lead to improved outcomes in the treatment for damaged cartilage and bone [1511]. Tissue engineering strategies that incorporate biocompatible scaffolds embedded with chondrocytes and/or stem cells have shown promise for the effective regeneration of articular cartilage tissue [1501, 1508]. Synthetically-prepared, layered scaffolds that mimic the structure of articular cartilage and subchondral bone have demonstrated preliminary success in repairs of osteochondral defects [1510].

Articular cartilage may be damaged or degenerate as a consequence of ageing, injury, or disease [1501, 1508]. Although damaged bone and cartilage will heal naturally in principle, functional tissue regeneration can be hampered by a wide range of pathological factors [1511, 1512]. As a result, there is a substantial need for clinical methods that promote the regeneration and repair of articular cartilage [1508], and cartilage rejuvenation has become a major topic in orthopedic research to mitigate osteoarthritis and joint degeneration [1513, 1514]. The drawbacks of current orthopedic treatments have motivated interest from biomedical science in the development of new polymers and methods for preparing customizable 3D scaffolds and implants for osteochondral tissue regeneration [1514].

Methods of tissue engineering have begun to emerge as potential treatments to fully restore damaged articular cartilage [1506]. Combining biomaterials, live cells, and growth factors to regenerate damaged tissues, tissue engineering aims to overcome limitations of conventional treatment strategies based on surgeries and transplants [1501, 1512]. Although hydrogel implants using natural and synthetic gel-type scaffolds have demonstrated positive outcomes in preliminary clinical studies [1501, 1515, 1516], osteochondral tissue itself remains the best scaffold to restore osteoarthritic defects [1517]. Although biomaterials are now used extensively in orthopedic work, their capacity to repair tissue is still limited by their failure to fully reproduce the mechanical, biochemical, and metabolic characteristics of the tissues they replace—highlighting the need for bioactive scaffolds that differentiate into bone and cartilage [1502].

Osteochondral tissue homeostasis is maintained by articular cartilage chondrocytes and subchondral bone osteoblasts according to physico-chemical cues which influence their generative functions [1518]. Those cells synthesize the structural components of the tissue matrix and establish key features of homeostasis via the corresponding metabolic, inflammatory, and immunologic pathways. Mesenchymal stromal cells are stem cells which also play a central role in maintaining the osteochondral matrix, secreting growth factors and signaling molecules which mediate tissue regeneration and repair [1512, 1519]. Despite their importance, the complex mechanisms and interactions that govern regeneration and stem cell differentiation in osteochondral tissue remain unclear [1512, 1520]. Nonetheless, it is broadly understood that biomechanical and electromagnetic factors provide key stimuli that regulating cartilage growth and ossification in line with skeletal development [1518]. Hence, the successful engineering of osteochondral tissue will require the strategic coordination of the interplay between cells, materials, and molecular factors [1521].

While the body has a natural ability to repair small or moderate sized injuries and tissue defects, substantial damage or damages to avascular tissue can become difficult to heal naturally [1511]. During injury, blood is typically the first tissue to respond by sealing the wound with an aggregation of platlets that are cross-linked with fibrin polymers to stop bleeding and promote wound repair [79]. Fibrin is is produced from fibrogen, a glycofibrous protein that is characterized its quadruple $\alpha$-helical coiled structure and ubiquitous presence in human blood [1522]. It is considered essential for blood coagulation, wound healing, inflammation, and the formation of new blood vessels, among other functions. As a soluble protein, it forms an insoluble clot upon its conversion to fibrin by the allosteric serine protease thrombin [1523]. The structure of the fibrin matrix provides a temporary home for cells to proliferate and organize a coordinated response to injury at an inflamed site, making it a promising candidate to facilitate tissue regeneration in surgical procedures as a hydrogel biomaterial [79, 1522].

Fibrin is increasingly used in biomedical applications where it can be applied to create scaffolds for tissue regeneration or targeted drug delivery systems, such as 3D hydrogels which recreate properties of the extracellular matrix (ECM) that support cell growth, differentiation, and migration [79]. Fibrin hydrogel systems are aqueous molecular networks with tunable properties which can be used to develop complex tissue mimetics that reproduce certain prototypical features and interactions of the cell matrix [1524]. Recreation of *in vivo*-like conditions *in vitro* are naturally suited for the development of native tissue models, disease mechanisms, pharmacology, and cell therapy—but hydrogel systems suffer from restricted diffusion and lack of vascularization which can result in cell hypoxia, stress, and death [1525]. Established uses of PBM for promoting cellular resistance to hypoxia and stress made it an immediate candidate method to help improve the viability of cells cultured in fibrin scaffolds and hydrogels [1525].

Fibrin scaffolds have been proposed for use in surgical procedures for neurology, orthopedics, periodontics, plastic surgery, implantology, and in oral and maxillofacial surgeries [79]. PBM is now being applied in the developing fields of regenerative medicine and tissue engineering, but the great diversity in PBM sources and parameters, cell types, scaffold compositions and geometries make it difficult to predict the outcome of any particular combination of experimental factors [1525–1527]. Red-to-NIR range PBM has been demonstrated effective at enhancing cell proliferation, differentiation, and survival in 3D scaffold and hydrogel systems *in vitro* [1525], and the use of specific frequencies of light to enhance specific cellular processes—such as the influence of 632.8 nm PBM on chondrogenesis [1528]—have also been explored. This work has also inspired proposals to design implantable PBM devices for the photo-stimulation of interior organs (*i.e.*, in the body cavity) *in vivo* [1483]. Despite its longstanding use and demonstrated effectiveness, there is still no clear understanding of photophysical mechanisms and chemical pathways that govern PBM. This means that research on PBM effects in 3D hydrogels, scaffolds, and implants may also ultimately provide new insights into the fundamental quantum mechanical processes of PBM itself.

# Chapter 18 – Morphogenetic Integration & Immunodynamics

Cells must integrate the information they received from their environment to coordinate their responses to environmental cues [1529]. Cell behaviors are central to all aspects of life, including development, homeostasis, reproduction, and regeneration [1530]. Beyond the conventional range of sensory-motor behavioral responses, cells and tissues exhibit a vast array of physico-chemical phenomena during complex phase transitions [1531]. These can incorporate multiple states of matter (solids, liquids, gases, and plasmas [1532]) while exhibiting diverse quantum optical, electrochemical, and magnetic properties [477, 1533, 1534]. To create an atlas of cell types, these characteristic cellular states of health and disease have been catalogued and benchmarked using various genomic, biochemical, and bioelectric signatures [1535, 1536].

Cells are sensitive to their electrochemical environment, responding to a wide variety of signaling molecules including growth factors, hormones, and neurotransmitters [1537, 1538]. Cellular mechanotransduction, the conversion of mechanical forces to biochemical signals, is believed to influence important biological processes involved in embryonic development, tissue repair, and regeneration [1539–1542]. For example, the viscosity of the ECM has also emerged as a key regulator of cell and tissue dynamics [1530], with demonstrations that cells respond to the viscoelasticity of their environment in their growth, migration, proliferation, differentiation, and development [1543]. Cells also exhibit diverse responses to their electromagnetic environment which influence respiration, photosynthesis, nutrient uptake, and other biochemical attributes related to concentrations of ROS, antioxidants, proteins, and metabolites [1544, 1545]. To form a complete picture of the mechanisms underlying the collective behaviors of multicellular systems [1546], it is necessary to extend the biochemical–mechanical model of biology to incorporate electrodynamic phenomena.

Although growth and regeneration are universal characteristics of life, the capacity for healing can vary greatly and is restricted in most multicellular organisms [1547–1549]. In mammals, in most cases, the regenerative ability of most tissue is limited to reparative processes that result in the permanent replacement of damaged tissue by collagenous connective tissue known as a scar [1547]. Only in certain highly-regenerative organisms is it possible to replace whole body parts, or in some cases, entire bodies [1549]. Free-swimming flatworms known as planaria are in the privileged category, making them a popular organism to research among biologists [1550]. The planarian regenerative capacity is largely attributed to the high density of stem cells found in their adult bodies, where a single type of pluripotent adult stem cell known as a "neoblast" can differentiate into the complete range of planarian cell types [1551, 1552].

The healing abilities of planaria have positioned them as a model system for research on regenerative medicine, where the focus is on developing a systems-level understanding of the morphological decision-making processes that enable planaria to regenerate their whole bodies [1553]. Planaria demonstrate complex responses to different colors of light [1554], and PBM has been shown to stimulate stem-cell proliferation in planaria during regeneration [1552, 1555]. WMFs less than 1 mT have been shown to influence tissue growth and regeneration in planaria where WMF exposure can increase or decrease the growth rate of new tissue [478]. Those studies revealed that ambient magnetic fields in the 100–500 $\mu$T range could induce statistically significant changes in planarian regeneration, due to altered stem cell proliferation and differentiation mediated by the accumulation of ROS and the expression of heat shock protein 70 [478, 1556]. These results highlight the potential to use non-ionizing electromagnetic radiation and static magnetic fields to control tissue growth and regeneration [1556, 1557].

ROS signals have also been shown to regulate new tissue growth in zebrafish where ROS production triggers the proliferation of cells necessary for regeneration to progress [1558]. Upon the amputation of a zebrafish tail fin, physiological ROS generation is tightly regulated at the wound site to initiate healing and coordinate fin regeneration [1559, 1560]. Zebrafish, salamanders, and newts are some of the only vertebrates that can persistently regenerate amputated limbs throughout the course of their lifespans [1558]. Similar to zebrafish, axolotl salamanders produce ROS at the site of a tail amputation to initiate regeneration and repatterning of the lost limb [1561]. Studies on tail regeneration in tadpoles show that the influx of $O_2$ caused by injury increase local ROS levels, in turn stabilizing hypoxia inducible factor (HIF)-1$\alpha$ and modulating heat shock protein 90 levels and electric currents critical to regeneration [1562]. ROS gradients at the wound site attract immune cells to the area [1563]. Regeneration-organizing cells at the wound site serve as a primary source of growth factors and regenerative signals, promoting the necessary proliferation in the underlying progenitor cells [1564]. These factors reveal the orchestrated interplay of biochemical and physical regenerative signals during the regeneration of limbs in vertebrates and invertebrates alike [1562, 1563].

Cell migration is necessary for physiological development, immune responses, and healing, but it is also a mechanism of cancer [1565]. During migration, cells navigate complex tissue environments, relying on chemical, electrical, and mechanical signals for direction, and traversing chemical gradients of chemokines and growth factors to reach their appointed destination(s) [1566]. While thousands of scientific works have explored biophysical mechanisms of cells migrating on flat, 2D substrates, less is known about cell navigation in 3D environments [1565, 1567]. Chemotaxis is the movement of cells in response to their chemical environment [1568, 1569]. Although chemotaxis has been studied for over a century, research on mechanotaxis—cell movement in response to mechanical cues—is only emerging now to characterize cell responses to substrate adhesions, stiffness, and density [1568, 1570–1573].

During embryogenesis, tissue repair, and cancer metastasis, cells undergo migration either as individuals or as coordinated clusters of cells [1574]. The collective dynamics of cell clusters are mediated by cell–cell junctions which transmit and respond to physiological signals between cells, as well as the single-cell interactions which mediate cells' senses of their environment [1575]. Embryonic cells within clusters may also adjust their stiffness in response to the stiffening of their native substrate when initiating collective cell migrations, where the softening cluster stiffness can trigger collective cell migration in soft substrates [1574]. Above and beyond cell migration, a wide range of dynamic morphological processes including cell division, phagocytosis, and stem cell differentiation are known to be modulated by the external cellular environment [1565, 1576]. A number of physiological factors linked to the ECM, including cell deformations, external mechanical forces, ECM mechanics (*e.g.*, stiffness, density, elasticity), and topographical features all may influence the regulation and outcome of stem cell differentiation [1576]. Studies of collective cell migrations *in vitro* have revealed cellular dynamics similar to those of a phase transition known as a glass transition [1577].

Cells undergoing tumorigenesis exhibit cytoskeletons with increased motility but reduced polarity [1578], and aberrant cell migration is also a characteristic of chromic inflammation, vascular disease, and metastasis [1579, 1580]. Although increased cell motility and reduced polarity are also observed in migrating cells during normal biological processes, subtle characterizations of cellular bioelectric signatures can predict the emergence of aberrant cellular behavior (*e.g.*, in incipient tumors in amphibians [1330]). Efficient metastasis requires the coordinated operation of the cytoskeleton to form the many cellular protrusions that enable cell migrations that lead to death in 90% of cancer patients [1581]. In cancer, the ECM also becomes highly disregulated, and the loss of proper homeostatic organization is a hallmark of solid tumours [1582]. The success of microtubule-targeting drugs has motivated a search for other pharamaceutical means of targeting mitosis to treat cancer, but this effort has been mostly unsuccessful [1583].

Circadian rhythms which govern the immune system synchronize endogenous oscillations in migration-promoting factors which in turn are mediated by the expression of cell adhesion complexes and chemokines [1476]. Whereas direct effects of cancer treatments such as PDT, CDT, and conventional chemo and radiation therapies can eradicate the bulk of a tumor, a native immune response is still needed to eliminate any surviving cells to prevent metastasis and cancer recurrence after remission [1354]. Collectively known as "immunotherapies" [1584, 1585], a growing range of treatments has become available which combat cancer by boosting the body's own native cancer-fighting mechanisms using pharmaceutical checkpoint inhibitors, *in vitro* T cell expansions, lab-created anti-bodies and cytokines, and cancer vaccines among others. However, applications of these treatments are still limited in scope and can include serious risks of side effects such as deadly cytokine storms [1585]. These complications have motivated research on the use of natural molecules, electromagnetic fields, and mechanical vibrations to enhance the natural human capacity for healing [1484].

Studies of wound healing provide insight into collective cell behaviours [1575], as local responses to injury can produce lasting consequences on the scale of the organism [1586]. Cellular metabolites such as ROS have been demonstrated to play an important role in healing in several highly regenerative species of animals, including planaria, zebrafish, tadpoles, and geckos [1562]. A growing body of literature shows that physiological ROS levels can be modulated by the application of static and dynamic electromagnetic fields [436], presenting an emerging picture of the essential role played by electromagnetic effects in the morphological development of living systems [1587]. Inadequate regeneration of organ function after injury is associated with many outstanding medical problems worldwide, raising the question of why some organisms have such greater regenerative capabilities than others [1562]. Planaria are living proof of the extent of anatomical regeneration that living systems are able to achieve [1553]. However, the complex interplay between electromagnetic, biochemical, and mechanical mechanisms that collectively regulate tissue response to injury, stem cell dynamics, regeneration, and cancer suppression remain unclear [1588].

Raised in the modern context by Schrödinger [39], the question of how living beings are able to generate biological structures while maintaining homeostasis in the intrinsically noisy cellular microenvironment remains one of the foremost issues of anatomy and physiology [1589, 1590]. To rationalize the high degree of functional organization found in biological systems, it has been proposed that the robust nature of living things ultimately relies on the performance of a complex ensemble of gene regulatory networks and maintenance ("clean-up") systems [1591, 1592]. Indeed, the genome specifies a cellular collective which is innately plastic, executing morphological rearrangements until a target morphology is achieved [1587]. Yet this simplistic picture of morphogenesis begs the question of how biological systems determine when to stop remodeling themselves once a preferred morphology has been reached.

The ECM is a highly responsive structure present in all tissues which is continuously remodeled in the process of maintaining homeostasis [1593, 1594]. It provides the essential structural foundation for multicellular organisms as well as many of the physiological cues that initiate and control cellular behavior [1595]. It regulates morphology during tissue development by governing key aspects of cell adhesion, differentiation, migration, and morphology [1596, 1597]. Distinctive cell characteristics give rise to different forms of cell migration. During mesenchymal cell migration, cells polarize and form leading-edge protrusions, which adhere to the ECM via integrin-mediated cell–matrix adhesions [1580, 1598]. These dynamical processes are actuated by vast cytoskeletal rearrangements and signaling cascades [1598].

Cells respond to mechanical forces during tissue development and countless disease processes as they regulate the mechanics of the ECM that maintain its physiological properties [1599]. Mechanical responses are mediated by the cell membrane, cell adhesion complexes, and the cytoskeleton itself [1600]. Cell migration, for example, is contingent on the cytoskeleton's mechanical and dynamical properties which achieve structural stability through dynamicity [1601]. As its main components, the cytoskeleton is composed of microtubules, actin, septin, and intermediate filaments which each have distinctive mechanical properties [1052, 1581, 1602]. Crosslinking plays an essential role mediating the dynamics of the cytoskeleton, where microtubule crosslinking facilitates organization and transportation, whereas actin crosslinking determines mobility [1603, 1604]. Despite the cytoskeleton's critical physiologial importance, there is still only a preliminary understanding of the mechanisms that govern its functionality [1605, 1606].

Animal cells can change shape at will, moving dynamically and adaptively in response to many different environmental cues, by virtue of the tightly-coupled biochemical and mechanical properties of the cytoskeleton which enable cell polarization, morphogenesis, and motility [1607]. Unlike long-lived cellular structures, an individual microtubule's lifetime may be as short as twenty seconds *in vivo* [1608], so microtubule networks are renewed constantly to maintain cytoskeletal structure and organization [1601]. Cytoskeletal self-organisation arises from the reaction mechanisms that control microtubule growth and shrinkage *in vitro* [1609–1611], and hence reaction–diffusion processes have been proposed to explain similar organizational behaviors *in vivo* [1612, 1613]. However, reaction–diffusion mechanisms are not sufficient to explain certain key aspects of cytoskeletal reorganization during cell migration [1614].

Mechanical signals affect many biological processes in both developing and adult organisms, including cell migration, differentiation, morphogenesis, and immune responses [1597, 1615]. For example, experimental research *in vitro* has shown that MSCs cultured on stiff substrates mimicked bone stiffness, whereas MSCs cultured on soft substrates mimicked the softness of brain tissue [1616]. Research on mechanical signaling in cells is relatively new, and detailed insights into the factors that influence mechanical transitions are rare [1615]. The mechanisms that regulate microtubule nucleation in frog eggs are largely unknown [1617]. Likewise, the nature of organizational waves in the microtubule networks of zebrafish embryos remain unclear, although researchers have concluded that microtubule network organization there cannot be described fully in static terms and must also include dynamic effects [1618]. The successful development of quantitatively predictive multiscale models of microtubule dynamics are expected to require deep, comprehensive insights into the structural and mechanical properties of tubulin [1018].

Microtubule networks control many aspects of cellular architecture, polarity, and proliferation in animal cells, but the mechanisms which govern the formation of those networks are only minimally understood [1054]. Tubulin is predicted to respond to short, intense electromagnetic pulses in simulations [1619], and Cifra *et al.*. have proposed that microtubules can themselves generate electric fields when excited electrochemically [678]. Electrical oscillations are intrinsic to bundles of microtubules found in brain cells, where they contribute to the fundamental oscillatory modes of neurons [723]. Although anti-cancer drugs such as taxanes are well known to have microtubule-stabilizing properties which perturb microtubule dynamics and interfere with microtubule growth, the exact nature of pharmaceutical action of those drugs remains poorly understood [727]. Proper microtubule function is crucial to cell health and thus microtubule system abnormalities can result in neurological disorder and disease [706, 1620].

Numerical studies have shown that quantum relaxation effects in microtubule networks are likely to be important and comparable in time with other dynamical processes occurring there [1621, 1622]. Real-time microtubule imaging using a scanning tunneling microscopy (STM) revealed how the application of an quantum mechanical tunneling current could be used to stimulate microtubule assembly from tubulin when the alternating tunneling current was tuned to distinct resonance frequencies of tubulin [754]. Advanced methods for simultaneously visualizing and manipulating the cytoskeleton that combine optical tweezers or atomic force microscopy with fluorescence methods are likely to provide further insights into the fundamental mechanical properties of cellular microtubule networks [1607]. Deep insights into the physics underlying cytoskeletal mechanics are needed to inform models of the cytoskeletal dynamics associated with cell motility, migration, differentiation, development, and disease [1542, 1598].

The immune system responds to biochemical and mechanical signals expressed as changes in the biophysical properties of tissue that reflect the many aspects of disease, inflammation, and ageing [1623]. Immune responses orchestrate the mobilization of millions of leukocytes at the sites of active inflammation, where swarms of immune cells enact complex behavioral strategies that are now being studied using high-dimensional datasets to describe vast multi-parameter "behavioural landscapes" of immunological activity [1624]. Immunodynamics are coordinated from the lymphoid organs of higher animals, where networks of fibroblastic reticular cells inhabit the collagen mesh found at each lymph node [1625]. Each reticular cell network provides the structure to support native immune cells while distributing essential molecules that enable immune cell recruitment, trafficking, and survival [1626].

Although reticular cell networks are primarily associated with the lymph nodes and spleen, they are also recognized for the more general biomechanical roles that they play in defining the physiological architectures of multiple organs to create structural continuity that transcends conventional organ boundaries [1627]. Dynamic in nature, the reticular network relaxes during immunological activation to allow for the influx of immune cells that occurs during clonal expansion [1625]. As a "small world" network [1628], it has important topological properties that help optimize communications between immune cells and provide resilience against structural damage and disruptions. This inherent topological robustness allows reticular networks to sustain losses of up to 50% of the fibroblastic reticular cells that inhabit it without compromising immune cell recruitment, migration, and activation—demonstrating the critical role played by network integrity in mounting and sustaining adaptive immune responses [1629, 1630].

Given the importance of its functions, surprisingly little is known about the structure of the reticular cell network (*e.g.*, compared to the brain [1626]). Nevertheless, the practical role that these networks play providing structural definition in biological systems has inspired efforts to develop customizable organ grafts by "re-seeding" decellularized fascia (rich in native collagen, elastin, proteoglycans, and glycoproteins) with the appropriate cells to create structurally-organized tissue scaffolds for use as synthetic organ grafts [1631, 1632]. Advances in the bioengineering of multicellular systems have driven efforts to test the limits of life's building blocks by exploring new methods of biological assembly that may reveal the fundamental biophysical principles that govern living matter [1633].

# Chapter 19 – Quantum Biotechnology: Universal Applications

Hormesis is a well-known effect in biology and medicine, whereby a low dose of a potential stressor induces a stimulating effect, whereas higher doses inhibit stimulation [1634–1638]. The study of hormesis has a colorful history dating back at least to the $16^{\text{th}}$ century when Paracelsus observed that "the dose makes the poison" [1634, 1639]. A classic example of hormesis is provided by vitamin $D_3$, which in low doses is necessary for proper brain development in rats [1640]—but is also administered in high doses as cholecalciferol found in rat poisons [1641]. The hormetic dose–response curve exhibits a characteristically "biphasic" rising and falling action, in contrast with simplified linear dose–response models which are serviceably accurate within certain dosage ranges and are "approximately correct for many environmental carcinogens" [1642]. Lately there has been renewed interest in hormetic dose–response models because of their relevance to biphasic treatments such as PBM therapy [1268, 1274].

The so-called "Goldilocks zone" of the hormetic dose–response curve (where the stimulating effect of the dose reaches maximum efficacy) is not limited to medicine and pharmacology, but is encountered as a characteristic of homeostasis at all levels of biology [1643, 1644]. It is likewise a feature of transport mechanisms in open quantum systems that harness the interplay between coherence and decoherence to increase transport efficiency [1645]. These synergetic effects of quantum dynamics are not limited to photosynthetic exciton transfer, and have been proposed for efficient energy and charge transport applications using a variety of nanoscale platforms including molecular electronics and organic photovoltaics [1646, 1647]. Contrary to the expectation that quantum effects should be restricted to low temperature systems, preliminary studies have revealed how quantum transport mechanisms which operate at intermediate or high temperatures can actually break down in the limit of low temperatures [1648].

Photosynthetic electron transfer rates in algae and plants become saturated at about one fourth of the full intensity of sunlight on Earth, due to the large optical cross section of light-harvesting antenna complexes which can capture photons an order of magnitude faster than the rate-limiting electron transfer step [1649]. To prevent electronic overloading, the photosynthetic apparatus dissipates excess excitation energy during intense light exposure by a process known as non-photochemical quenching [1650]. Although the exact details of this quenching mechanism are not known, it is believed to be controlled by a switching effect triggered by a proton gradient across the photosynthetic membrane [656]. Regardless of the exact mechanistic details, both natural and artificial light-harvesting systems rely on photo-excitation mechanisms to produce a light-activated charge-transfer state which can subsequently decay (non-photochemically) back to its ground state, produce free charge carriers (in photovoltaics), or generate radicals for photochemistry or photosynthesis [1651]. Studies of quantum coherence effects in artificial light-harvesting nanotube (LHN) systems have shown how excitonic delocalization and vibronic coherences can be controlled to increase fluorescence quantum yields by up to 30% by manipulating system disorder [656, 660].

The fundamental understanding of these photophysical processes is essential, not only for the development of efficient photocatalytic and photovoltaic light-harvesting systems [302, 1652, 1653], but also for important biotechnology and agricultural applications [1654]. Light-driven redox catalysis is now a viable method of synthesizing organic compounds using solar energy [1655], and emerging photoredox mechanisms have inspired proposals for photocatalytic hydrogen production, clean $CO_2$-to-fuel conversion, and photoactivated anaerobic organic syntheses [1656]. However, it remains a challenge to achieve satisfactory photo-exciton lifetimes and diffusion lengths, despite scientific progress developing artificial light-harvesting structural scaffolds and photo-active metal nanoclusters [1657, 1658], even in light of growing recognition that both quantum coherence and dephasing effects are necessary to achieve optimal energy transduction [137, 1659].

Single-atom photocatalysts have been proposed to overcome challenges associated with photoactive nanoparticles, nanoclusters, and bulk materials, due to the exceptionally-high photoactivity and selectivity of the electronic structures associated with atomic coordination centers [1660]. Individual atoms have the added benefits of reduced environmental and fiscal costs associated with the use of catalytic metals, highly tunable electronic properties, and comparatively simple reactive sites to ease the analysis of reaction mechanisms and structure–performance relationships [1661]. Of course, single atom catalysts are already found ubiquitously in biological systems, where porphyrin rings and their derivatives carry individual atoms of iron (in heme), cobalt (in vitamin $B_{12}$), magnesium (in chlorophyll), and nickel (in coenzyme $F_{430}$) [1662]. Biological metal–porphyrin complexes are critical to countless life-sustaining processes involving electron transfer, and enzyme catalysis, oxygen transport, and photosynthesis, and hence have provided inspiration for research on the development of artificial metalloenzymes [1663, 1664].

The many versatile physico-chemical, electronic, and spectroscopic properties of metal–porphyrin complexes have made them a subject of numerous biological studies, inspiring discoveries in multiple research areas with applications for antimicrobial drugs, biomedicine, catalysts, optoelectronics, sensing, and solar cells [1665–1668]. Yet porphyrin is only one among many examples of catalytic sites found in biological systems, be they metallic [1669] or organic [1670]. Many of these are involved in or related to the biological electron transfer processes that mediate the electronic excitations that fundamentally enable all known cellular activities [1655]. Ultrafast spectroscopy experiments are now uncovering a wealth of information about bioelectronic excitation dynamics [159, 1654, 1671].

In contrast with the canonical donor–bridge–acceptor model of electron transfer that models the electronic coupling in terms of a simple distance-dependent exponential decay function [1672]), detailed quantum mechanical studies indicate electron transfer reactions cannot be modelled accurately by considering the electronically active site in isolation [1673]. This complexity is reflected in quantum biological treatments such as PBM therapy, which can exhibit complicated dose–response effects due to the diverse selection of treatment parameters involved. For example, PBM is nuanced by a wavelength-dependent dose–response function characterized by a variable range of excitory or inhibitory responses that can depend on the exact colors of light used for treatment [1674].

Growing understanding of biomolecular interactions with light bring the myriad benefits of engineering biomolecular interactions with light. Genetic engineering techniques are already being applied to native molecular targets such as cytochrome enzymes to enhance photosynthetic yield [574]. Likewise, optogenetic techniques can be used to implement optical control of metabolism in bioengineered yeasts by triggering the assembly and disassembly of metabolically active enzyme clusters to control the formation of desired metabolic products [1675]. Light-based interactions have also been used to initiate chemical signaling between synthetic and natural cells, demonstrating the use of light to trigger the delivery of small molecules to live cells for biomedical applications [1676]. Synthetic cells have also been used to demonstrate luminescent signaling within and between cells, initiating photo-induced conidiation (*i.e.*, spore forming) in neighboring fungal cells [1677]. These results define preliminary steps toward the design and construction of "bottom-up" synthetic biology by using photochemical reaction networks, biological compartmentalization, and cellular communication pathways to orchestrate biological dynamics [1678].

Biological light-harvesting systems challenge the present understanding of chemical dynamics with sophisticated arrangements of chromophores that optimize light capture with highly-tuned chromophore site energies and subtle manipulations of electronic excitations [307]. Respiratory electron transport chains perform similar energy transformations, transferring electrons long distances through intricate molecular machineries that defy a detailed mechanistic description using modern biochemical research [103]. These typical examples of biomolecular networks represent the vast array of astonishingly complex systems of electronically-active molecules found in biology.

Living systems achieve a level of control over the electronic states of their internal structures that has long been sought after by researchers in the field of molecular electronics [1679–1681]. From that perspective, it would appear that the native biomolecular reaction networks of living organisms offer timeless solutions to the problem of harnessing strongly-interacting many-body nonequilibrium open quantum system dynamics [1682]—a highly-complex quantum mechanical problem that has driven intense research on the development of molecular electronic devices for several decades [1681, 1683]. Native bioelectronic systems are resistant to the intrinsic disorder of the biomolecular microenvironment. This inherent robustness has motivated growing interest in prospects for the development of advanced quantum biotechnologies for use in quantum communication and computation [1684–1686].

One topic that has captured the imagination of biocomputing researchers is the prospect of molecular computing with DNA [1687–1690]. Inspired by Feynman's imperative to engineer submicroscopic computers [1389], Adleman proposed the idea of using molecular computation to solve hard combinatorial problems in 1994 [1691, 1692]. Today, computational DNA reaction networks have been programmed to perform arithmetic, digital logic, and even neural network computations for artificial intelligence applications [1693]. DNA-based neural networks designed for pattern recognition have been used in disease profiling and diagnostics [1693, 1694], motivated by the need for precise, early cancer diagnoses and rapid, inexpensive, non-invasive disease screening [1695, 1696]. As the global production of digital data grows exponentially, costs associated with conventional (*i.e.*, magnetic and semiconductor) digital storage media are becoming inviable. For this reason, DNA-based data storage technology has been identified as a promising candidate to enable cost-effect, dense, and durable data storage solutions [1697–1699].

The applications of programmable DNA are not limited to computation, and proposals to develop DNA as programmable matter in the construction of autonomous molecular systems have emerged in line with the expanding complexity of DNA "circuit" dynamics [1693]. Programmable DNA scaffolds have been proposed to actuate spatial control over densely-packed clusters of cyanine dyes with tunable absorption spectra and strongly-coupled exciton dynamics similar to chromophore networks found in natural light-harvesting systems [532]. These efforts are representative of an emerging class of molecular-level studies intended to help mimic biological implementations of light sensing and harvesting systems [1700, 1701], and stimuli-responsive photonic materials [1702–1704].

Controlled assemblies of physiological crystal structures produce diverse, responsive optical effects for coloration and vision widely across the animal kingdom [1705]. For example, investigations of the vivid patterns found on butterfly wings have revealed key aspects of the self-organizing structural dynamics that generate vibrant 3D biophotonic nanostructures there [1706]. Although the formation of organic biocrystals are ubiquitous in plants and animals, the fundamental organizational mechanisms underlying these structures remain mysterious [1705, 1706]. Research on the mechanisms that govern biological nanophotonics *in vivo* therefore have great potential to enable applications of biomimetic optical devices using photonic crystal engineering [1706].

Research on quantum nanophotonics has experienced a recent renaissance of interest in theory, implementations, and applications of quantum biotechnologies [1707]. Biological applications are heralded by the emergence of "soft" quantum bionanophotonic devices [1708], biofunctionalized nanoparticles, and quantum dots [918]. Strong light–matter interactions between nanophotonic enclosures and quantum emitters have unveiled unprecedented prospects for modifying chemical landscapes and reaction cascades with applications for biomolecular sensing, protein engineering, and control of hybrid light-matter states of living cells [1709]. Recent studies have demonstrated dynamical photocontrol of microtubule integrity, migration and mitosis using fluorescent protein-imaging-compatible photoswitching reagents *in vivo* [1445]. Optogenetic tools have also been used to control physiological development by patterning cell ion currents and signaling cascades with high spatiotemporal precision [1462]. These optogenetic techniques can be used to leverage precise control over selected cell types—even in heterogeneous tissues [1456].

Biosynthetic nanostructures and assemblies have also been proposed in strategies to manipulate cell organelles *in vivo* for biomedical applications [1710]. These strategies hold great promise for proposals to carry out chemical or genetic engineering of cells *in situ*, enabling the administration of practical, cost-effective bioengineering tools to strengthen inherently-favorable cell traits or even to confer cells with new functionalities [1711–1713]. This "top down" approach has been used to endow bacteria with new abilities for food and biofuel production, cancer treatment, chemical synthesis, environmental remediation, light harvesting, and fundamental biological research [1714, 1715]. For the time being, living cells remain too complex to redesign completely, and not all new functions may be compatible with an intended host cell. These limitations motivate an alternative "bottom up" approach [1714].

In the "bottom up" approach to synthetic biology [1716], cells and processes are dismantled and reassembled to build up new forms of life or life-like creations. This work is enabled by advances in DNA sequencing and synthesis developed in the context of systems biology [1717]. "Bottom up" synthetic biology has been used to reconstitute many natural biomolecules, cell components, whole cells, and even primitive designer organisms [1714]. It has been used to create a wide assortment of molecular-scale motors [1718, 1719] and robots [1720–1722], as well as new fundamental building blocks for life in the form of noncanonical amino acids [1723]. By merging the capabilities of "top down" and "bottom up" synthetic biology [1714], bioengineers aim to create new forms of devices, materials, sensors, and therapeutics out of living or life-like components [1712]. These emerging biotechnologies may be expected to have profound implications for agriculture, biomedicine, electronics, environmental science, and robotics [1724].

Proteins play countless essential roles in nature, where they are critical for the survival and reproduction of living organisms by enabling them to adapt to continually changing environmental forces. Bioengineered protein materials have emerged recently as multi-functional media that are durable with respect to heat, light, humidity, magnetic fields, and other factors [1725]. Protein-based materials are naturally responsive and adaptable, making them choice materials for use in the development of synthetic tissues [1501], shape-changing robots [1726], and reconfigurable organisms [1727]. Efforts to re-shape living organisms for the purposes of biomedicine or bioengineering will require precision control of the structure and function of an organism at all scales [1728, 1729]. Many problems in medicine, especially, could be addressed by answering the primary question of how cells and organelles cooperate to build and maintain essential anatomical structures [1730]. More fundamentally, the prospect of building, repairing, and remodeling biological structures is attendant to the idea of generating bespoke organisms.

Bioelectric gradients are increasingly recognized as control factors that govern growth and form in living organisms [1731]. Fundamental advances in bioengineering have presented the possibility of disassembling, assembling, and recombining biological structures at multiple scales to contrive living organisms in new configurations that defy many traditional conceptualizations of life [1732]. Although self-assembly is found prevalently in inanimate and biological system alike, in biological systems it primarily takes the form of self-limiting processes characterized by the assembly of smaller subunits into larger, yet finite-sized, anatomical structures [1733]. Biological assembly and organizational processes are as common as they are vital, yet by-and-large poorly understood [1271, 1734].

Although eukaryotic cells contain many specialized organelles to perform specific tasks in concert, the mechanisms that coordinate their positions in the cell with respect to their functions remain largely unknown [1271]. In fact, even the relatively unstructured cytoplasm of bacteria exhibits glassy dynamics indicative of memory effects which may be altered dramatically during cellular metabolism, suggesting a highly complex matrix that is not well approximated as a simple fluid medium [1734]. Despite the progress in the development of nonequilibrium thermodynamics to predict the emergence of self-organized behaviors in classical complex systems, the task of extrapolating those classical principles to the quantum scale remains daunting [37]. One approach, in line with theories of mitochondrial retrograde signaling [1271] and membrane dynamics [1735, 1736], hypothesizes that strong electromagnetic fields generated by the mitochondrial membrane are a key factor regulating global cellular activity [1737, 1738].

Collective oscillations are found ubiquitously in biological systems where they govern many developmental processes [126]. In multicellular systems, cells, tissues, and organisms are nested functionally as well as structurally, prompting the idea that collective cellular decisions are coordinated primarily by biochemical, bioelectric, and biomechanical signals [1739, 1740]. Membrane potentials are believed to play critical roles in biological patterning processes [1741], where bioelectrical signals complement chemical and mechanical signals governing embryonic development, morphogenesis, and healing [1731]. Cellular ion channels, known for regulating tissue morphology in planarian flatworms [1553, 1742], have also been characterized using non-Hermitian quantum theory [1743]. This presents the possibility that the inherent topological features of living organisms may emerge from underlying principles of quantum mechanics which characteristically give rise to forms of biological synchronization [1744].

These observations raise questions about the fundamental forces or constraints that drive living systems to converge toward certain recognizable forms and functions [1745], and the emergent properties (such as multicellularity) that arise from universal self-organizing principles [1633]. Existing efforts to synthesize bionic systems (*e.g.*, out of silicon) are prone to numerous structural and functional limitations, falling far short of the fluid and facile capabilities of native biological systems [1746]. Nevertheless, today's rapid interdisciplinary advances in atomistic fabrication have been characterized as the beginning of a new industrial revolution [1747], and artificial nanofluidic systems are now beginning to approach a level of functionality similar to that of the biological channels that enable sensation, transport, and neurotransmission in living systems [1748].

Unlike conventional optomechanical systems which required sophisticated cooling mechanisms [1749], living systems must operate with energies comparable to ambient thermal noise [1748]. As concurrent advances in quantum optics facilitate controlled manipulations of excitonic states at room temperature in strongly-coupled quantum systems [1750], recent research has revealed how the efficiency of quantum transport can be enhanced by ambient noise in natural and engineered systems designed for light-harvesting, superconduction, and ionic confinement [1751]. This raises prospects for the development of radically new biotechnologies that harness noise-assisted, non-Hermitian quantum effects at the vanguard of physics research [604, 1752]. This new class of self-repairing, replicating entities may be expected to possess powerful capabilities that harbor critical ethical implications [1716]. The advent of novel living technologies will afford new and unprecedented opportunities for biomedicine and health science that engage the innate, collective intelligence of living tissues and organs [1753]. Yet, like past scientific revolutions, these new technologies may also herald increasingly dynamic, circumstantial, and adaptive new forms of socio-economic challenges [1754].

# Chapter 20 – Quantum Biology: Essential Further Research

Living systems effectively bridge the gap between the quantum realm of atoms and molecules and the macroscopic world of daily life. Unlike quantum theory, which requires making the formal (albeit artificial) distinction between the observer and the observed—the classical and the quantum—living organisms contain no such boundary. This presents the scientific challenge of providing a consistent description of the self-referential systems that are so characteristic of biology. Although there is little room to doubt the importance of quantum effects in determining the properties of tunneling-mediated electron transfer processes, photochemical reactions, enzyme active sites, metallorganic clusters, and molecular van der Waals forces (just to name a few), the length and time scales over which biological quantum effects should be deemed relevant remain to be established. In this respect, quantum effects can be subtle, and even the processes which govern quantum mechanical decoherence (*i.e.*, dephasing and spontaneous relaxation) are themselves fundamental quantum dynamical effects.

The following considerations highlight the importance of research on quantum biology, not only to shed light on the governing principles of anatomy and physiology, but to illuminate the physical principles which give rise to living matter. In the interest of developing a unified theoretical framework with which to pursue rigorous investigations in quantum biology, it may be necessary to clarify certain aspects of the axiomatic quantum formalism which have obscured the fundamental quantum nature of living structures.

As anticipated by Bohr and Schrödinger, these research questions challenge the assumption that biology may be reduced to matters of conventional physics and chemistry as they are understood today. These questions furthermore raise the prospect that it may be necessary to revisit certain limitations of the established quantum formalism to properly account for living phenomena. Unlike the artificial philosophical constructs imposed by the postulates of quantum theory, living systems do not unambiguously distinguish between the quantum and the classical, the system and the apparatus, or the observed and the observer. Further research in quantum biology is essential to define the relationship between quantum and classical effects.

Quantum biology begins wherever the energetics of life begin, be it in the photosynthetic networks of phototrophs or the chemosynthetic and electrosynthetic networks of lithotrophs, all forms of life rely on oxidative metabolism enabled by biological electron transfer. Oxidative chains of electron transporting chromophores are practically universal in living organisms, and the electronically active reaction centers that make up these chains have physiologically important quantum mechanical characteristics by virtue of their functional roles in mediating electron transfer. Chromophores tend to be photoactive, as their name implies, with tunable photo- and electrochemical properties that are often functionally dependent on environmental conditions. These properties require further study. For example, many details of the interactions that control flavin biophysics remain unclear even after decades of flavoenzyme research.

Many structural and functional properties of respiratory and photosynthetic electron transport chains have now been characterized. There is still much work to be done to account for many-body and self-synchronization effects in those systems, and further research will be needed to fully account for the high quantum yield and thermodynamic efficiency of biological electron transporters, proton pumps, and synthases. Electron transfer reactions are often accompanied by proton transfer steps that must also be accounted for quantum mechanically in model simulations. Even developing the correct physical formalism can be challenging when modeling complex many-body systems of this kind. Further research is needed to characterize the exact roles of quantum coherence and dissipation in the mechanisms that govern photosynthesis and cellular respiration in living systems.

Biophotons have attracted interest for their potential use in evaluating redox physiology, metabolism, and cell stress for medical assessments and health monitoring. More commonly known as ultraweak photon emissions, these faint cellular light emissions are primarily a product of high-energy oxidative cellular processes. As a result, their production is linked to the reactive dynamics of various species of oxygen, and therefore also closely tied to certain aspects of mitochondrial and cytoskeletal function. Experimental observations have consistently associated weak magnetic field effects with the production of cellular reactive oxygen species, which are implicated in mechanisms of morphogenesis and regenerative healing, although these physiological phenomena also remain poorly understood.

Light–matter interactions are critical to many aspects of physiology including vision, photosynthesis, and vitamin D synthesis. Many optically-active molecules play important roles in redox chemistry, electron transfer, oxygen transport, and other catalytic processes. The flash of a firefly is just one example of the many photochemical reactions that occur in living systems. Although the importance of biological photochemistry has been recognized since photosynthetic pigments were discovered in the $18^{\text{th}}$ century, the role of optical-frequency excitations in essential cellular regulatory processes such as mitochondrial retrograde signaling have only recently become topics of medical research interest. This may be due to growing interest in the development of nanomedical interventions such as photobiomodulation and photodynamic therapy, and the emerging use of optically-active nanoparticles in medical treatments. Growing use of optogenetic techniques in biological research has also brought attention to the role of optical effects in living systems, and more important discoveries may be expected in these areas.

The exact mechanism(s) underlying the effects of weak magnetic fields on biological processes remain unclear and require extensive further research. Although radical pair dynamics are implicated in many possible mechanisms that have been proposed to mediate interactions between weak magnetic fields and living cells, the nature of their role is not yet definitive. Significant efforts have also been made to characterize the chemical amplification of weak fields in the presence of radical scavenging molecules. More research is needed to assess the general role of magnetic fields in biochemistry and the physiological processes that enable sensing, metabolism, and cellular growth. More research is especially needed to determine the influence of Earth-strength ($\sim 50\,\mu$T) magnetic fields on physiology, particularly in the context of cellular magnetoreception and its implications for health. Biomagnetic theory remains to be formulated to rationalize all observed effects of electromagnetic fields on cellular radicals, and the nature of the influence of magnetic fields on cellular oxygen metabolism remains to be conclusively determined.

The microscopic mechanisms that govern the formation of the complex hierarchical structures found in eukaryotic cells are practically unknown. However, microtubules are implicated in most if not all aspects of cellular organization due to their important structural roles. Once characterized primarily in terms of its branching structural motifs, the microtubule cytoskeleton is now gaining attention for its central role in mediating the sensory-motor faculties of cells. As key integrators of biochemical, electromagnetic, and optomechanical signals, microtubules are thought to generate oscillating electric fields that coordinate core cellular functions. Nevertheless, a clear picture of the interactions that govern polarization effects in cytoskeletal networks is absent, and there is need for fundamental research to explain the growing body of evidence for photonic and quantum optical effects in microtubules.

The relationship between microtubule dynamics and mitochondrial function—including their interactions with radical species and their electron transfer dynamics—requires further attention. Microtubule assembly dynamics can be controlled by weak magnetic fields and tunneling electron currents, suggesting the importance of subtle quantum mechanical driving effects in the mechanisms that govern these systems. Tubulin is a birefringent protein with unique optical and optoelectronic material properties. It deserves to be characterized comprehensively under resting (ground state) and electronically excited circumstances, which will be essential to determine its fiber-optic, optomechanical, and chemomechanical properties as a sub-cellular structural and information conduit. Theoretical and experimental characterizations of coherent quantum optical effects in microtubules require further attention to discern the role of the geometrically complex networks of biological chromophores implicated there.

Terahertz frequency radiation has been shown to influence physiological processes in living cells, including primary cellular activities, amino acid release, cell proliferaction, and apoptosis. Although the mechanisms underlying Thz cellular interactions are unclear, these processes are likely to be related to other long-range electrodynamic interactions that have recently been discovered between biomolecules. Electrodynamic diagnostic techniques such as THz imaging and ultraweak photon emission measurements are likely to provide access to important metabolic information that can be used to help identify the presence and type of cancer for medical applications. However, the effects of electromagnetic fields on different types of cancerous, neoplastic, and normal tissues are not yet fully understood, and more research is needed to determine the safety, specificity, and efficacy of electromagnetic field-based therapies and diagnostics.

Although mechanical forces are traditionally in the domain of classical physics, the transduction of mechanical stimuli into electrochemical signals (and vice versa) are inherently quantum mechanical and universally present in biological systems. The systematic inventory of the various adiabatic and nonadiabatic quantum effects involved in biological processes will be useful for the ongoing characterization of biological structure–function relationships, beginning with photochemical and photophysical processes, internal conversion mechanisms, intersystem crossing effects, the Jahn-Teller effect, and various forms of vibrationally-assisted energy, electron, and proton transfer.

The future of nanomedicine depends on understanding the biomolecular basis for the regulation and control of all major cellular functions. Homeostasis requires the systematic regulation of a host of cellular processes and the associated metabolites which include reactive oxygen and nitrogen species, metal ions, lipids, peptides, and nucleic acids. Cellular regulation involves many complex interactions between governing organelles such as mitochondria, the cytoskeleton, and the cell nucleus. The complex relationship between microtubule dynamics and mitochondrial function requires further elucidation, particularly concerning the interplay between biochemical, bioelectric, and biomagnetic effects. Mitochondrial retrograde signaling, as it is implicated in the mechanism of photobiomodulation therapy, is also likely to play a critical role in cellular health and disease processes of broad biomedical interest.

Just as biological cofactors involved in enzyme catalysis respond to the electrochemical and electromagnetic properties of their environment, so do organelles and whole cells respond to their respective cellular and extracellular environments. The successful development and implementation of emerging nanomedicines and nanotheranostic tools, such as chemodynamic and sonodynamic therapies, requires in-depth research into fundamental mechanisms that determine their efficacy and efficiency. This research will also be required to realize potential applications that incorporate the full electromagnetic spectrum into treatments for wound healing and regenerative medicine.

Although the density of any stationary electronic system is quantum mechanically “trivial” in the sense that in principle it can be predicted using semiclassical theory alone, in practice the problem of finding the semiclassical approximation that replicates the exact many-electron density function of a quantum mechanical system has been recognized to be as difficult as solving any of the hardest verifiable computational problems in quantum complexity theory. This fundamental theoretical difficulty represents the ongoing quest to find the fabled universal exchange-correlation potential function of quantum chemistry. The problem is central to studies of structural biology, where functionally-critical molecular cofactors and enzyme active sites often contain highly-complex quantum mechanical electronic structures that can be intractable to solve on present-day computing hardware. Research directed toward improving approximations and numerical methods to tackle these problems is ongoing, requiring interdisciplinary efforts from molecular biologists, quantum chemists, numerical physicists, and computer scientists.

Electronically-active biomolecular systems can incorporate both local dynamic and nonlocal static quantum mechanical correlations which are highly-nontrivial to predict even using advanced methods and approximations. Model simulations of biomolecules present additional challenges because of the ubiquitous presence of charge transfer and weak binding effects in these systems, which require the extension of conventional methods in quantum chemistry to account for the dynamical phenomena involved in these processes. Charge transfer is accompanied by quantum spin transfer, and spin-mediated effects are crucial to many biochemical and physical phenomena including radical pair and scavenging dynamics, spin crossing reactions, lipid peroxidations, and chirality-induced spin-selective electron transport. Yet the exact physical mechanisms underlying these processes are often unclear and subject to debate, reflecting the need for experimental and theoretical developments to resolve current ambiguities.

For example, the significance of radical pair dynamics in cryptochrome magnetosensitivity remains inconclusive. The role of radical electron dynamics is more clear in the photoactivated mechanism of the DNA-repair enzyme photolyase, where the mechanism itself is complicated and worthy of further investigation. Photosensitized processes in native biological macromolecules such as photolyase and hemoglobin have made them prototypes for biomedical research in the emerging field of nanomedicine, with the goal of developing increasingly safe, targeted, effective treatments for a broad spectrum of acute and chronic diseases. The development of novel medical treat-ments based on DNA-repair enzymes, for example, will benefit from comprehensive models of the complex sequence of structural rearrangements and functional dynamics that enable rapid DNA photo-catalysis in those systems. Quantum biology research heralds the possibility of dynamic control of vital spin-chemical processes for a plethora of potential biomedical applications geared toward restoring cellular regulation, communication, and control.

Nonlocal, wavelike quantum mechanical fluctuations play an important role in defining the form and structure of many biologically-important nanosystems. Foremost among these is the structure of DNA, where the characteristic stacking of base pairs into the iconic double helix is enabled quantum mechanical van der Waals interactions and subtle solvent stabilization effects that are not predicted consistently using simplified classical models. The multifunctional properties of DNA have made DNA-based information processing and data storage a topic of intense and growing interest. DNA computing is representative of the broader scientific field of bioelectronic computation, where great strides are being made to develop truly molecular-scale components to use to build the next generation of miniaturized computing machines. However, current efforts to integrate molecular transistors, diodes, and logic gates suffer from various fabrication issues which may prove difficult to overcome using semiconductor devices.

Comprehensive models of the functional properties of cellular ultrastructure require analyses at all levels of approximation. However, full scale biomolecular simulations of enzymatic systems present insurmountable difficulties to computational methods due to the comparatively large sizes of biological macromolecules and the ubiquitous presence of static and dynamic quantum correlations in systems of this kind. Substantial progress has been made by introducing many-body dispersion effects into simulation methods using both *ab initio* quantum and parameterized classical mechanical approaches. Computational techniques based on multi-scale models and quantum embedding schemes have had some success in simulations of enzyme-catalyzed biological processes, but these methods still entail the use of complicated approximations and the arbitrary partitioning of the system under study into distinct classical and quantum subsystems. Even small catalytic sites, such as the transition metal atom centers found in the molecular cofactors of many enzymes, can have extraordinarily complex electronic structures that defy the use of classical simulation methods even on the large supercomputers. Quantum computing has been proposed as a means of overcoming the difficulty of simulating structural and chemical properties of complex biomolecules.

In many cases, the true significance of underlying quantum mechanical effects remain unknown, such as during protein folding, wherein it is still unclear how the newly-translated primary amino acid sequence can relax into its optimal folded conformation in finite time. The problem of folding a protein into its minimum free energy conformation amidst an exponentially-large number of folding possibilities is reminiscent of comparable fundamental problems in computer science and combinatorial optimization. This has motivated interest in the use of quantum computing and combinatorics to solve hard problems in molecular biology—and also the possibility of using molecular biology to solve many of the most difficult "NP-complete" problems in computer science.

Living systems overcome the problem of negotiating charge transport through an insulating medium by using tunneling-mediated electron transfer through a series of electronically active sites in the electron transport chains of the assorted photosynthetic, respiratory, and immunological engines of biology. These catalytic engines carry out the numerous electronic transformations needed to enable all aspects of anabolic and catabolic metabolism. Broadly speaking, these catalytic transformations are enabled by the many biomolecular cofactors found at the biochemical active sites of enzymes. Biomolecular cofactors are essential to carry out the countless electronic transformations that are needed to sustain growth and maintain homeostasis, and as a result they remain prototypical chemical agents in health science, pharmacology, and medical chemistry. Despite their central importance in physiology, the structural properties underpinning the functional characteristics of biocatalytic cofactors are obscured by the same highly-complex electronic configurations that enable their seemingly-endless functionalities. Addressing fundamental questions about the relationship between electronic structure and function in biology will set the stage for further research on the processes that generate, organize, and replicate living cells, tissues, and organisms.

# Chapter 21 – Conclusion: A Quantum Framework

Quantum biology would not exist without quantum mechanics, but quantum mechanics itself is framed by the premise that quantum measurements are performed using classical measuring apparatus. This premise is enshrined in the measurement postulate of quantum theory, made famous by the paradox that accompanies it—the so-called

"measurement problem" of quantum mechanics. Yet, physical paradoxes are in the eye of the beholder, a consequence only of ill-conceived assumptions. In the case of quantum theory, the paradox is that the measurement postulate requires that quantum measurements are implemented using classical measuring devices, in apparent contradiction with the fact that laboratory apparatus are made of atoms like everything else. This conundrum was not lost on the founders of quantum mechanics, who went to great lengths to formulate a quantum–classical correspondence principle.

Championed by Bohr, the correspondence principle provided the cornerstone for the orthodox "Copenhagen" interpretation of quantum physics by posing the premise that the predictions of quantum mechanics should converge to recover classical physical phenomena in some statistical limit. However, what *exactly* this statistical limit should be is no more agreed upon by physicists now than it was in Bohr's time. The correspondence principle today is not so much a principle of nature as it is a standing philosophical conjecture. Efforts pioneered by Dirac to derive quantum mechanics as a modification to classical physics—while physically appealing—remain both philosophically and mathematically unsatisfying, given that the goal of such "canonical quantization" procedures is to derive a more fundamental theory from a less fundamental one.

Living systems are unlike either the inanimate quantum systems or the experimental apparatus found in the physics laboratory, and for this reason they may offer a unique advantage as a reference with which to consider the true nature of the quantum–classical correspondence. Rather than beginning from the assumption that living beings are essentially like other inanimate classical objects, quantum biology may begin with the founding principle that quantum theory is the rule, rather than an exception. This principle is important to adopt when considering biological systems, which are not simply chemical entities made from organic molecules, but living beings that can exert control over their own internal biochemistry, often at will.

In this work, we have provided a broad overview of the many nontrivial quantum effects and related processes found ubiquitously in biology, including electron tunneling during cellular respiration, light-harvesting during photosynthesis, and cofactor/ligand coordination during protein translation and enzyme catalysis. Although the scientific basis to recognize these as quantum effects has been established for many years, their biological implications are not widely appreciated due to the complicated nature of the quantum mechanics involved and the resulting technical difficulties associated with applying quantum mechanics to functional aspects of biology. This suggests a need to revise the current biophysical perspective to update classical assumptions about the living world.

Rather than beginning with a classical viewpoint which is then modified by quantum corrections, we propose that it is appropriate to understand the role of quantum mechanics in biological processes from first principles. This may also help to clarify the emergence of a correspondence between quantum and classical phenomena, without immediately invoking the existence of artificial classical measuring devices, yet with the final goal of explaining how seemingly-classical and quantum processes can co-exist in living systems. This approach defines a prototype framework for studies in quantum biology, where the openness of quantum systems may be accounted for using a non-Hermitian Hamiltonian formalism according to the unitary projection operator techniques developed by Fano and Feshbach (without recourse to the conventional Lüders–von Neumann projective measurement postulate).

A scientific approach that focuses on the quantum physics and maintains a quantum perspective ensures the coherence of the methodology by applying a unified set of founding assumptions throughout. This provides a natural framework in which to represent and understand quantum electrodynamic interactions in biological systems. This framework is well-posed to incorporate the influence of environmental factors on open quantum systems, which are critical to describe the complex quantum dynamics of living matter. By accounting for the interplay between the nonlocal electronic structures of biological cofactors, the nonadiabatic quantum dynamics of excited electronic processes, and the environmental effects of dissipation, external driving, and spontaneous decay, the approach we describe here aims to capture the essential physical features of life itself.

One goal of this shift in fundamental research perspective is to ensure that classical reasoning is not spuriously imposed in models of fundamentally quantum phenomena, so that effectively-classical model approximations are only applied where methodologically warranted. This approach has the added benefit of acknowledging the possibility of discovering new fundamental phenomena in these systems, while maintaining a consistent theoretical framework in which to form theories, pose hypotheses, and design experiments. It emphasizes the pursuit of knowledge and understanding with the secondary intention of developing long-term applications which could be quite different from present-day technologies.

As science moves toward harnessing direct quantum control over living processes, it will enable the creation of whole new types of living systems not anticipated by previous technologies. This poses potentially-unforeseeable societal and ethical challenges. The potential risks associated with the unregulated development of these emerging technologies indicates a strong need to establish the overarching bioethical consultation, regulation, and governance needed to respond to new challenges. The implications of new forms of biotechnology may be dramatically different from their predecessors in terms of their magnitude and scope, with important consequences for biomedicine and bioengineering.

Living systems, known in physics as active matter, are extraordinarily dynamic. The complex phenomena associated with these systems poise quantum biology as a highly interdisciplianary field of research encompassing biology, physics, chemistry, medical science, and nanotechnology. Developing shared scientific terms, language, and methods will pose a core challenge to overcome. For this reason, it is all the more imperative to develop a common understanding of the foundational physics underlying the field. It is hoped that formalizing these shared intentions will provide the theoretical foundation needed to properly ground the work while propelling research forward. The field of quantum biology has unlimited potential for discovery and we invite you to approach the possibilities emerging from this new scientific horizon, starting with the physical processes and principles shared herein.

# Glossary of Terms

**ab initio**: Latin (trasl. "from the beginning"), a method of approaching a subject or performing calculations from first principles, using basic physical constants without empirical parameters.

**biphasic**: Refers to a system or process that involves two distinct phases, such as a solid phase and a liquid phase, or a rising phase and a falling phase, often in the context of chemical or physical dynamics.

**chemotaxis**: The movement of a cell or organism in response to chemical stimuli, often in the study of cell signaling.

**chirality**: Describes asymmetrical structures that cannot be superimposed onto their mirror images, often in the context of molecular interactions; sometimes known as (left or right) "handedness."

**circadian rhythms**: Biological cycles that display an endogenous, entrainable oscillation of about $\sim 24$ hours, influenced by external cues like light, especially in biological and ecological physics. See also "Hermitian,"

**closed quantum**: A system where quantum states are defined, and interactions with the environment are negligible, allowing for the isolation of quantum phenomena.

**coherent quantum effect**: Quantum phenomena where particles or states exhibit a definite phase relationship, crucial in the description of quantum interference and superposition.

**configuration (electronic)**: The specification of the quantum state of an electron or electrons, in terms of a specific set of quantum numbers.

**configuration interaction**: A quantum mechanical method used to describe the electronic state of a molecule or other quantum system by considering interactions between various electron configurations.

**conjugated organic molecule**: A molecule with alternating single and multiple bonds, resulting in the delocalization of electrons which significantly affects its optical and electronic properties.

**decoherent tunneling**: A quantum mechanical process wherein a particle moves through an otherwise-insurmountable barrier while losing its quantum coherence, typically due to interactions with the environment.

**degeneracy-breaking effect**: A phenomenon where a previously degenerate set of quantum states of a system split into distinct energy levels due to an external perturbation or field, to impact the system's behavior.

**density functional theory**: A computational quantum mechanical modeling method used to investigate the electronic structure of many-body systems by using electron density rather than wave function; typically pertains to Kohn–Sham density functional theory.

**dispersion forces**: Weak forces arising from fluctuations in transient dipole moments of molecules, significant in understanding molecular interactions.

**dynamic instability**: A condition in which a system is sensitive to initial conditions, leading to unpredictable changes in state or behavior over time; especially in microtubules.

**electron bifurcation**: A process in which the pathway of one or more electrons diverges into two separate routes and/or energetic levels, studied in the context of biochemical reactions and energy transduction.

**electron delocalization**: The distribution of electron density in space, often over several atoms rather than being localized to a single atom, essential to quantum mechanical models of chemical bonding and reactivity.

**electron transfer**: The movement of electrons from one molecule or atom to another, fundamental in processes like redox reactions and energy conversion; usually specific to tunneling-mediated transfer according to Marcus theory.

**electron transport chain**: A series of electronically active molecules in a protein complex that transfer electrons, usually through a membrane, critical in cellular respiration, photosynthesis, and immune function.

**electron valence**: The number of electrons in the outermost quantum mechanical energy level of an atom that are available for bonding, influencing chemical reactivity and properties.

**electrostatic**: Relates to stationary arrangements of electric charge, fundamental in understanding forces between charged particles and molecular interactions.

**endergonic**: Describes a reaction or process that requires energy input to proceed, often used in thermodynamics to describe non-spontaneous reactions.

**endogenous**: Refers to processes or substances that originate from within an organism, system, or process; significant in biological and medical physics.

**endothermic**: Pertains to a reaction or process that absorbs heat from its surroundings, impacting thermodynamic behavior and energy transfer.

**entanglement**: A quantum phenomenon where two or more particles become correlated in such a way that the state of one cannot be described independently of the state of the other, regardless of separation.

**exchange-correlation**: A term in density functional theory that describes the effective interaction between electrons, taking into account the particle-exchange symmetry and correlated motion of identical pairs of electrons.

**extracellular matrix**: A complex network of proteins and carbohydrates outside cells that provides structural and biochemical support, studied in biophysics.

**Fermi's "Golden Rule"**: Refers to the second of two formulae popularized by Enrico Fermi, used for calculating the probability of transitions between quantum states, essential in quantum mechanics and statistical physics.

**Gibbs free energy**: A thermodynamic potential that measures the maximum reversible work that can been obtained from a thermodynamic system at constant temperature and pressure.

**Hartree-Fock theory**: A method for approximating the wave function and energy of a quantum many-body system in a stationary state using a single Slater determinant, foundational in quantum chemistry.

**Heisenberg cut**: A conceptual boundary between quantum and classical systems, typically discussed in the context of quantum measurement.

**heliotherapy**: The use of sunlight for therapeutic purposes involving the absorption of light by living systems, studied in biophysics and medicine.

**Hermitian symmetry**: A property of quantum mechanical operators, traditionally used in quantum mechanics to ensure real eigenvalues and orthogonal eigenstates of operators; pertains to the closure of operaters on a Hilbert space with respect to the standard Dirac inner product.

**homeostasis**: The process by which biological systems maintain stability while adjusting to changing external conditions, used in physiology and medicine.

**hormesis**: A dose-response phenomenon characterized by low-dose stimulation and high-dose inhibition, showing non-linear, typically biphasic responses in biological systems.

**H-theorem**: A theorem in statistical mechanics that describes the approach to equilibrium of a thermodynamic system, demonstrating the second law of thermodynamics.

**in situ**: Latin (transl. "in place" or "on site"), refers to processes or measurements conducted in the original place or context, significant in experimental physics, biology, and medicine.

**in vitro**: Latin (transl. "in glass"), refers to experiments conducted outside of a living organism, often in controlled environments, important for understanding biological and biochemical processes.

**in vivo**: Latin (transl. "in life"), refers to experiments or processes that take place within a living organism, crucial for studies in biology and medicine.

**incoherent tunneling**: See decoherent tunneling.

**inelastic**: Referring to processes where kinetic energy is not locally conserved, typically involving energy loss during interactions due to heat, light, and noise; important in collision physics.

**interference**: The phenomenon that occurs when two or more waves superpose to form one resultant wave, significant in optics, electromagnetism, and quantum mechanics.

**intersystem crossing**: A process in which a molecule transitions between different spin states, important in photochemistry and photophysics.

**inverted effect**: A phenomenon where a system behaves oppositely to expectations, often discussed in optics and quantum mechanics; especially in Marcus theory when an increase in the force driving a reaction decreases the reaction rate.

**ion**: An atom or molecule that has gained or lost one or more electrons, resulting in a net electrical charge, fundamental in chemistry and physics.

**Jahn-Teller effect**: A distortion of molecular or ligand geometry that occurs in certain degenerate electronic states, impacting the properties of the molecule.

**Langevin dynamics**: A computational method used to simulate the behavior of systems influenced by both deterministic and stochastic forces, typically used to reproduce the effect of temperature in thermodynamically-realistic simulations.

**Lennard-Jones potential**: A semiclassical mathematical model that is used to approximately account for interactions between pairs of neutral atoms or molecules, significant in molecular dynamics simulations.

**Marcus theory of electron transfer**: A theoretical framework that describes the rates of electron transfer using quantum mechanics and the principle of conservation of energy, crucial in physical chemistry and biochemistry.

**mean field theory**: An approximate method used in statistical mechanics to analyze phase transitions and describe collective behavior of systems by assuming that collective interactions between particles is properly approximated by an averaged interaction.

**metabolic biophoton emissions**: Faint light emissions produced by biological organisms as a result of metabolic processes, studied in biophysics. See also ultraweak photon emissions.

**Mitchell's Q cycle**: A theoretical model explaining the mechanism of energy conversion in the electron transport chain during cellular respiration; the Q stands for quinol.

**molecular orbital theory**: A method for determining molecular structure by considering the distribution of electrons in molecular orbitals (rather than atomic orbitals).

**multi-scale models**: Approaches that integrate different scales of analysis (from atomic to macroscopic) to understand complex systems in physics and biology, often used in computational physics and numerical modeling.

**nanocrystal**: A crystal with dimensions on the nanoscale, exhibiting unique physical and chemical properties, studied in materials science and nanotechnology.

**nanotheranostic**: A field combining diagnostic and therapeutic applications at the nanoscale, significant in medical physics and drug delivery.

**nonadiabatic**: Referring to processes that occur without the assumption of slow changes, where electronic and nuclear motions are coupled.

**nonlocality**: A property of quantum mechanics where particles exhibit spatial distributions and correlations that cannot be accounted for by local phenomena, crucial in organic chemistry and quantum information theory.

**non-Hermitian**: Referring to operators that do not satisfy Hermitian symmetry, often leading to complex-valued eigenvalues and implications for open quantum systems.

**one-particle theory**: A theoretical framework that describes a many-body system of interacting particles in terms of individual non-interacting particles, effectively trivializing those interactions.

**open quantum systems**: Quantum systems that interact with their environment, leading to phenomena like decoherence and loss of information.

**optical**: Pertaining to light and its properties, often studied in the context of wave phenomena and quantum optics; can refer to wavelengths of light over the infrared, visible, and ultraviolet spectrum.

**optoelectronic**: Pertains to the combination of optics and electronics, focusing on devices that source, detect, and control light using electronics.

**orbital angular momentum**: The angular momentum associated with the motion of electrons in their orbitals, significant in atomic, molecular, and optical physics.

**photobiomodulation**: The use of light to stimulate biological processes, often studied in medical physics for therapeutic applications.

**photoelectric effect**: The phenomenon where electrons are ejected from a material when it is exposed to light of a sufficiently high frequency, fundamental in quantum mechanics and solid-state physics.

**photoentrainment**: The process by which light influences biological rhythms, particularly circadian rhythms, significant in chronobiology.

**photon**: A quantum of electromagnetic radiation, fundamental in the study of light and its interactions with matter using quantum electrodynamics.

**photoswitchable**: Referring to a molecule that can change its structure or function upon exposure to light, important in materials science and photochemistry.

**phototropism**: The growth response of plants to light direction, studied in biological physics and investigations of plant behavior.

**Planck's constant**: A fundamental constant that relates the energy of a photon to its frequency, crucial in quantum mechanics.

**polarizable media**: Materials that can develop an electric dipole moment in response to an external electric field, significant in electromagnetism.

**polaron**: A quasiparticle that consists of an electron and its associated polarization field, important in condensed-matter physics.

**quanta**: Discrete packets of energy or matter, fundamental in quantum theory, representing the smallest possible units of a physical quantity.

**quantization**: The process of constraining a physical quantity to discrete values rather than continuous ones, fundamental in quantum mechanics.

**quantum adiabatic theorem**: A principle stating that a quantum system remains in its instantaneous eigenstate if a given perturbation is applied slowly enough.

**quantum correlations**: Non-classical correlations between quantum systems that cannot be explained by classical physics, significant in quantum information theory; can be static or dynamic.

**quantum dots**: Nanoscale semiconductor particles that have quantum mechanical properties, important in optoelectronics and nanotechnology.

**quantum dynamics**: The study of time-dependent behavior of quantum systems, focusing on the evolution of quantum states over time.

**quantum electrodynamics**: A fundamental theory describing how light and matter interact at the quantum level, combining quantum mechanics and electrodynamics.

**quantum embedding theory**: A framework that allows for the study of a small quantum system within a larger environment, useful in computational chemistry.

**quantum field theory**: A theoretical framework combining quantum mechanics and special relativity to describe how matter and energy interact.

**quantum information processing**: The study and application of quantum mechanics to process and transmit information, fundamental in quantum computing.

**quantum jump**: A sudden transition between quantum states, often observed in the context of atomic or molecular systems; also colloquially referred to as a quantum leap.

**quantum spin**: A property of electrons, photons, and other quantum mechanical particles that arises as a consequence of intrinsic angular momentum; plays a critical role in chemical and magnetic properties of materials.

**quantum tunneling**: A quantum phenomenon where particles pass through potential barriers that would be classically impenetrable, significant in nuclear and solid-state physics.

**resonant tunneling**: A quantum tunneling process that occurs at specific energy levels, often utilized in electronic and optical devices.

**self-consistent field**: A method used in quantum chemistry to find a solution where the field created by electrons is consistent with their distribution, often used in conjunction with mean field theory.

**singlet-triplet interconversion**: A process where a singlet state (paired spins) transitions to a triplet state (unpaired spins) and vice versa, relevant in photochemistry and spin dynamics. See also inter-system crossing.

**small world network**: A type of network characterized by short path lengths and high clustering, often studied in statistical physics and network theory.

**solitons**: Self-reinforcing wave packets that maintain their shape while traveling at constant speeds, important in nonlinear dynamics.

**spacetime**: The four-dimensional continuum in which events occur, combining three dimensions of space with one dimension of time, fundamental in relativity.

**spatiotemporal**: Relating to both space and time, often used in physics to describe phenomena that evolve over both spatial and temporal dimensions.

**spin-exchange**: A quantum mechanical phenomenon involving the exchange of spin states between particles, significant in magnetic and spin chemical systems.

**static magnetic fields**: Magnetic fields that do not change over time, important in various applications including imaging techniques and magnetic resonance.

**stochastic resonance**: A phenomenon where the presence of noise enhances the detection of weak signals in nonlinear systems, studied in statistical mechanics.

**subradiant**: A state of a quantum mechanical system of light emitters where the emission of radiation is suppressed compared to the normal radiation rate, important in quantum optics.

**superexchange**: An interaction mechanism in condensed-matter physics that allows for indirect coupling between spins in a material through virtual excitations.

**superposition**: A fundamental quantum mechanical effect whereby a physical system can exist simultaneously in multiple quantum states.

**superradiance**: A collective emission phenomenon where a group of excited atoms or molecules emits light more intensely than they can individually, studied in quantum optics.

**symmetry-breaking effect**: A situation in which a system that is symmetric under certain transformations becomes asymmetric, leading to distinct states or phases; can be spontaneous.

**system**: A defined set of interacting components, which can be isolated or part of a larger environment.

**theranostic**: The combination of therapeutic and diagnostic methods, particularly in medicine and nanotechnology.

**time-dependent perturbations theory**: A mathematical framework in quantum mechanics used to systematically analyze physical systems subject to small time-varying influences.

**time-independent perturbations theory**: A mathematical framework in quantum mechanics used to analyze physical systems subject to small time-invariant influences.

**trivial**: Referring to a system or result that is straightforward or without complexity, often used in contrast to non-trivial findings; in physics this typically refers to cases when the correlations, dynamics, or interactions in a system become negligible.

**ultraweak photon emissions**: Extremely low-level light emissions from biological systems that are produced by oxidative or metabolic processes, often studied in biophysics and medical physics.

**uncertainty principle**: A fundamental principle in quantum mechanics stating that certain pairs of physical properties cannot be simultaneously known with arbitrary precision.

**vacuum field**: The lowest energy state of a quantum field, which can still exhibit fluctuations and is important in quantum field theory.

**vibronic coupling**: The interaction between electronic and vibrational states in molecules, significant in describing molecular spectra and dynamics.

**Warburg effect**: A metabolic process in which cancer cells break down glucose to produce energy even when oxygen is plentiful, also known as aerobic glycolysis.

**weak magnetic field**: A magnetic field that is significantly lower in strength compared to typical magnetic fields used in experiments, usually less than one milliTesla, often studied in quantum mechanics and biophysics.

# Sample Syllabus

**Class Structure:** Twenty-two (22) lecture-style classes, 90 minutes each, additional classes during semester can be designated for research paper assistance, student presentations, question and answer periods, special topics, and review for the mid-term and final exam.

**Prerequisites:** For college students enrolled in physics, biology, biochemistry, biotechnology, medical science, or related studies. Specific mathematics pre-requisites are not required for this course.

**Class 1 – Introduction to Quantum Biology**

I. Introduction to Quantum Biology (15 minutes)
    A. Definition and Scope of Quantum Biology
        - Overview of quantum biology as an interdisciplinary field
        - Importance of studying quantum effects in biological systems
    B. Historical Context
        - Brief history of the development of quantum biology
        - Key milestones and breakthroughs in the field

II. Course Objectives and Structure (15 minutes)
    A. Learning Outcomes
        - Discussion of key concepts and skills students will gain
        - Overview of the relevance of quantum biology to various scientific disciplines
    B. Course Structure
        - Outline of topics to be covered throughout the semester
        - Explanation of the integration of theoretical and practical components

III. Syllabus Review (20 minutes)
    A. Course Schedule
        - Detailed overview of the weekly topics and class activities
        - Important dates for assignments, exams, and projects
    B. Grading Policy
        - Explanation of grading criteria and assessment methods
        - Discussion of participation, assignments, and exam formats

IV. Instructor Introduction (10 minutes)
    A. Instructor Background
        - Brief presentation of the instructor's academic and research background
        - Overview of the instructor's interests in quantum biology
    B. Teaching Philosophy
        - Discussion of the instructor's approach to teaching and learning
        - Importance of student engagement and collaboration in the course

V. Introduction to Key Concepts (20 minutes)
    A. Quantum Mechanics Basics
        - Overview of fundamental principles of quantum mechanics relevant to biology
        - Discussion of wave-particle duality, superposition, and entanglement
    B. Applications of Quantum Principles in Biology
        - Examples of quantum mechanical influences on biological processes (e.g., photosynthesis)
        - Introduction to the concept of quantum coherence and its implications for living systems

VI. Conclusion and Q & A (10 minutes)
    A. Summary of Key Points
        - Recap of the course structure, objectives, and key concepts introduced
    B. Open Floor for Student Questions
        - Opportunity for students to ask questions about the course or the topics covered

**Initial Assessment:** Completion of a short introductory assignment (e.g., a reflection on student interests in quantum biology) to be submitted by the next class.

**Class 2 – Quantum Theory and the New Observables**

I. Introduction to Quantum Theory and Biological Systems (15 minutes)
  A. Overview of Quantum Theory's Historical Context
    - Discussion of early challenges in physics and the emergence of quantum theory
    - Key figures and events leading to the development of quantum mechanics
  B. Hormesis in Biology
    - Definition and examples of hormesis in biological systems
    - Importance of the biphasic dose-response curve

II. Quantum Effects and Biological Processes (20 minutes)
  A. Quantum Transport Mechanisms
    - Explanation of coherence and decoherence in biological systems
    - Role of quantum effects in photosynthesis and energy transfer
  B. Applications of Quantum Biology
    - Overview of how quantum dynamics influence biological processes
    - Examples of quantum phenomena in living organisms

III. The Transition from Classical to Quantum Mechanics (20 minutes)
  A. Limitations of Classical Physics
    - Discussion of the inadequacies of classical theories in explaining biological phenomena
    - The gap between quantum mechanics and classical observations
  B. Advances in Quantum Field Theory
    - Introduction to quantum field theory and its implications for open quantum systems
    - Challenges in unifying quantum mechanics with relativity

IV. Implications of Quantum Biology in Technology (20 minutes)
  A. Innovations in Quantum Biotechnology
    - Overview of applications in biomedicine, energy production, and nanotechnology
    - Examples of artificial light-harvesting systems and photocatalysis
  B. The Future of Quantum Computing and DNA Technology
    - Discussion of molecular computing with DNA and its potential applications
    - Exploration of synthetic biology and programmable matter

V. Ethical Considerations and Future Directions (10 minutes)
  A. Ethical Implications of Quantum Biotechnology
    - Discussion of the societal and ethical challenges posed by advancements in quantum biology
    - Potential impacts on health, environment, and technology
  B. Current Research Opportunities
    - Summary of emerging research areas in quantum biology
    - Importance of interdisciplinary collaboration

VI. Conclusion and Q & A (5 minutes)
  A. Summary of Key Concepts
  B. Open floor for student questions

**Class 3 – Quantum Electrodynamics: Lighting Up Life**

I. Introduction to Quantum Electrodynamics and Biology (15 minutes)
  A. Overview of Quantum Electrodynamics (QED)
    - Definition and significance of QED in understanding biological processes
    - Historical context of electron transfer reactions in biology
  B. The Role of Light in Biological Systems
    - Explanation of how photons excite electrons and drive biological functions
    - Importance of the $E/f$ relationship and Planck's constant in biology

II. Quantum Tunneling and Electron Transfer (20 minutes)
  A. Quantum Tunneling Explained
    - Comparison between classical mechanics and quantum tunneling
    - Discussion of resonant vs. decoherent tunneling in biological systems
  B. Biological Applications of Electron Tunneling
    - Overview of how electron tunneling is utilized in cellular respiration and metabolism
    - Examples of quantum electrodynamics in electron transfer between proteins

III. Photosynthesis and Energy Conversion (20 minutes)
  A. Quantum Mechanics in Photosynthesis
    - Examination of light absorption and electron transfer in photosynthetic organisms
    - Discussion of the efficiency of energy conversion in natural systems
  B. Advances in Artificial Light-Harvesting Systems
    - Overview of developments in artificial photosynthesis and photocatalysis
    - Importance of vibronic coupling in enhancing charge transfer efficiency

IV. Proton-Coupled Electron Transfer (20 minutes)
  A. Understanding Proton-Coupled Electron Transfer (PCET)
    - Explanation of the relationship between electron and proton transfers in biochemical systems
    - Role of PCET in cellular respiration and metabolic processes
  B. Mechanisms of Electron Transfer in Enzymes
    - Examination of cytochrome P450 and its role in electron transfer reactions
    - Discussion of challenges in modeling electron transfer dynamics

V. Quantum Spin and Biological Processes (10 minutes)
  A. The Role of Chirality and Spin in Biology
    - Overview of chirality-induced spin selectivity (CISS) and its biological significance
    - Implications of spin polarization in electron transport and cellular processes
  B. Applications of CISS in Biochemistry
    - Discussion of CISS effects in radical species and their relevance to biological reactions

VI. Conclusion and Q & A (5 minutes)
  A. Summary of Key Concepts
  B. Open floor for student questions

**Main Assignment #1:** Short 2-3 page research paper on quantum biology topic of interest related to the students field of study to be completed before mid-term study window begins.

**Class 4 – Definitions of Non-Triviality for Quantum Biology**

I. Introduction to Quantum Triviality and Biology (15 minutes)
  A. Defining Triviality in Physics
    - Overview of triviality as a benchmark in physical sciences
    - Explanation of quantum triviality and its implications for biological systems
  B. Importance of Non-Trivial Quantum Effects
    - Discussion of how non-trivial effects impact biological structure and function
    - Examples of trivial vs. non-trivial entanglement in biological systems

II. Quantum Effects in Biological Chemistry (20 minutes)
  A. Quantum Chemistry and Electron Transfer
    - Overview of electron transfer reactions and their significance in biology
    - Introduction to quantum tunneling and its role in biological processes
  B. Challenges in Modeling Biological Molecules
    - Discussion of the complexities of simulating biomolecules in quantum chemistry
    - Limitations of semiclassical models in accounting for quantum effects

III. The Role of Density Functional Theory (20 minutes)
  A. Understanding Density Functional Theory (DFT)
    - Explanation of DFT and its application in quantum chemistry
    - How DFT simplifies the modeling of many-electron systems
  B. Non-Trivial Challenges in DFT Applications
    - Issues with long-range dispersion effects and their biological significance
    - Overview of advancements in molecular orbital methods for biomolecular calculations

IV. Exploring Proton-Coupled Electron Transfer (20 minutes)
  A. Mechanisms of Proton-Coupled Electron Transfer (PCET)
    - Explanation of the PCET process and its relevance in cellular respiration
    - Role of quantum coherence and electron tunneling in PCET dynamics
  B. Implications for Metabolic Processes
    - Discussion of how PCET impacts energy transduction and metabolic functions
    - Examples of PCET in respiratory enzymes and their mechanisms

V. Quantum Biology: Bridging Concepts (10 minutes)
  A. The Interplay of Quantum Effects in Biology
    - Overview of how quantum effects such as coherence and entanglement influence biological systems
    - Discussion of the significance of trivial quantum effects in understanding non-trivial phenomena

VI. Conclusion and Q & A (5 minutes)
  A. Summary of Key Concepts
  B. Open floor for student questions

**Class 5 – Photosynthesis & Open Quantum System Dynamics**

I. Introduction to Photosynthesis and Quantum Dynamics (15 minutes)
  A. Importance of Photosynthesis
    - Overview of photosynthesis as the basis of life on Earth
    - The role of sunlight in energy conversion
  B. Quantum Coherence and Dissipation
    - Explanation of how quantum coherence influences photosynthetic efficiency
    - Discussion of the interplay between coherence and dissipation in biological systems

II. Mechanisms of Photosynthetic Energy Transfer (20 minutes)
  A. Role of the FMO Complex
    - Introduction to the Fenna–Matthews–Olson (FMO) complex and its function in light-harvesting
    - Overview of exciton energy transfer and its significance in photosynthesis
  B. Coherent Oscillations and Vibronic Couplings
    - Discussion of how coherent oscillations facilitate energy transfer
    - Explanation of hybrid vibronic states and their implications for energy dynamics

III. Quantum Effects in Photosynthetic Processes (20 minutes)
  A. Quantum Tunneling in Electron Transfer
    - Overview of quantum tunneling and its relevance to biological electron transport
    - Discussion of the Marcus inverted effect in photosynthetic efficiency
  B. Implications of Quantum Coherence
    - Examination of how quantum coherence enhances electron transfer and energy transport
    - Examples of quantum dynamics in different photosynthetic organisms

IV. Environmental Interactions and Photosynthesis (20 minutes)
  A. Role of Environmental Noise
    - Overview of how environmental noise can enhance light-harvesting efficiency
    - Discussion of superradiance and superabsorption in photosynthetic systems
  B. Engineering Quantum Resources
    - Exploration of how dissipative processes can be harnessed for quantum information tasks
    - Implications for developing advanced technologies based on quantum biology

V. Challenges and Future Directions in Quantum Biology (10 minutes)
A. Current Challenges in Modeling Photosynthesis
- Overview of difficulties in simulating photosynthetic mechanisms using classical models
- Discussion of the need for comprehensive quantum models in biological research
B. Future Research Opportunities
- Exploration of these emerging research areas in quantum biology and their potential applications
- Importance of interdisciplinary collaboration in advancing the field

VI. Conclusion and Q & A (5 minutes)
A. Summary of Key Concepts
B. Open floor for student questions

**Class 6 – Light Receptors, Spin Chemistry, & Cryptochrome**

I. Introduction to Light Sensing in Biology (15 minutes)
A. Overview of Light Receptors
- Importance of light sensing in various biological processes (vision, circadian rhythms, etc.)
- Introduction to cryptochrome as a light receptor and magnetoreceptor
B. Fundamental Mechanisms of Photosensing
- Explanation of the radical pair model of magnetoreception
- Overview of how photons influence biological signaling

II. Photoreceptors and Visual Processes (20 minutes)
A. Types of Photoreceptor Cells
- Discussion of rod and cone cells in vertebrate vision
- Overview of opsins and their roles in phototransduction
B. Mechanisms of Light-Induced Responses
- Explanation of the photoisomerization process in rhodopsin and melanopsin
- Role of GPCRs in light-sensing and signal transduction

III. Cryptochrome and Magnetoreception (20 minutes)
A. Role of Cryptochrome in Circadian Regulation
- Overview of how cryptochrome functions in circadian light entrainment
- Discussion of the relationship between cryptochrome and magnetoreception
B. Radical Pair Model of Magnetoreception
- Explanation of how cryptochrome generates reactive radical pairs
- Insights into the experimental evidence supporting the radical pair model

IV. Quantum Effects in Light Sensing (20 minutes)
A. Spin Chemistry and Quantum Dynamics
- Overview of the significance of electron spin in biological processes
- Discussion of how quantum coherence influences radical pair dynamics
B. Implications for Biological Function
- Examination of how magnetic fields affect biological responses
- Overview of the RPM and its relevance to animal magnetoreception

V. Current Research and Future Directions (10 minutes)
A. Challenges in Understanding Magnetoreception
- Overview of ongoing debates and challenges in the field
- Discussion of alternative mechanisms for magnetoreception
B. Potential Applications of Quantum Biology
- Exploration of implications for biotechnology and health sciences
- Importance of understanding these effects in biological systems

VI. Conclusion and Q & A (5 minutes)
A. Summary of Key Concepts
B. Open floor for student questions

**Class 7 – Dynamic Control of DNA Repair in Photolyase**

I. Introduction to Photolyase and DNA Repair (15 minutes)
A. Overview of Photolyase
- Definition and function of photolyases as UV and blue-light photoreceptor proteins
- Importance of DNA repair in cellular processes
B. Mechanisms of DNA Repair
- Explanation of how photolyase repairs DNA using light energy
- Role of radical pair dynamics in the repair process

II. Radical Pair Dynamics in Photolyase (20 minutes)
A. The Role of Magnetic Fields
- Overview of how magnetic fields influence radical pair dynamics during DNA repair
- Discussion of magnetochemical control in photolyase activity
B. Photo-Activation and Electron Transfer
- Explanation of the electron transfer process initiated by blue light
- Description of the radical pair formation and its significance in DNA repair

III. Structural and Functional Dynamics of Photolyase (20 minutes)
A. Key Steps in DNA Repair Mechanism
- Detailed overview of the catalytic steps in photolyase activity
- Discussion of the role of flavin adenine dinucleotide (FAD) and tryptophan residues
B. Experimental Techniques
- Overview of spectroscopic methods used to study photolyase dynamics
- Insights from time-resolved absorption spectroscopy and molecular dynamics simulations

IV. Quantum Effects in DNA Repair (20 minutes)
A. Quantum Spin Dynamics and Biological Processes
- Explanation of how quantum spin dynamics influence enzymatic reactions
- Discussion of the relationship between electron spin and metabolic processes
B. Implications for Disease Treatment
- Exploration of how understanding photolyase mechanisms can inform therapeutic applications
- Overview of potential treatments for DNA damage-related diseases

V. Future Directions in Quantum Biology Research (10 minutes)
A. Novel Applications of Photolyase Mechanisms
- Discussion of potential biotech and biomedical applications of photolyase
- Importance of quantum dynamics in future research and enzyme design
B. Challenges and Opportunities
- Overview of ongoing challenges in studying and applying photolyase mechanisms
- Exploration of new research avenues in quantum biology

VI. Conclusion and Q & A (5 minutes)
A. Summary of Key Concepts
B. Open floor for student questions

**Class 8 – Enzyme Catalysis: Quantum Fundamentals**

I. Introduction to Flavins and Enzyme Catalysis (15 minutes)
A. Overview of Flavins
- Definition and importance of flavin molecules in biological systems
- Role of flavins in various cellular processes (e.g., electron transfer, signaling)
B. Quantum Effects in Enzyme Catalysis
- Explanation of electron bifurcation and its significance in energy transduction
- Introduction to the role of quantum effects in enzyme function

II. Mechanisms of Electron Bifurcation (20 minutes)
A. The Process of Electron Bifurcation
- Detailed explanation of how enzymes utilize electron bifurcation
- Role of flavins in facilitating electron transfer pathways
B. Quantum Dynamics and Radical Pair Formation
- Overview of radical pair dynamics in flavin-containing proteins
- Discussion of how magnetic fields influence bifurcation processes

III. Photophysical Properties of Flavins (20 minutes)
A. Structural Characteristics of Flavins
- Description of the isoalloxazine ring structure and its electronic properties
- Role of flavin in photo-absorption and radical pair generation
B. Experimental Techniques in Flavin Research
- Overview of computational methods and spectroscopic techniques used to study flavins
- Insights gained from time-resolved spectroscopy and molecular dynamics simulations

IV. Quantum Control in Biological Processes (20 minutes)
A. Quantum Effects on Cell Function
- Discussion of how quantum dynamics influence cellular redox states and signaling pathways
- Examples of quantum control in metabolic processes and energy generation
B. Implications for Biomedicine
- Exploration of potential applications of flavin-based enzymes in medical treatments
- Discussion of how understanding quantum mechanisms can lead to advancements in biotechnology

V. Future Directions in Quantum Biology Research (10 minutes)
A. Challenges in Understanding Flavin Biophysics
- Overview of current research gaps and unanswered questions in flavin studies
- Discussion of the need for improved modeling techniques in quantum biology
B. New Avenues for Investigation
- Exploration of how advancements in quantum biology can impact various scientific fields
- Potential for developing new biomolecular systems leveraging quantum properties

VI. Conclusion and Q & A (5 minutes)
A. Summary of Key Concepts
B. Open floor for student questions

**Class 9 – Ultraweak Photon Emission & Collective Cellular Behavior**

I. Introduction to the Role of Light in Biology (15 minutes)
A. Historical Context
- Ancient treatments using sunlight
- Development of modern photobiology
B. Importance of Light in Regulating Biological Processes
- Circadian rhythms
- Metabolic rates
- Cell growth

II. Types of Light and Their Biological Effects (20 minutes)
A. Classification of Light
- Visible, Infrared (IR), and Ultraviolet (UV)
B. Beneficial Effects of Infrared Light
- Mitochondrial function and ATP production
C. Visible Light and Its Impact on Health
- Effects of artificial light on oxidative stress and chronic diseases
D. UV Light: Risks and Benefits
- Cancer induction vs. moderate beneficial effects

III. Historical Discoveries in Light and Life (15 minutes)
  A. Gurwitsch's Experiments and Ultraweak Photon Emission (UPE)
    - Introduction to biophotons and UPE
    - Mitogenetic effect and cell replication
  B. Influential Figures in Phototherapy
    - Finsen's contributions to modern phototherapy
    - Development of heliotherapy by Rollier

IV. Quantum Effects of Light in Biological Systems (20 minutes)
  A. Quantum Mechanics and Biology
    - Bohr's lecture on light therapy and its implications
    - Schrödinger's influence on genetics and life
  B. Current Research Trends
    - UPE in different life forms
    - The relationship between UPE, oxidative stress, and cellular functions

V. Collective Behavior and Cellular Interactions (15 minutes)
  A. Microtubules and Photodynamics
    - Role of microtubules in cellular structure and function
    - Effects of UV and IR light on microtubule dynamics
  B. Long-range Electrodynamic Interactions
    - Cellular responses to near-infrared light (NIR)
    - Implications for cancer research and therapy

VI. Applications and Future Directions (5 minutes)
  A. Medical Diagnostics and Therapeutics
    - UPE as a tool for monitoring cellular health
  B. Biotechnological Innovations
    - Potential for manipulating cytoskeleton dynamics in various fields

VII. Conclusion and Q & A (5 minutes)
  A. Summary of Key Points
  B. Open floor for student questions

**Class 10 - Electromagnetic Oscillations in Living Systems**

I. Introduction to Microtubules and Their Significance (15 minutes)
  A. Overview of Microtubules
    - Structure and composition of microtubules
    - Role in the cytoskeleton: interaction with actin and intermediate filaments
  B. Mechanical and Electrical Properties
    - Ion transport and charge storage capabilities
    - Signal amplification and electromagnetic computing potential

II. Dynamic Instability of Microtubules (20 minutes)
  A. Concepts of Dynamic Instability
    - Mechanisms of self-assembly and depolymerization
    - Role of GTP hydrolysis in dynamic instability
  B. Cellular Functions of Microtubule Dynamics
    - Importance in cell motility, mitosis, and differentiation
    - Impact on tissue homeostasis and disease (e.g., carcinogenesis)

III. Microtubules as Biological Electromagnetic Systems (20 minutes)
  A. Electrical Oscillations in Microtubules
    - Spontaneous electrical oscillations and their implications
    - Microtubules as biological transistors for signal amplification
  B. Interaction with Electromagnetic Fields
    - Effects of oscillating electric fields on microtubule dynamics
    - Coupling of microtubule oscillations with cellular electric fields

IV. Quantum Optical Properties of Microtubules (20 minutes)
A. Cavity Quantum Electrodynamics (cQED) Model
- Microtubules as quantum optical cavities
- Mechanisms for generating coherent quantum states
B. Refractive Index Controversies
- Discrepancies in the refractive index of tubulin
- Implications for electromagnetic mode confinement and excitonic interactions

V. Exciton-Polaron Interactions and Their Biological Relevance (10 minutes)
A. Formation of Exciton-Polaron Pairs
- Mechanisms and effects on microtubule properties
- Implications for structural dynamics and transport properties
B. Enhancing Microtubule Behavior Through Electromagnetic Fields
- Applications of MHz-band oscillating fields in microtubule growth

VI. Applications and Future Research Directions (5 minutes)
A. Potential Applications in Biotechnology and Medicine
- Modifying reactivity and conductivity in biological systems
B. Challenges and Opportunities

VII. Conclusion and Q & A (5 minutes)
A. Summary of Key Concepts
B. Open floor for student questions

**Class 11 – Chemical Control in Biological Systems**

I. Introduction to Mitochondrial Functions in Cellular Signaling (15 minutes)
A. Overview of Mitochondrial Roles
- Integration of metabolism, redox chemistry, and apoptosis
- Mitochondria as critical cell signaling hubs
B. Mitochondrial Dysregulation in Cancer
- Structural and dynamic changes during carcinogenesis
- Impact on metabolic regulation and energy transduction

II. Mitochondrial Dynamics and Reactive Oxygen Species (ROS) (20 minutes)
A. Mitochondrial ROS as Oncogens
- The role of ROS in cancer metabolism and the Warburg effect
- Consequences of aberrant mitochondrial morphology
B. ROS and Cellular Stress Responses
- Effects of external factors (e.g., magnetic fields) on ROS levels
- Mechanisms of high ROS stress in cancer cells

III. The Role of GTP Hydrolysis and Microtubules in Cellular Processes (15 minutes)
A. GTP Hydrolysis and Microtubule Dynamics
- Mechanisms of microtubule self-organization and dynamics
- Influence of microtubules on calcium signaling and ROS dynamics
B. Feedback Loops in Signal Transduction
- Interplay between microtubules, calcium channels, and ROS

IV. Electromagnetic Fields and Their Biological Effects (15 minutes)
A. Effects of Magnetic Fields on Cellular Function
- Exploration of magnetic resonance and its relation to calcium oscillations
- Evidence from studies on low-frequency magnetic fields
B. Stochastic Resonance in Ion Channels
- Role of ion channels in amplifying signals and magnetic field interactions

V. The Interconnectedness of ROS, Calcium, and Cellular Health (15 minutes)
  A. Calcium Homeostasis and Mitochondrial Function
    - Importance of calcium in cellular signaling and mitochondrial dynamics
    - Mechanisms of calcium-induced calcium release and implications for health
  B. ROS Regulation and Disease Associations
    - Connection between ROS dysregulation and various diseases
    - Therapeutic potential of modulating ROS levels

VI. Therapeutic Implications and Future Directions (5 minutes)
  A. Targeting Mitochondrial Function in Cancer Therapy
    - Mitochondria-targeted therapeutic strategies
  B. Challenges in ROS Amplification for Treatment
    - Balancing efficacy and side effects in therapeutic interventions

VII. Conclusion and Q & A (5 minutes)
  A. Summary of Key Concepts
  B. Open floor for student questions

**Class 12 – Magnetic Biomodulation: Biodynamic Control**

I. Introduction to Magnetic Fields and Animal Behavior (15 minutes)
  A. Overview of Animal Magnetoreception
    - Examples of species exhibiting magnetic field perception (e.g., birds, insects, mammals)
    - Importance of Earth's magnetic field in animal migration and behavior
  B. Physiological Effects of Geomagnetic Field Deprivation
    - Impact on neurogenesis, cognition, and developmental processes

II. Mechanisms of Magnetic Field Influence on Biological Systems (20 minutes)
  A. Effects of Hypomagnetic Fields (HMFs)
    - Impairment of nerve growth and cognition in mammals
    - Role of reactive oxygen species (ROS) in mediating these effects
  B. Proposed Mechanisms of Magnetic Field Effects (MFEs)
    - Radical pair mechanism (RPM), voltage-gated ion channels, and kinase signaling pathways
    - The significance of flavoprotein photoreceptors in ROS generation

III. Magnetic Fields and Cellular Metabolism (20 minutes)
  A. Impact on Redox Metabolism
    - Effects of magnetic fields on cancer-like metabolism in cells
    - Balance between ROS production and metabolic processes
  B. Wound Healing and Tissue Repair
    - Influence of magnetic fields on ROS and calcium ion concentrations
    - Applications of magnetic fields in regenerative medicine

IV. Experimental Findings on Magnetic Field Effects (20 minutes)
  A. Studies on Planarians and Fibroblasts
    - Effects of static and oscillatory magnetic fields on regeneration and collagen production
    - Autophagy responses in fibroblasts exposed to oscillating magnetic fields
  B. Neutrophil ROS Production and Metabolic Resonance
    - Synchronization of NADPH oscillations with pulsed magnetic fields
    - Regulation of nitric oxide and ROS production in neutrophils

V. Quantum Effects in Biological Systems (10 minutes)
  A. Microtubules as Quantum Electrodynamical Cavities
    - Role of microtubules in cognition and consciousness
    - Potential for quantum information processing in biological systems
  B. Interaction of Magnetic Fields with Light
    - Magneto-optic interactions and their implications for ROS modulation and cell growth

VI. Implications for Health and Therapy (5 minutes)
A. Potential Applications of Magnetotherapy
- Overview of current research and therapies involving magnetic fields
B. Considerations for Future Research
- Need for unified theories and mechanistic understanding of MFEs

VII. Conclusion and Q & A (5 minutes)
A. Summary of Key Concepts
B. Open floor for student questions

**Assessment:** Mid Term Exam

**Class 13 – Intermolecular Forces: Solvent Effects and Dispersion**

I. Introduction to Collective Synchronization in Biological Systems (15 minutes)
A. Overview of Collective Behavior in Cells
- Coordination of molecular activities: differentiation, mobilization, motility, and replication
- Emergent phenomena and spontaneous synchronization in biology
B. The Problem of Protein Folding
- Challenges in understanding protein folding dynamics
- Role of long-range quantum interactions in life processes

II. Density Functional Theory (DFT) in Quantum Chemistry (20 minutes)
A. Historical Context and Development
- Overview of Hohenberg and Kohn's work on electronic structures
- Kohn-Sham equations and their implications for solving many-body problems
B. Limitations of Conventional DFT
- Challenges in accounting for long-range electron correlations
- The need for effective classical potentials to model quantum effects

III. Non-Covalent Interactions and Their Biological Importance (20 minutes)
A. Types of Non-Covalent Interactions
- Overview of van der Waals, hydrogen bonds, and $\pi$–$\pi$ interactions
- Their roles in molecular assembly and biological processes
B. Solvation Effects and Molecular Dynamics
- Impact of solvent environments on biomolecular behavior
- Role of hydrogen-bonding networks in aqueous solutions

IV. Quantum Effects in Biological Systems (15 minutes)
A. Dispersion Forces and Their Role in Biomolecules
- Importance of dispersion forces in protein folding and enzyme pathways
- Examples of dispersion forces in DNA and RNA interactions
B. Quantum Transport in Nucleic Acids
- Charge transport mechanisms in DNA and RNA
- Implications for DNA repair and integrity

V. Noise and Signal Amplification in Biological Systems (10 minutes)
A. Role of Noise in Biological Decision-Making
- Noise as a facilitator for detecting weak signals
- Examples of stochastic resonance in cellular contexts
B. Microtubules and Noise-Driven Processes
- Microtubules as responsive structures to biochemical and mechanical stress
- Implications for learning and memory

VI. Calcium Signaling and Mitochondrial Function (5 minutes)
A. Importance of Calcium Ions in Cellular Processes
- Role of calcium in signaling pathways and homeostasis
- Interaction between calcium signaling and mitochondrial function

B. Advances in Light-Based Calcium Regulation
- Photosensitizer-based mechanisms for controlling calcium levels
- Implications for cellular homeostasis and signaling

VII. Conclusion and Q & A (5 minutes)
A. Summary of Key Concepts
B. Open floor for student questions

**Class 14 – Multiscale Modeling of Biomolecular Systems**

I. Introduction to Quantum Mechanics in Biological Systems (15 minutes)
A. Overview of Quantum Effects in Biology
- Historical context: "warm, wet, and noisy" systems
- Misconceptions about quantum mechanics and biological coherence
B. Schrödinger's Contributions
- The central dogma of molecular biology and the order-disorder problem
- The importance of thermodynamic free energy in living systems

II. Classical and Quantum Approaches to Biological Modeling (20 minutes)
A. The Role of Classical Molecular Mechanics (MM)
- Overview of MM simulations in studying molecular conformations
- Applications to proteins, nucleic acids, and macromolecular complexes
B. Challenges of Traditional DFT Methods
- Limitations of DFT in accounting for long-range interactions
- The need for classical potentials and their construction

III. Multiscale Modeling Techniques (20 minutes)
A. Introduction to QM/MM Methods
- Combining quantum mechanics with classical mechanics
- Applications in enzymatic reactions and charge transfer
B. Historical Development of QM/MM Techniques
- Key milestones in the development of multiscale models
- Recognition of pioneers in the field (Karplus, Levitt, Warshel)

IV. Applications of Multiscale Modeling in Biochemistry (20 minutes)
A. Case Studies in Enzymatic Reactions
- Mechanistic insights from QM/MM simulations (e.g., lysozyme reaction)
- Impact of electrostatics on catalytic efficiency
B. Photodynamics and Quantum Processes
- Investigating vitamin D photosynthesis and phototropin photocycle
- Role of nonadiabatic processes in biological systems

V. Challenges and Future Directions in Multiscale Modeling (10 minutes)
A. Limitations of Current QM/MM Approaches
- Issues with strong quantum correlations and non-local effects
- Need for advanced computational methods and techniques
B. Emerging Techniques and Machine Learning Applications
- Potential of machine learning in enzyme engineering and drug discovery
- Multi-level embedding strategies for enhanced simulations

VI. Conclusion and Q & A (5 minutes)
A. Summary of Key Concepts
B. Open floor for student questions

**Main Assignment #2:** Student presentation on quantum biology topic of interest to be given before final exam study period begins.

**Class 15 – Quantum Correlations in Biological Cofactors**

I. Introduction to Quantum Effects in Biological Systems (10 minutes)
  A. Overview of Cellular Metabolism and Quantum Mechanics
    - Historical context: Advances in understanding cellular processes
    - Misconceptions about the relevance of quantum effects in biology
  B. Quantum Information Processing Applications
    - Use of pulsed electron paramagnetic resonance (EPR) in DNA studies
    - Quantum synchronization in DNA–enzyme interactions

II. Role of Metal Ions in Biological Functions (20 minutes)
  A. Importance of Iron and Metal Clusters
    - Functions of FeS clusters in DNA repair and electron transport
    - Overview of heme proteins and their versatile roles
  B. Quantum Mechanics in Metal-Containing Enzymes
    - Influence of quantum correlations on enzyme behavior
    - Challenges in studying iron-containing enzymes

III. Quantum Mechanical Models and Their Applications (25 minutes)
  A. Approaches to Modeling Quantum Effects
    - Limitations of DFT and local density approximation (LDA)
    - Importance of correlated electronic effects in metalloproteins
  B. Advanced Techniques in Quantum Chemistry
    - Overview of methods like DFT+DMFT and their applications
    - Examples of modeling enzymatic reactions and electron transfer

IV. Challenges in Studying Quantum Dynamics (15 minutes)
  A. Limitations of Conventional QM/MM Methods
    - Issues with multi-reference methods and quantum correlations
    - Need for sophisticated techniques to model complex biochemical systems
  B. Emerging Computational Strategies
    - Hybrid quantum/classical computing approaches
    - Applications in drug discovery and materials science

V. Quantum Dynamics in Biological Processes (15 minutes)
  A. Relevance of Quantum Dynamics in Biochemistry
    - Importance in photosynthesis, light-harvesting, and photocatalysis
    - Nonadiabatic dynamics and their implications for molecular reactions
  B. Insights from Recent Quantum Simulations
    - Examples of trapped ion quantum computers in studying molecular dynamics
    - Exploration of photoinduced processes and their significance

VI. Conclusion and Q & A (5 minutes)
  A. Summary of Key Concepts
  B. Open floor for student questions

**Class 16 – Photobiomodulation & Electromagnetic Therapies**

I. Introduction to Light-Based Therapy (15 minutes)
  A. Historical Context of Phototherapy
    - Finsen's contributions to light therapy and early applications
    - Rediscovery of red light therapy in the mid-1960s
  B. Overview of Photobiomodulation (PBM)
    - Definition and significance of PBM in medical science
    - Transition from low-level laser therapy (LLLT) to PBM terminology

II. Mechanisms of Photobiomodulation (20 minutes)
  A. Photophysical Forces and Their Effects
    - Overview of how light at specific frequencies influences biological processes
    - Action spectrum for DNA and RNA synthesis
  B. Mitochondrial Function and Retrograde Signaling
    - Role of mitochondria in PBM and energy metabolism
    - Mechanisms of signaling from mitochondria to the nucleus

III. Applications of PBM in Health (20 minutes)
  A. Therapeutic Effects on Musculoskeletal and Neurological Conditions
    - Use of PBM for pain reduction, inflammation, and tissue regeneration
    - Effects on neurological diseases such as stroke and Parkinson's disease
  B. PBM and Cancer Treatment
    - Potential applications of PBM in cancer therapy and chemotherapy-related neuropathies
    - Evidence of PBM effects on cancer cell metabolism and growth

IV. Electromagnetic Therapies and Biological Responses (20 minutes)
  A. Overview of Electromagnetic Field (EMF) Therapies
    - Introduction to pulsed electromagnetic field (PEMF) therapy and its applications
    - Mechanisms by which EMFs influence cellular functions
  B. Research on Electromagnetic Fields and Health
    - Effects of extremely low frequency (ELF) EMFs on wound healing and inflammation
    - Potential benefits and risks associated with EMF exposure

V. Future Directions and Challenges in PBM and EMF Research (10 minutes)
  A. Need for Standardization of Treatment Parameters
    - Challenges in defining optimal PBM protocols
    - Importance of understanding physiological mechanisms behind PBM and EMF
  B. Emerging Research Areas
    - Exploration of new therapeutic applications and technologies
    - Integration of PBM and EMF with existing treatment modalities

VI. Conclusion and Q & A (5 minutes)
  A. Summary of Key Concepts
  B. Open floor for student questions

**Class 17 – Photodynamic Therapy & Nanotheranostics**

I. Introduction to Photodynamic Therapy (PDT) (15 minutes)
  A. Historical Context of PDT
    - Early photochemical transformations and the role of photosensitizers
    - Key developments in PDT from the 20th century to present
  B. Mechanism of Action
    - Overview of how PDT generates reactive oxygen species (ROS)
    - Role of light and photosensitizers in targeting pathological tissues

II. Photosensitizers and Their Applications (20 minutes)
  A. Types of Photosensitizers
    - Flavins, porphyrins, and their roles in biological systems
    - Development of hematoporphyrin derivatives and their clinical significance
  B. Clinical Applications of PDT
    - Use in oncology, dermatology, and treatment of infections
    - Overview of therapeutic effects and patient tolerance

III. Nanomedicine and Its Integration with PDT (20 minutes)
  A. Concept of Nanomedicine
    - Definition and significance of nanoparticles in medical applications
    - Role of nanotechnology in enhancing PDT effectiveness

B. Emerging Nanotherapeutic Approaches
- Overview of chemo dynamic therapy (CDT) and its mechanisms
- Introduction to photothermal therapy (PTT) and its applications

IV. Mechanisms of Action in Nanomedicine (15 minutes)
A. Interaction of Nanoparticles with Biological Systems
- Mechanisms of ROS generation and cellular responses
- Importance of targeted delivery and specificity in treatment
B. Examples of Nanomaterials in Therapy
- Functionalized magnetic nanoparticles and their applications
- Combination therapies and their potential benefits

V. Advanced Techniques in PDT and Nanomedicine (15 minutes)
A. Innovations in Photodynamic and Nanotherapeutic Approaches
- Use of optogenetics and light-activated molecules in therapy
- Development of nanotheranostic systems for diagnostics and treatment
B. Challenges and Future Directions
- Limitations in current PDT and nanomedicine practices
- Need for further research into mechanisms and treatment parameters

VI. Conclusion and Q & A (5 minutes)
A. Summary of Key Concepts
B. Open floor for student questions

**Class 18 – Regenerative Mechanisms: Cells, Tissues, & Organs**

I. Introduction to Circadian Rhythms and Physiological Regulation (15 minutes)
A. Overview of Diurnal Cycles
- Influence of light and darkness on physiology and behavior
- Importance of circadian rhythms in biological functions
B. Role of Light in Physiological Processes
- Effects of light intensity, duration, and color on cellular functions
- Introduction to chronobiology

II. Photobiomodulation (PBM) and Regeneration (20 minutes)
A. Mechanisms of PBM
- Overview of PBM therapy and its regenerative effects on cells
- Role of ROS in PBM and tissue healing
B. Applications of PBM in Stem Cell Therapy
- Enhancing stem cell proliferation and differentiation
- Potential for improving outcomes in tissue engineering

III. Stem Cell Engineering and Tissue Regeneration (20 minutes)
A. Overview of Stem Cell Therapies
- Techniques for isolating and directing stem cell differentiation
- Challenges and risks associated with stem cell treatments
B. Innovations in Tissue Engineering
- Use of biomaterials and scaffolds for tissue repair
- Advances in creating lab-grown tissues for clinical applications

IV. Challenges in Cartilage and Bone Regeneration (15 minutes)
A. Difficulties in Cartilage Healing
- Characteristics of articular cartilage and its regenerative limitations
- Overview of current therapeutic interventions
B. Advances in Osteochondral Tissue Engineering
- Strategies for effective cartilage regeneration
- Role of scaffolds and biocompatible materials in repair

V. Fibrin Scaffolds and Their Applications (15 minutes)
A. Role of Fibrin in Wound Healing
- Mechanisms of fibrin formation and its role in tissue repair
- Applications of fibrin hydrogels in biomedical fields
B. Integration of PBM with Fibrin Scaffolds
- Enhancing cell viability and function within hydrogel systems
- Potential for optimizing outcomes in regenerative therapies

VI. Conclusion and Q & A (5 minutes)
A. Summary of Key Concepts
B. Open floor for student questions

**Class 19 – Morphogenetic Integration & Immunology**

I. Introduction to Cell Behavior and Environmental Integration (15 minutes)
A. Overview of Cellular Responses to Environmental Cues
- Importance of cell behaviors in development, homeostasis, and regeneration
- Role of physico-chemical phenomena in cellular dynamics
B. Mechanotransduction and its Biological Significance
- Impact of mechanical forces on cellular processes
- Importance of the extracellular matrix (ECM) in cell behavior

II. Regenerative Mechanisms and Stem Cell Therapy (20 minutes)
A. Overview of Regeneration in Multicellular Organisms
- Comparison of regenerative capacities across species
- Focus on planaria and their unique regenerative capabilities
B. Applications of Stem Cell Therapy
- Mechanisms of stem cell differentiation and proliferation
- Challenges and risks associated with stem cell treatments

III. Role of Reactive Oxygen Species (ROS) in Regeneration (20 minutes)
A. Mechanistic Insights into ROS Production
- Role of ROS in tissue growth and healing
- Evidence from zebrafish and salamander regeneration studies
B. Influence of Electromagnetic Fields on ROS Dynamics
- Effects of electromagnetic fields on cellular regeneration processes
- Research findings on electromagnetic radiation and tissue repair

IV. Cell Migration and Its Importance in Physiology (15 minutes)
A. Mechanisms of Cell Migration
- Overview of chemotaxis and mechanotaxis
- Role of the cytoskeleton in migration and morphology
B. Implications for Development and Disease
- Relationship between cell migration and cancer metastasis
- Influence of the ECM on cell migration dynamics

V. Immune Responses and Tissue Regeneration (15 minutes)
A. Role of the Immune System in Healing
- Overview of immune cell behavior during inflammation and tissue repair
- Interaction between immune cells and tissue microenvironments
B. The Fibroblastic Reticular Cell Network
- Structure and function of reticular cell networks in lymphoid organs
- Implications for tissue engineering and organ grafts

VI. Conclusion and Q & A (5 minutes)
A. Summary of Key Concepts
B. Open floor for student questions

**Class 20 – Quantum Biotechnology: Universal Applications**

I. Introduction to Hormesis and Dose-Response Relationships (15 minutes)
A. Overview of Hormesis
- Definition and historical context of hormesis in biology and medicine
- Examples of hormesis in vitamin $D_3$ and other biological systems
B. Importance of Biphasic Dose-Response Curves
- Comparison with linear dose-response models
- Implications for treatments such as photobiomodulation (PBM)

II. Quantum Effects in Biological Systems (20 minutes)
A. Quantum Transport Mechanisms
- Introduction to coherence and decoherence in biological systems
- Applications in energy and charge transport in photosynthesis
B. Role of Photosynthetic Systems
- Mechanisms of non-photochemical quenching and energy dissipation
- Importance of excitonic delocalization in artificial light-harvesting systems

III. Advances in Photocatalysis and Nanotechnology (20 minutes)
A. Light-Driven Redox Catalysis
- Overview of photocatalytic methods for organic compound synthesis
- Emerging techniques for clean energy production
B. Single-Atom Photocatalysts
- Benefits of single-atom catalysts in enhancing selectivity and efficiency
- Comparison with traditional nanoparticle systems

IV. Applications of Metal-Porphyrin Complexes (15 minutes)
A. Biological Significance of Metal-Porphyrin Complexes
- Roles in electron transfer and enzyme catalysis
- Applications in biomedicine and catalysis
B. Development of Artificial Metalloenzymes
- Importance and applications of synthetic metal–porphyrin complexes

V. Innovations in Synthetic Biology (15 minutes)
A. Overview of Synthetic Biology Approaches
- Comparison of "top-down" and "bottom-up" strategies
- Applications in molecular computing and programmable matter
B. Potential of DNA-Based Systems
- Use of DNA for data storage and computational applications
- Advances in programmable DNA for therapeutic purposes

VI. Conclusion and Q & A (5 minutes)
A. Summary of Key Concepts
B. Open floor for student questions

**Class 21 – Essential Further Research**

I. Importance of Research in Quantum Biology (5 minutes)
A. Discussion of how quantum effects influence various biological phenomena
- The challenge of bridging quantum and classical biology

II. Overview of Quantum Effects in Biological Processes (10 minutes)
A. Key Quantum Effects
- Overview of electron transfer, photochemical reactions, molecular interactions, and the role of van der Waals forces in biological systems
B. Quantum Coherence and Decoherence
- Explanation of quantum coherence in biological systems and its implications
- Discussion of decoherence processes and their significance in living organisms

III. Research Area: Electron Transport in Living Systems (20 minutes)
  A. Biological Electron Transfer Mechanisms
    - Overview of oxidative metabolism and electron transport chains
    - Importance of chromophores in electron transfer processes
  B. Proton Transfer and Quantum Mechanics
    - Discussion of how proton transfer accompanies electron transfer in biological reactions
    - Implications for modeling complex biochemical systems

IV. Research Area: Photonic and Quantum Optical Effects (20 minutes)
  A. Biophotons and Cellular Function
    - Review of ultraweak photon emissions and their biological significance
    - Role of light-matter interactions in cellular signaling and physiology
  B. Applications in Medical Research
    - Overview of photobiomodulation and photodynamic therapy
    - The emerging role of optogenetic techniques in biological research

V. Challenges and Future Directions in Quantum Biology Research (30 minutes)
  A. Current Research Gaps
    - Identification of unresolved questions in quantum biology and its implications for health
    - Discussion of the need for interdisciplinary approaches in research
  B. Future Research Opportunities
    - Exploration of potential applications in nanomedicine and biotechnology
    - The importance of understanding quantum mechanisms for developing innovative treatments

VI. Conclusion and Q & A (5 minutes)
  A. Summary of Key Takeaways
  B. Open Floor for Student Questions

**Class 22 – Important Considerations for the Future of Quantum Biology**

I. The Quantum-Classical Relationship (5 minutes)
  A. Overview of Quantum Theory
    - Definition and significance of quantum biology in understanding living systems
    - Discussion of the historical context and evolution of quantum theory
  B. The Quantum-Classical Distinction
    - Examination of the measurement problem in quantum mechanics
    - Overview of the classical measuring devices and their implications for quantum theory

II. Review of Quantum Effects in Biological Processes (15 minutes)
  A. Key Quantum Phenomena
    - Discussion of tunneling, electron transfer, and light-harvesting in biological systems
    - Examples of nontrivial quantum effects and their relevance to biology
  B. Revising the Physical Perspective
    - Argument for understanding biological processes from a quantum perspective
    - Importance of moving beyond classical approximations to capture true quantum behavior

III. Theoretical Framework for Quantum Biology (30 minutes)
  A. Non-Hermitian Hamiltonian Formalism
    - Explanation of the framework developed by Fano and Feshbach
    - Discussion of how this framework accounts for environmental influences on open quantum systems
  B. Incorporating Quantum Dynamics
    - Overview of dissipation, external driving, and relaxation in the context of biological systems
    - Implications for understanding the complex dynamics of life

IV. Interdisciplinary Nature of Quantum Biology (15 minutes)
A. Active Matter Systems
- Review of the concept of active matter and its relevance to quantum biology
- Discussion of symmetry-breaking effects and their significance in living organisms
B. Collaboration Across Disciplines
- Importance of integrating knowledge from biology, physics, chemistry, and medical science
- Challenges in developing a common language and methods in quantum biology research

V. Ethical Considerations and Future Directions (15 minutes)
A. Ethical Framework for Quantum Biology
- Discussion of the potential societal and ethical challenges associated with emerging technologies
- Importance of establishing bioethical guidelines for research and applications
B. Future Research Opportunities
- Exploration of the potential implications of quantum biology for biomedicine and bioengineering
- Encouragement for students to engage with the emerging possibilities in the field

VI. Discussion (10 minutes)
A. Discussion on the Implications of Quantum Biology
- Group discussion on how quantum biology might change our understanding of life and technology

**Assessment:** Final Exam